\documentclass[aps,prb,twocolumn,english,showpacs,superscriptaddress,
amsfonts,nofootinbib]{revtex4-2}

 \usepackage{framed}
\usepackage{graphicx}
\usepackage{epsfig}
\usepackage{color}
\usepackage{amsmath}
\usepackage{amsfonts}
\usepackage{mathrsfs}
\usepackage{amsthm}
\usepackage{txfonts}
\usepackage{subfig}
\usepackage{ulem}
\usepackage{comment}
\usepackage{tabularx}
\usepackage{makecell}
\usepackage{algorithm}
\usepackage{listings}

\usepackage[breaklinks=true]{hyperref} 
\hypersetup{colorlinks=true, linkcolor=blue, citecolor=blue, filecolor=blue, urlcolor=blue}

\newcommand{\bea}{\begin{eqnarray}}
\newcommand{\eea}{\end{eqnarray}}

\newcommand{\trace}[1]{\text{Tr}~ \left[ #1 \right] }

\newcommand{\bpm}{\begin{pmatrix}}
\newcommand{\epm}{\end{pmatrix}}

\newcommand{\kket}[1]{| #1 \rangle\!\rangle}
\newcommand{\bbra}[1]{\langle\!\langle #1 |}
\newcommand{\brakket}[2]{\left\langle\!\langle #1 | #2 \right\rangle\!\rangle}
\newcommand{\Hom}{\text{Hom}}
\newtheorem{theorem}{Proposition}
\theoremstyle{empty}
\begin{document}

\title{Nonequilibrium topological response under charge dephasing}
	\author{Shuangyuan Lu}
    \author{Lucas Q Silveira}
 \author{Yizhi You}
\affiliation{Department of Physics, Northeastern University, Boston, MA, 02115, USA}
\date{\today}

\begin{abstract}
We explore nonequilibrium topological responses of symmetry-protected topological (SPT) states in open quantum systems subject to decoherence. For SPT wavefunctions protected by a product symmetry $G\times S$, where $G$ defects are decorated with $S$ charge, we show that local dephasing of the $S$ charge density generically induces spontaneous strong-to-weak symmetry breaking (SWSSB) of $G$ in the resulting mixed-state ensemble. We extend this mechanism to SPT phases protected by higher-form and spatially modulated symmetries, and further to gapless SPT states, demonstrating that dephasing-induced SWSSB persists well beyond conventional gapped $0$-form settings. Our results provide a qualitative, channel-defined fingerprint of SPT order that is intrinsic to open-system dynamics and goes beyond equilibrium linear response.
\end{abstract}
\maketitle

\section{Motivation}

Topological phases of matter \cite{wen03}, distinguished by long-range entanglement in many-body wavefunctions, have been a central focus of condensed matter physics over the past few decades. Symmetry can further enrich this landscape, giving rise to \textit{symmetry-protected topological} (SPT) phases \cite{chen11a,chen11b,chen2011two,Chen:2011pg,pollmann2012a,pollmann2012symmetry,else2014classifying}, which exhibit exotic boundary phenomena and quantized topological responses.
To date, most studies of SPT phases have emphasized ground-state properties and linear response in closed, thermal-equilibrium systems. In realistic settings, however, unavoidable coupling to an environment leads to decoherence and dissipation, motivating an extension of topological concepts to open quantum systems \cite{lee2023quantum,ma2023topological,sarma2025effective,sala2025entanglement,wang2024anomaly,wang2023intrinsic,xu2024average,you2024intrinsic,zang2024detecting}. This viewpoint has spurred growing interest in symmetry-protected topology beyond equilibrium, including dynamics under noisy quantum channels \cite{gu2024spontaneous,guo2024new,guo2025strong,moharramipour2024symmetry,salo_steady_state_strong_sym_2025,liu2025parent,lu2025holographic}, statistical ensembles generated by quenched disorder, and the characterization and classification of mixed-state SPT phases in nonequilibrium settings \cite{bao2023mixed,fan2023diagnostics,LeeYouXu2022,zhang2022strange,lessa2024mixedstate,huang2025interaction,ding2024boundary,hsin2024anomalies,sala2024spontaneous,kawabata2024lieb,zhou2025reviving,you2024intrinsic,zang2024detecting,wang2024anomaly,sun2024holographic,hsin2024anomalies,gu2024spontaneous,sala2025entanglement,lu2025holographic}.

In open system settings, an equally important question is how a wavefunction evolves under decoherence. Can one define a \textit{topological response} that captures how an initial SPT wavefunction reacts to local noise \cite{ma2022average,lee2025symmetry,sarma2025effective,zhang2022strange, ma2024topological,guo2025strong,ando2024gauge,zhang2025probing,sala2025entanglement,murciano2023measurement,garratt2023measurements}? Concretely, under the \textit{same} local quantum channel, do SPT wavefunctions evolve into mixed states with salient features that are qualitatively distinct from those decohered from a trivial symmetric state, and can such distinction be used to distinguish different SPT phases? 
Meanwhile, since SPT wavefunctions can serve as resource states for quantum simulator and measurement-based quantum computation, it is natural to ask which of these capabilities persist in the presence of decoherence \cite{briegel2009measurement,lee2022measurement,murciano2023measurement,raussendorf2003measurement,Stephen_2017}. 

In this work, we scrutinize the topological response of an SPT wavefunction subject to dephasing noise, focusing on the following claim:
\begin{theorem}
For an SPT wavefunction protected by a $G\times S$ symmetry, where $G$-defects are decorated by $S$ charge, applying a decoherence channel that measures (or dephases) the local $S$-charge density generically drives strong-to-weak spontaneous symmetry breaking (SWSSB) of the $G$ symmetry in the resulting mixed state.\end{theorem}
Meanwhile, the remaining strong $S$ symmetry, together with $G$ (now realized only as a weak symmetry due to SWSSB), can still protect a nontrivial mixed-state SPT structure of the mixed-state ensemble \cite{guo2025strong,ando2024gauge}.
We interpret this phenomenon as an out-of-equilibrium topological response that distinguishes an SPT wavefunction from a trivial symmetric state.\footnote{Unlike equilibrium linear response, where one perturbs the system infinitesimally and measures the induced change, here the response is defined by how the initial wavefunction is transformed into a mixed ensemble under a strong charge-dephasing channel, and by the qualitative features of that resulting state.}
Once the $S$-charge density is strongly dephased, the $G$ symmetry generically becomes vulnerable to SWSSB: $G$ charge can effectively tunnel into the environment, so $G$ survives only as a weak symmetry of the mixed-state dynamics.
This behavior contrasts sharply with the trivial case. Starting from a trivial paramagnet, dephasing a local charge density typically produces a featureless mixed-state ensemble. The qualitative difference in how an SPT wavefunction responds to a decohering noise channel therefore provides a natural fingerprint of SPT order in nonequilibrium settings.

Notably, while SWSSB is broadly observed in open-system dynamics including generic quantum channels or Lindblad evolutions, its occurrence and mechanism can depend strongly on both the initial state and the type of noise applied \cite{Lee_2023,lessa2025,guo2025strong,song2025strong,zhang2024strong,song2025strong,sala2024spontaneous}. In particular, if we start from a trivial state that is $G\times S$ symmetric, SWSSB of $G$ typically arises only when the noise channel explicitly enables $G$-charge fluctuations, such as $G$-charge pair creation/annihilation. By contrast, a dephasing channel that measures only the local charge density does not, on its own, trigger SWSSB when the initial state lies in the trivial disordered phase.

\subsection{Decoherence-induced SWSSB}

For mixed-state density matrices, it is useful to distinguish strong (exact) symmetry from weak (average) symmetry. A strong symmetry acts on one side of the density matrix and leaves it invariant (up to a global phase):
\begin{align}
U(g)\rho=e^{i\theta}\rho.
\end{align}
This implies that every pure-state component of $\rho$ is an eigenstate of $U(g)$ with the same eigenvalue. By contrast, a weak symmetry preserves the density matrix only when applied on both the left and right:
\begin{align}
U(g)\rho\neq e^{i\theta}\rho,\quad U(g)\rho U(g)^\dagger=\rho.
\end{align}
Equivalently, $\rho$ can be brought to a block-diagonal form in a basis adapted to the weak symmetry, with each block carrying a definite symmetry charge (which may differ across blocks).

A mixed quantum state can exhibit spontaneous strong-to-weak symmetry breaking (SWSSB), where a \textit{strong} symmetry is spontaneously reduced to a \textit{weak} symmetry.
To formalize this, we use the Choi-Jamiołkowski isomorphism, which maps a density matrix to a bilayer pure state in the doubled Hilbert space:
\bea \label{eq:choidou}
\mathcal{J}: \rho \rightarrow \kket{\rho} = \sum_{i, j} \rho_{ij} ~|i\rangle_1 \otimes|j\rangle_2
\eea
where $|\rangle_1$ and $|\rangle_2$ denote the ket and bra indices of $\rho$, respectively. In this doubled space, a strong symmetry can be perceived as $G_1 \times G_2$ symmetry, with $G_1$ ($G_2$) acting only on the upper (lower) layer. Equivalently, each layer carries its own conserved $G$ charge. SWSSB occurs when the doubled state $\kket{\rho}$ spontaneously breaks $G_1 \times G_2$ down to its diagonal subgroup $G$: the bilayer state remains invariant under the diagonal action, while charge is allowed to tunnel between the two layers.

To diagnose this SWSSB, we consider the spatial correlation between operators that are charged under $G_1 \times G_2$ but neutral under the diagonal subgroup $G$,
\bea \label{eq:dou}
C(r, r^\prime) = \frac{\bbra{\rho} O_{1,r} \otimes O_{2,r}^* \left(O_{1,r^\prime} \otimes O^*_{2,r^\prime}\right)^\dagger \kket{\rho}}{ \brakket{\rho}{\rho}}
\eea
where $O$ is an operator charged under the $G$ symmetry. If $C(r,r^\prime)$ exhibits long-range order as $|r-r^\prime|\to\infty$, it indicates that the strong symmetry $G_1 \times G_2$ is spontaneously broken to its diagonal subgroup, i.e., SWSSB occurs.
The doubled-space correlator in Eq.~\eqref{eq:dou} admits an equivalent expression directly in terms of the physical density matrix $\rho$, which can be written as a R\'enyi-2 correlator \cite{lee2023quantum,lee2025}:
\bea\label{eq.renyi_2}
R^{(2)}(r, r^\prime) = \frac{\trace {\rho O_r O_{r^\prime}^\dagger \rho O_{r^\prime}O_r^\dagger} }{\trace {\rho^2}}
\eea
Meanwhile, to confirm that the remaining \textit{weak} symmetry is not broken, we also examine the usual charge correlation of the mixed state $\trace {\rho O_{r^\prime}O_r^\dagger}$ , which should decay superpolynomially with spatial separation.

Besides the R\'enyi-2 correlator, a more universal and robust diagnostic of SWSSB is the fidelity correlator \cite{lessa2025}:
\bea \label{eq:fid}
F(r, r^\prime) = \trace { \sqrt{\sqrt{\rho}~ O_r O^\dagger_{r^\prime} \rho O_{r^\prime} O_r^\dagger \sqrt{\rho}}}
\eea
In the SWSSB phase, the fidelity correlator approaches a finite value in the long-distance limit, whereas it vanishes in phases without SWSSB. The fidelity correlator is often viewed as a more genuine probe because it is stable under local quantum channels: Long-range ordered fidelity correlator implies that the mixed state is not locally recoverable, corresponding to a divergent Markov length \cite{sang2023mixed,sang2025mixed,zhang2024strong} and non-vanishing conditional mutual information (CMI). Nonetheless, many numerical studies still focus on the R\'enyi-2 correlator, whose main advantage is computational convenience, since it is quadratic in the density matrix. In this work, we will primarily use the R\'enyi-2 correlator, and later present detailed CMI data for our model to further substantiate the presence of SWSSB.

\subsection{Dephasing the SPT Wavefunction}
In this work, we study local quantum channels (equivalently, finite-time Lindblad evolutions) that dephase an initial pure state into an ensemble of mixed states. We focus on a class of dynamics in open quantum systems in which the jump operators in the Lindblad master equation (or, equivalently, the Kraus operators of the channel) are built from local charge densities $s_i$ (with $s_i$ being the onsite charge density of the symmetry group $S$). Concretely, we consider the dephasing (measurement) channel
\bea \label{eq:chargeden}
\mathcal{E}(\rho)=(1-2p) ~\rho + 2p \sum_{\{s_i\}} \hat{P}(\{s_i\} )\rho \hat{P}(\{s_i\} )  .
\eea
Here $\hat{P}(\{s_i\})$ is a projector onto a specified configuration of local charge patterns $\{s_i\}$, and $2p$ is a measurement rate that interpolates between projective measurement ($p=1/2$) and no measurement ($p=0$).
This channel can be viewed as a weak measurement of the local charges on all sites, performed without post-selection: the measurement outcomes are not recorded, and one averages over them. Because the channel only dephases in the $s_i$ basis, it admits an extensive number of degenerate steady states (in particular, any state diagonal in that basis is stationary). Consequently, the steady state reached under repeated application of $\mathcal{E}$ can depend strongly on the initial state.

If we start from a trivial symmetric wavefunction in which all charges are localized, applying the charge-dephasing channel in Eq.~\eqref{eq:chargeden} does not generate any nontrivial structure: the evolution simply produces a featureless mixed state. The situation can change qualitatively for more exotic initial wavefunctions. Previous work has shown that, starting from a fermionic ground state with a Fermi surface or a Dirac semimetal structure, dephasing of the fermion charge density effectively implements a Gutzwiller-type projection in the Choi-doubled description~\cite{su2025spin}. As a result, the corresponding Choi-doubled wavefunction can be interpreted as a spin-liquid state in the doubled Hilbert space.

Of particular interest, we examine how charge-dephasing channels act on a class of SPT wavefunctions protected by a $G\times S$ symmetry, in which $S$ defects are decorated by $G$ charge \cite{chen2014,you2016decorated,li2024,ma2023average,SpectralSequence}. Due to this defect-charge decoration structure, adding dephasing noise that measures the local $S$ charge (equivalently, proliferating $S$ defects in the Choi-doubled description) simultaneously drives $G$-charge-pair condensation in the doubled Hilbert space, resulting in SWSSB.
To make this connection explicit, we first recall that for an SPT wavefunction $|\phi\rangle$, if we project (equivalently, measure with postselection) the local $S$ charge on every site onto a fixed configuration of eigenvalues $\{s_i\}$, then the postselected state $P_{\{s_i\}}|\phi\rangle$ generically exhibits spontaneous breaking of the $G$ symmetry, signaled by long-range order in $G$-charged observables. This mechanism has also been exploited for quantum state preparation in measurement-feedback circuits \cite{lu2022measurement,verresen2021prediction,verresen2021efficiently,tantivasadakarn2024,lee2022}: measuring the $S$ charge in an SPT wavefunction (which itself can be prepared by a shallow circuit) can drive the post-measurement state into a $G$-symmetry-broken phase with long-range correlations.
If, instead, we do not record the measurement outcomes and simply average over them, the resulting mixed state is a classical mixture of symmetry-broken sectors with random relative orientations. Consequently, conventional $G$-charged correlators vanish after averaging over measurement outcomes, but a glassy long-range order persists. This glassy order can be diagnosed by an Edwards-Anderson type correlator or, equivalently, by a R\'enyi-2 correlator.

In the remainder of the paper, we test and sharpen this principle through a sequence of explicit settings, including 1d bosonic SPT chains (Sec.~\ref{sec:1d_zxz}--\ref{sec:haldane}) and a 2d quantum spin Hall state in Sec.~\ref{sec:qsh}. We then broaden the scope to charge dephasing in higher-form SPTs (Sec.~\ref{sec:higherform}) and in intrinsically gapless SPTs (Sec.~\ref{sec:gapless}). Finally, in Sec.~\ref{sec:general}, we provide a general discussion of the mechanism underlying SWSSB in SPT wavefunctions, including an MPS-based argument and connections to decorated domain-wall structures.

\section{ZXZ Cluster Model}
\label{sec:1d_zxz}

A paradigmatic example we begin with is the one-dimensional SPT chain protected by $\mathbb{Z}_2\times \mathbb{Z}_2$ symmetry, whose ground state admits a simple decorated domain-wall structure: domain walls of $\mathbb{Z}^a_2$ are bound to charges of the other $\mathbb{Z}^b_2$. A microscopic realization of this construction is provided by the 1d cluster (ZXZ) Hamiltonian \cite{chen2014,you2016decorated,li2024,ma2023average,SpectralSequence},
\bea\label{eq.hamiltonian_zxz}
\hat{H} = -\sum_{i} Z_{i-1} X_i Z_{i+1} - h \sum_{i} X_i
\eea
The $\mathbb{Z}_2 \times \mathbb{Z}_2$ symmetry is generated by independent spin flips on the even and odd sublattices:
\begin{equation}\label{eq.zxz_symmetry}
\begin{aligned}
\mathbb{Z}^a_2 &:~ \prod_i X_{2i} \\
\mathbb{Z}^b_2 &:~ \prod_i X_{2i+1}
\end{aligned}
\end{equation}
When $h=0$, the Hamiltonian admits an exactly solvable ground state with a transparent decorated-domain-wall interpretation: it is an equal weight superposition of all possible $\mathbb{Z}_2^{a}$ domain-wall configurations, where each domain wall is decorated with a $\mathbb{Z}_2^{b}$ charge. Turning on the transverse field $h$ interpolates between the SPT phase and a trivial paramagnet via a continuous quantum phase transition \cite{zonzo2018}.
The bulk gap closes at the critical point $h_c=1$: for $h<1$ the ground state realizes the $\mathbb{Z}_2\times\mathbb{Z}_2$ SPT state, whereas for $h>1$ it is adiabatically connected to a trivial product state.

Although the SPT phase is fully symmetric and admits no local order parameter, it can be sharply distinguished from the trivial phase by a nonlocal string order parameter. For the ZXZ cluster model, this order parameter is defined as \cite{pollmann2012symmetry}:
\begin{align}
    &O_{\text{str}} (r, r^\prime) = \langle \psi | \hat{O}_{\text{str}} |\psi \rangle \nonumber\\
    &\hat{O}_{\text{str}} = Z_{2r-1} \left( \prod_{i = r} ^ {r^\prime} X_{2i}\right) Z_{2r^\prime+1} \label{eq:zxz_string}
\end{align}

In the SPT phase with $h<1$, the string order approaches a finite value at large separation, while in the trivial phase with $h>1$ it vanishes in the thermodynamic limit. Thus, the nonlocal string order parameter serves as the diagnostic distinguishing the $\mathbb{Z}_2 \times \mathbb{Z}_2$ SPT phase from the trivial product state. Notably, this string order reflects the decorated domain-wall structure. The $\mathbb{Z}_2^{a}$ charge string can be viewed as a domain-wall creation operator that creates $\mathbb{Z}_2^{a}$ domain walls at the two string endpoints, and these endpoints necessarily carry a $\mathbb{Z}_2^{b}$ charge.

Starting from the ground state $|\psi(h)\rangle$ of the Hamiltonian in Eq.~\eqref{eq.hamiltonian_zxz}, we apply a local charge-dephasing channel that measures the $\mathbb{Z}_2^{a}$ charge density on the even sublattice,
\bea \label{eq.zxz_mixed_state}
\rho(h) = \mathcal{E}[\rho_0(h)], \qquad \rho_0(h)=|\psi(h)\rangle\langle \psi(h)| .
\eea
The channel factorizes into onsite maps acting on the even sites,
\bea\label{eq.zxz_channel}
\mathcal{E} = \prod_{i} \mathcal{E}_{2i}, \qquad
\mathcal{E}_{2i}[\rho] = \frac{1}{2}\Big(\rho + X_{2i} \rho X_{2i}\Big),
\eea
so each $\mathcal{E}_{2i}$ measures the local $\mathbb{Z}_2^{a}$ charge density on site $2i$ without recording the outcome, thereby fully dephasing coherences between different $X_{2i}$ eigenstates.
Such a channel $\mathcal{E}$ preserves the strong $\mathbb{Z}^a_2 \times \mathbb{Z}^b_2$ symmetry.

In what follows, we show that when the initial state $|\psi(h)\rangle$ lies in the SPT phase ($h<1$), dephasing the local $\mathbb{Z}_2^{a}$ charge generically induces SWSSB of the $\mathbb{Z}_2^{b}$ symmetry.
To characterize the SWSSB, we evaluate the R\'enyi-2 correlator:

\bea\label{eq.zxz_renyi_2}
R^{(2)}(r, r^\prime) = \frac{\trace{~\rho Z_{2r-1} Z_{2r^\prime+1}\rho Z_{2r -1} Z_{2r^\prime + 1} }}{\trace{~\rho^2}}
\eea

\begin{figure}[h]
    \centering
    \includegraphics[width=0.8\linewidth]{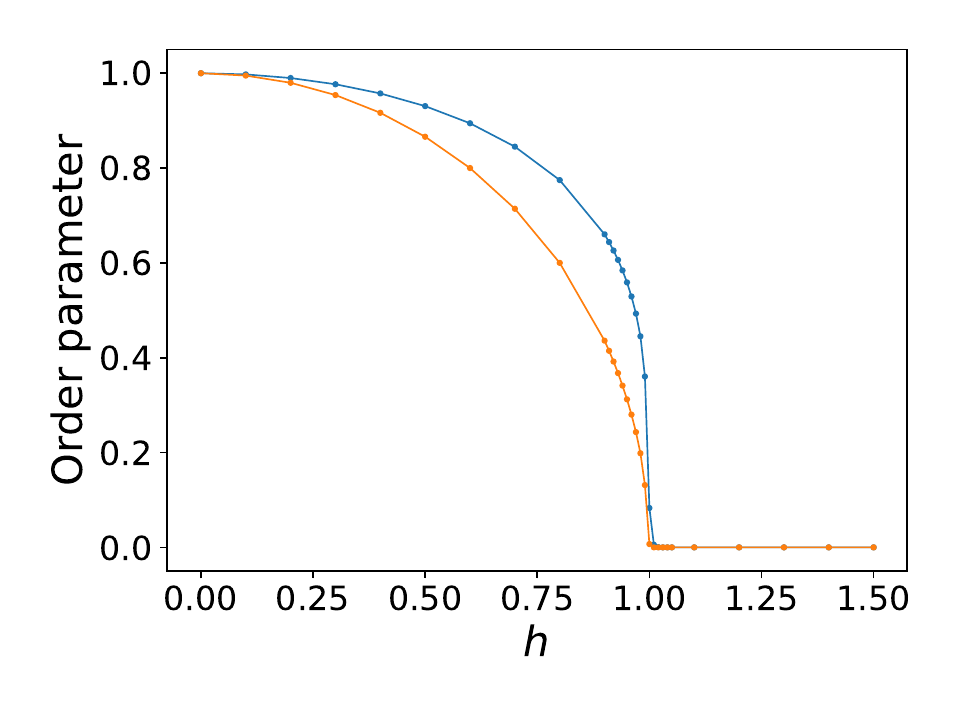}
    \caption{String order parameter $O_{\text{str}}(r, r^\prime)$ (blue) for ZXZ model ground state $|\psi (h)\rangle$ and R\'enyi-2 correlator $R^{(2)}(r, r^\prime)$ (orange) for the decohered mixed state $\rho(h)$ \eqref{eq.zxz_mixed_state} versus magnetic field strength $h$ . DMRG parameters: Open boundary condition, system size $L = 1000$, bond dimension $\chi=100$, up to 100 sweeps. $2r = L / 4, 2r^\prime = 3L/4$.}
    \label{fig:zxz_correlator}
\end{figure}

Using DMRG, we compute the ground state $|\psi(h)\rangle$ of the cluster Hamiltonian with a varying transverse field as Eq.~\eqref{eq.hamiltonian_zxz}, focusing on (i) the string order parameter of the ground state $|\psi(h)\rangle$ before decoherence and (ii) the R\'enyi-2 correlator of the corresponding mixed state $\rho(h)$ obtained by applying the dephasing channel in Eq.~\eqref{eq.zxz_channel}. We pay particular attention to the vicinity of the critical point $h=1$ (see Fig.~\ref{fig:zxz_correlator}).
As expected, the string order parameter of the pure state vanishes at $h=1$. In addition, we find that the R\'enyi-2 correlator remains finite throughout the SPT regime $h<1$, but vanishes for $h>1$. This shows that, provided the initial wavefunction is in the SPT phase, dephasing the local $\mathbb{Z}_2^{a}$ charge generically induces SWSSB of the $\mathbb{Z}_2^{b}$ symmetry.

For a trivial $\mathbb{Z}_2 \times \mathbb{Z}_2$ paramagnet, dephasing the local $\mathbb{Z}_2^{a}$ charge has little impact on the other symmetry. In contrast, in the SPT phase the two symmetries are nontrivially intertwined: dephasing the $\mathbb{Z}_2^{a}$ charge generically triggers SWSSB of the $\mathbb{Z}_2^{b}$ symmetry.
The emergence of SWSSB can be understood intuitively by relating the R\'enyi-2 correlator of the mixed state to the string order parameter in Eq.~\eqref{eq:zxz_string}. The dephasing channel projects $\rho$ onto the subspace satisfying
$X_{2i}\rho X_{2i}=\rho$,
so the R\'enyi-2 correlator can be rewritten as a correlator of two string operators:
\bea
R^{(2)}(r, r^\prime) = \frac{\trace{~\rho ~\hat{O}_\text{str} ~\rho ~ \hat{O}_\text{str}}}{\trace{~\rho^2}}
\eea
Moreover, the string operator acting on a density matrix commutes with the quantum channel:
\bea
\mathcal{E}[\hat{O}_\text{str} ~\rho ~\hat{O}_\text{str}] = \hat{O}_\text{str} ~\mathcal{E} [\rho] ~\hat{O}_\text{str}. 
\eea
Hence the string operators can be moved inside the channel, giving
\bea
R^{(2)}(r, r^\prime) = \frac{\trace{~\mathcal{E}[|\psi \rangle \langle \psi|] ~\mathcal{E}[\hat{O}_\text{str} ~ |\psi \rangle \langle \psi|~ \hat{O}_\text{str}]}}{\trace{~\mathcal{E}[|\psi \rangle \langle \psi|]^2}}
\eea
Since the string order parameter is nonvanishing for the initial SPT state \cite{pollmann2012a}, it follows that the R\'enyi-2 correlator $R^{(2)}(r,r^\prime)$ is also guaranteed to remain nonzero after dephasing.

While the mixed state exhibits SWSSB of the $\mathbb{Z}_2^{b}$ symmetry, its SPT character does not disappear entirely. After SWSSB, a weak $\mathbb{Z}_2^{b}$ symmetry remains on the odd sites, and together with the strong $\mathbb{Z}_2^{a}$ symmetry on the even sites it is still sufficient to protect a mixed-state SPT (mSPT) phase. The resulting mixed-state SPT order can be diagnosed using the strange correlator.

\begin{figure}
    \centering
    \includegraphics[width=0.8\linewidth]{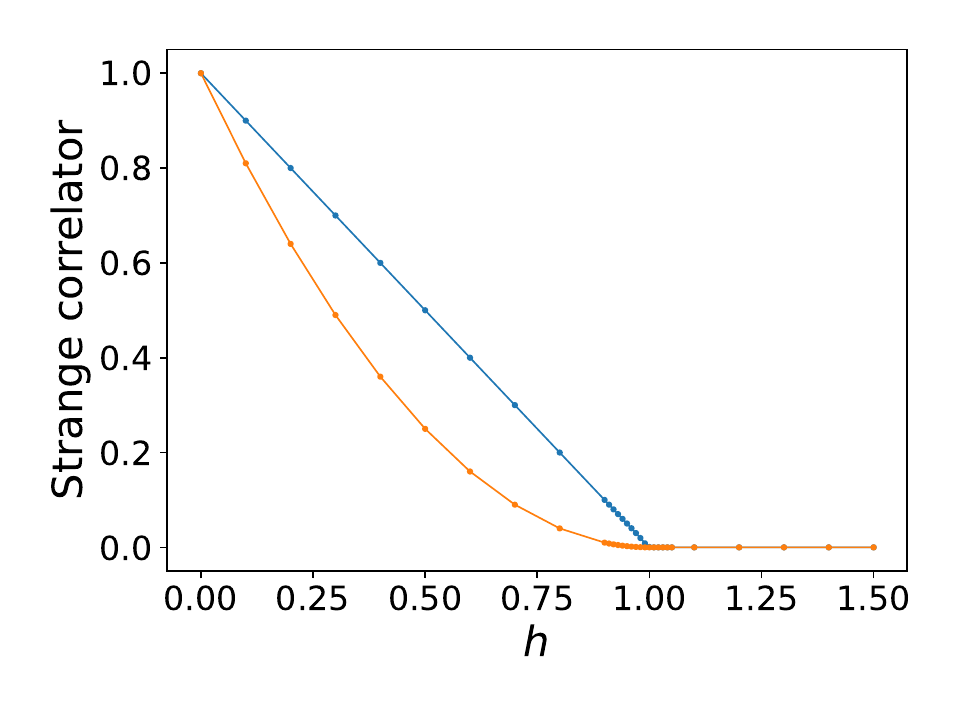}
    \caption{Strange correlator (blue) for ZXZ model ground state $|\psi(h)\rangle$ and type-II strange correlator for decohered state $\rho(h)$ \eqref{eq.zxz_mixed_state} versus field strength $h$. DMRG parameters: Open boundary condition, system size $L = 1000$, bond dimension $\chi=100$, up to 100 sweeps. $2r = L / 4, 2r^\prime = 3L/4$.}
    \label{fig:zxz_strange_correlator}
\end{figure}

Recall that a pure SPT state can be detected by the strange correlator \cite{you2014,zhang2022strange,lepori2023strange,lee2025symmetry,sala2025entanglement}:
\bea
C_{\text{strange}}(r, r^\prime) = \frac{\langle \psi_0| Z_{2r} Z_{2r^\prime} | \psi\rangle}{\langle \psi_0| \psi\rangle} 
\eea
where $|\psi_0\rangle$ is a symmetric trivial product state.
Similarly, for the mixed state $\rho$, we can define the type-II strange correlator introduced in Ref.~\cite{lee2025}:
\bea
C^{\text{II}}_{\text{strange}}(r, r^\prime) = \frac{\trace{~ \rho_0 Z_{2r} Z_{2r^\prime}~\rho~Z_{2r}Z_{2r^\prime}}}{\trace{ ~ \rho_0 \rho }}
\eea
Here $\rho_0$ denotes the density matrix of a trivial product state with all spins polarized along the $X$ direction, i.e., $X_i=1$ for every site. We choose the operator $Z_{2r}$ to be charged under the strong symmetry $\mathbb{Z}_2^{a}$, which is expected to exhibit LRO for the type-II strange correlator in the mSPT phase. We numerically evaluate (i) the strange correlator of the ground-state wavefunction $|\psi(h)\rangle$ before applying the channel and (ii) the type-II strange correlator of the mixed state $\rho(h)$ after decoherence. As shown in Fig.~\ref{fig:zxz_strange_correlator}, both correlators remain finite in the SPT phase ($h<1$) and vanish in the trivial phase ($h>1$).

\subsubsection{Conditional Mutual Information and Entanglement Negativity}\label{sec:entanglement}

In our previous discussion, we demonstrated that for a 1d SPT wavefunction with $\mathbb{Z}_2\times \mathbb{Z}_2$ symmetry, dephasing the $\mathbb{Z}_2^{a}$ charge density can trigger SWSSB of the $\mathbb{Z}_2^{b}$ symmetry. This effect can be viewed as an out-of-equilibrium topological response for SPT: a local charge-density measurement (without recording outcomes) induces SWSSB of the complementary symmetry. At the same time, this reveals a quantum phase transition in an open system under decoherence: under the same \(\mathbb{Z}_2^{a}\) dephasing channel, the resulting mixed state can be tuned from an SWSSB phase to a featureless phase by varying a parameter of the initial wavefunction.

We further investigate this transition from an information-theoretic perspective.
A defining feature of the SWSSB phase is nonlocal recoverability under local quantum channels: after applying an arbitrary local operation on a finite region $A$, its effect cannot be undone by any recovery channel supported on a finite neighborhood of $A$.
This failure of local recovery is quantified by a nonvanishing conditional mutual information (CMI) \cite{lessa2025,sang2023mixedstate} across a tripartition of the system into regions $A$, $B$, and $C$.
\bea
I(A: B|C) = S(AC) + S(BC) - S(C) - S(ABC)
\eea
where $S(R)$ denotes the von Neumann entropy of the reduced density matrix $\rho_R$.
CMI measures the amount of correlation shared between $A$ and $B$ after conditioning on region $C$ and is always positive.
In the SWSSB phase, the conditional mutual information $I(A : B | C)$ remains finite when $A$ and $B$ are widely separated, reflecting a divergent Markov length. Notably, for mixed states exhibiting long-range order in the fidelity correlator (Eq.~\eqref{eq:fid}), the CMI is necessarily nonvanishing. Thus, the CMI provides an information-theoretic diagnostic for mixed states exhibiting SWSSB.

\begin{figure}[h]
    \centering
    \includegraphics[width=0.48\linewidth]{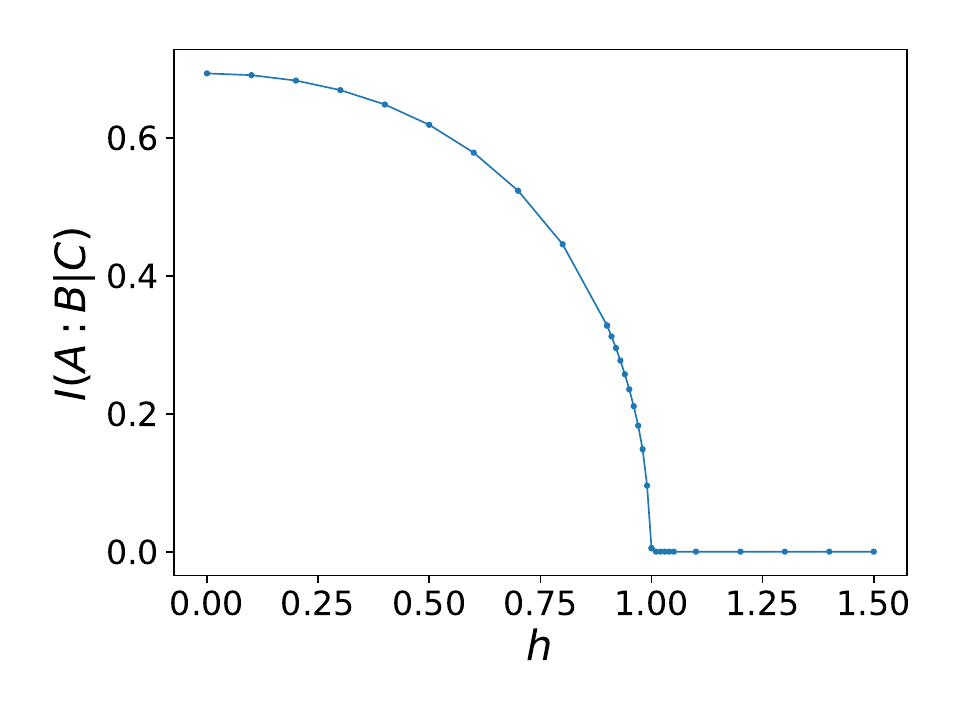}
    \includegraphics[width=0.48\linewidth]{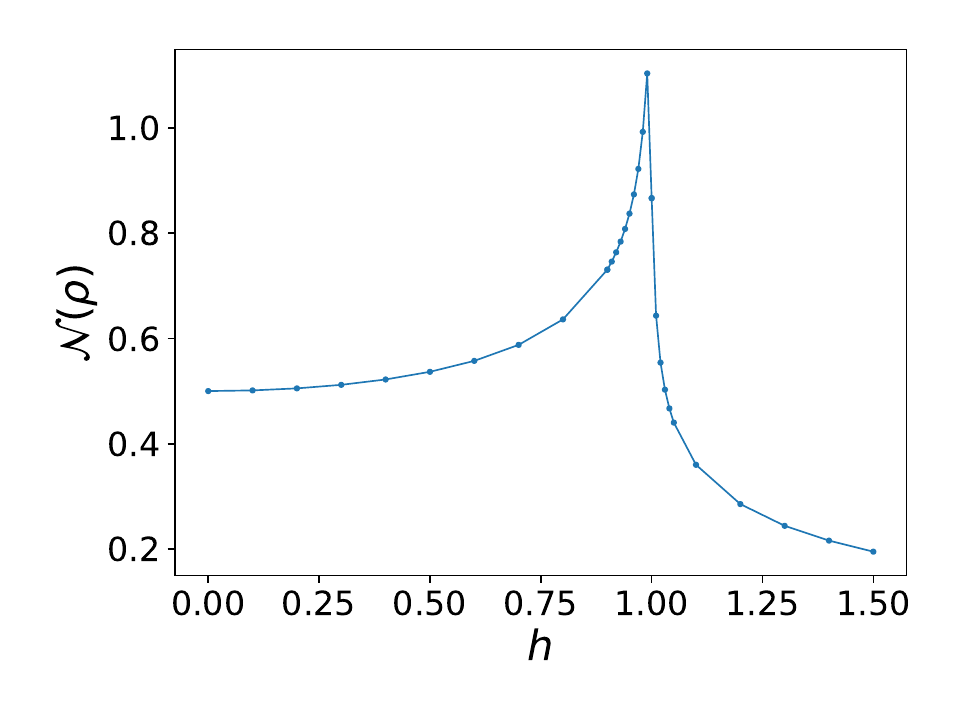}
    \caption{(a) CMI $I(A:B|C)$ and (b) Negativity $\mathcal{N}(\rho)$ for the decohered mixed state $\rho(h)$ \eqref{eq.zxz_mixed_state} versus field strength $h$. DMRG parameters: system size $L = 1000$, bond dimension $\chi=100$, up to 100 sweeps.}
    \label{fig:zxz_cmi_negativity}
\end{figure}

In our numerical simulation, we choose $A$ and $B$ to be the single sites at positions $L/4+1$ and $3L/4+1$ with $L/4$ even. The region $C$ is the remainder of the chain.
We compute the CMI of $\rho(h)$ obtained by applying the dephasing channel in Eq.~\eqref{eq.zxz_mixed_state} to the ground state of Eq.~\eqref{eq.hamiltonian_zxz} with periodic boundary conditions. As shown in Fig.~\ref{fig:zxz_cmi_negativity}(a), the CMI remains finite when the initial state lies in the SPT phase ($h<1$), indicating that $\rho(h)$ exhibits SWSSB. By contrast, for $h>1$, where the decohered pure state is in the trivial phase, the CMI vanishes.

We also evaluate the entanglement negativity of the mixed state $\rho(h)$ as an additional probe across the SWSSB transition. The negativity quantifies quantum entanglement in mixed states and is given by
\bea
\mathcal{N}(\rho) = \frac{||\rho^{T_A}||_1 -1 }{2}
\eea
where the chain is bipartitioned into left and right halves $A$ and $B$, and $T_A$ denotes the partial transpose on region $A$. $||\cdot||_1$ denotes the trace norm: $\|X\|_1 = \mathrm{Tr}\left[\sqrt{X^\dagger X}\right]$. Negativity provides a more practical way to detect entanglement in mixed states, since it is not affected by purely classical correlations.

In earlier studies of 1d mixed-state SPT phases, many mSPT density matrices can be written as convex mixtures of classical ensembles \cite{chen2023separability,ma2022average}, and consequently exhibit vanishing entanglement negativity. Likewise, for mixed states that display SWSSB \cite{chen2023symmetryenforced,sala2024spontaneous,lessa2024strong,lessa2024mixedstate,lu2025holographic}, the underlying ensemble can still be entirely unentangled, so the negativity can remain zero \footnote{Here we focus on global onsite symmetries. For SWSSB involving fermionic higher-form symmetries or non-onsite symmetries (e.g., CZX symmetry), the density matrix can still carry entanglement.}.

In the example of Eq.~\eqref{eq.zxz_mixed_state}, the coexistence of SWSSB and mSPT enforces a nonzero entanglement negativity.
We first confirm this numerically, as shown in Fig.~\ref{fig:zxz_cmi_negativity}. Specifically, we compute the entanglement negativity of the dephased state $\rho(h)$. For $h<1$, $\rho(h)$ lies in the SWSSB phase while simultaneously retaining mSPT features, and its entanglement negativity remains nonvanishing.
This nonzero negativity can be understood from the interplay between SWSSB and the decorated domain-wall structure. To make this explicit, we take $h=0$ as an example.
\bea
\rho(0)=\sum_{\{x_{2i}\}} (|\{x_{2i}\}\rangle \langle\{x_{2i}\}|)_{\text{even}}\otimes (|\text{GHZ}(\{x_{2i}\})\rangle\langle \text{GHZ}(\{x_{2i}\})|)_{\text{odd}}\nonumber\\
\eea
$\{x_{2i}\}$ denotes a specific configuration of even-site spins in the $X$ basis. The density matrix is an incoherent mixture over all even-spin product states $|\{x_{2i}\}\rangle$. For each configuration $\{x_{2i}\}$, the odd spins form a corresponding symmetric GHZ state constrained by $Z_{2i-1}Z_{2i+1}=x_{2i}$.
Accordingly, the density matrix is block-diagonal in the $\{x_{2i}\}$ basis: each block with fixed $x_{2i}$ pattern contains a long-range entangled state $|\mathrm{GHZ}(\{x_{2i}\})\rangle$ living on the odd sublattice. Since each $|\mathrm{GHZ}(\{x_{2i}\})\rangle$ has nonzero negativity, the total entanglement negativity is simply the sum of the negativity contributions from the individual diagonal blocks. Away from $h=0$, as long as the system remains in the regime where mSPT coexists with SWSSB ($h<1$), the strong dephasing ensures that the density matrix remains block-diagonal in the $\{x_{2i}\}$ basis. Moreover, within each block labeled by a fixed $\{x_{2i}\}$ pattern, the odd sublattice retains a long-range-ordered state inherited from the decorated-domain-wall construction (although it is no longer an exact GHZ state). Since each such block has nonvanishing negativity, the total entanglement negativity, obtained by summing the contributions from all blocks, remains nonzero. We will present the details of the entanglement negativity calculation in Appendix~\ref{app:cmi_negativity}.

\subsubsection{Physical Interpretation}

Physically, the mechanism by which local $G$-charge dephasing acting on an SPT wavefunction induces SWSSB is self-explanatory in the decorated domain-wall picture: a 1d SPT protected by $G\times S$ can often be viewed as a state in which domain walls of one symmetry are embedded by charges of the other. A key consequence is that if one measures the local $S$ charge and post-selects an outcome (or implements a feedback unitary based on the outcome), the resulting conditional wavefunction generically exhibits spontaneous symmetry breaking of $G$. This fact can be demonstrated directly from the MPS structure of 1d SPTs, which we will elaborate on in Sec.~\ref{sec:gene}. Meanwhile, it is worth noting that the R\'enyi-2–correlator diagnosis of SWSSB in the mixed state produced by dephasing an SPT wavefunction is always lower-bounded by the disorder-averaged strange correlator computed in the original (undephased) wavefunction; we defer the detailed proof to Appendix~\ref{sec:sc}.

In the ZXZ example introduced in Sec.~\ref{sec:1d_zxz}, measuring the even site $\mathbb{Z}_2^a$ charge and post-selecting the outcome $X_{2i}=1$ triggers the spins on the odd sublattice forming a GHZ state, thereby spontaneously breaking the $\mathbb{Z}_2^b$ symmetry. 
For the mixed state, implementing the strong dephasing channel of the local charge density on even sites in Eq.~\eqref{eq.zxz_channel} can be viewed, in the Choi-doubled space, as a projection onto $X^1_{2i}X^2_{2i}=1$, where $1,2$ label the two layers originating from the ket and bra indices. This projection can be interpreted as a domain-wall bound-state condensation: the $\mathbb{Z}_2^a$ domain walls on the first and second layers are forced to form an interlayer bound state, and these bound states then proliferate. Since this domain-wall bound state carries $\mathbb{Z}_2^b$ charge, condensing it inevitably triggers SWSSB. 
Notably, this mechanism is precisely what makes 1d SPT wavefunctions valuable resource states for measurement-based quantum computation (MBQC) \cite{lu2022measurement,lee2022measurement,raussendorf2003measurement,verresen2021efficiently}. It has also been exploited as a practical route to prepare long-range entangled states using measurement-and-feedback protocols, which have been implemented experimentally in a variety of quantum-simulation platforms \cite{piroli2021quantum,tantivasadakarn2021long,bravyi2022adaptive,tantivasadakarn2023hierarchy,iqbal2023creation,foss2023experimental,iqbal2023topological,satzinger2021realizing}.

Likewise, one can interpret this effect from the viewpoint of non-invertible symmetry \cite{seiberg2024majorana,seifnashri2024cluster,kim2025noninvertible}. For the ZXZ cluster state with $h=0$ in Sec.~\ref{sec:1d_zxz}, the model and its ground state realize a non-invertible symmetry relating the operators
\bea
Z_{2i}Z_{2i+2}\ \rightarrow\ X_{2i+1},\qquad
Z_{2i+1}Z_{2i-1}\ \rightarrow\ X_{2i}.
\eea
With this identification, implementing the dephasing channel that measures the charge $X_{2i}$ is effectively akin to dephasing the composite charge $Z_{2i+1}Z_{2i-1}$, which naturally tends to promote SWSSB.

So far, we have focused on SPT wavefunctions subject to strong dephasing channels, which are equivalent to measuring the local charge density and then averaging over (but not recording) the outcomes. A natural question is what happens if we start from the same state but instead apply a weak measurement (partial dephasing) channel,
\bea
\mathcal{E} = \prod_{i} \mathcal{E}_{2i}, \qquad
\mathcal{E}_{2i}[\rho] = (1-p)\rho + p X_{2i}\rho X_{2i}.
\eea
Here $p$ is the measurement rate that tunes the dephasing strength. The strong-measurement (fully dephasing) limit is recovered at $p=\tfrac{1}{2}$. 
For a 1d SPT wavefunction with a finite correlation length (i.e., the ground state of a gapped local Hamiltonian), we do not expect partial dephasing with $p<\tfrac{1}{2}$ to induce SWSSB; instead, SWSSB can only emerge at the singular `strong-measurement' point $p=\tfrac{1}{2}$, appearing as a crossover rather than a genuine transition for $p<\tfrac{1}{2}$.

The logic is as follows: A gapped 1d ground state can be prepared by imaginary-time evolution and therefore admits a Euclidean path-integral representation in $(1+1)$-d \cite{lee2023quantum,garratt2023measurements}:
\begin{align}
    |\phi\rangle \propto \lim_{\tau \to\infty} e^{-\tau H}|0\rangle
\end{align}
which is akin to the path integral on a semi-infinite strip in Euclidean time $\tau\in[0,\infty)$. The initial pure-state density matrix $\rho=|\phi\rangle\langle\phi|$ maps, in the Choi-double construction, to a bilayer Euclidean path integral with separate ket and bra sheets. A local dephasing channel leaves the interior of each sheet unchanged and instead couples the two layers only at the final time boundary $\tau=\infty$. Concretely, the quantum channel induces an effective interlayer boundary interaction localized at $\tau=\infty$, while the bulk region $(0<\tau<\infty)$ remains that of the original gapped theory.
Because the bulk is gapped, it can be integrated out to produce a local effective action for the boundary degrees of freedom at \(\tau=\infty\), i.e., a \(0+1\)d (quantum-mechanical) problem. In this regime, tuning \(p<\tfrac{1}{2}\) merely changes the strength of a local boundary coupling. But a \(0+1\)d theory cannot host a symmetry-breaking phase transition driven by any finite local boundary perturbation. Consequently, there is no SWSSB transition for \(p<\tfrac{1}{2}\); the qualitative change can occur only at the maximally dephasing point \(p=\tfrac{1}{2}\), where the boundary term becomes an exact projection.

This argument relies crucially on $|\phi\rangle$ being gapped. If $|\phi\rangle$ is critical, the $1+1$d bulk path integral is already critical, and it could generate effectively long-ranged correlations along the boundary \cite{lee2022measurement,garratt2023measurements,guo2025quantum}. In that case, varying $p$ can produce a bona fide boundary criticality and potentially drive SWSSB even for $p<\tfrac{1}{2}$. We will return to this point in Sec.~\ref{sec:gapless} for the dephasing effect in intrinsic gapless SPT.

\section{Decohere the 1d Haldane chain}\label{sec:haldane}

Similar to the ZXZ model, we consider another one-dimensional bosonic SPT: the spin-1 Heisenberg chain. The nearest-neighbor spin-1 Heisenberg model is well known to realize the Haldane phase, a nontrivial SPT protected by spin-rotation symmetry \cite{haldane1983, haldane1983a}. For our purposes, it is sufficient to retain only the discrete dihedral subgroup $D_2$, generated by $\pi$ rotations of the spins about the $x$, $y$, and $z$ axes, $\prod_i \exp(i\pi S_i^\alpha)$ with $\alpha=x,y,z$. We then introduce an on-site anisotropy term $(S_i^z)^2$, which tunes the system from the Haldane SPT phase into a trivial product phase \cite{hu2011}.
We begin with a microscopic Hamiltonian for the spin-1 chain:
\bea\label{eq.spin_1_hamiltonian}
\hat{H} = \sum_{i=1} \mathbf{S}_i \cdot \mathbf{S}_{i+1} + D \sum_{i=1} (S^z_i)^2
\eea
For small $D$, the ground state is in the Haldane phase, a nontrivial SPT protected by the $D_2 \cong \mathbb{Z}_2 \times \mathbb{Z}_2$ subgroup of spin rotations. As the on-site anisotropy $D$ is increased, the system undergoes a quantum phase transition into a trivial product state with $S_i^z=0$ on every site, occurring at $D \approx 0.968$ \cite{hu2011}.
This SPT phase can be distinguished from the trivial phase by a nonlocal string order parameter:
\bea
O_\text{str} = \langle \psi | S^x_{r-1} \left( \prod_{i=r}^{r^\prime} e^{i \pi S^x_i} \right) S^x_{r^\prime+1} | \psi \rangle
\eea
This order parameter takes a finite value in the Haldane phase and vanishes in the trivial phase, thus serving as the diagnostic for SPT order in the spin-1 chain.

Begin with the ground state of the Hamiltonian in Eq.~\eqref{eq.spin_1_hamiltonian}. We then apply a local dephasing channel that measures the on-site $\mathbb{Z}_2^a$ charge parity associated with a $\pi$ rotation about the $x$ axis. The resulting mixed state is
\bea\label{eq.spin_1_mixed_state}
\rho(D) = \mathcal{E}[|\psi(D)\rangle\langle\psi(D)|]
\eea
where the quantum channel is a product of on-site maps,
\bea
\mathcal{E} = \prod_i \mathcal{E}_i, \quad
\mathcal{E}_i[\rho] = \frac{1}{2}\left(\rho + e^{i \pi S^x_i} \rho e^{-i \pi S^x_i}\right).
\eea
Equivalently, $\mathcal{E}_i$ dephases $\rho$ in the eigenbasis of the $\pi$-rotation operator $e^{i\pi S^x_i}$. The full channel preserves the strong $D_2$ symmetry.

After dephasing, the mixed state $\rho$ can exhibit SWSSB of the $\mathbb{Z}_2^b$ symmetry generated by a $\pi$ rotation about the $z$ axis. We diagnose this using the R\'enyi-2 correlator:
\bea
R^{(2)}(r, r^\prime) = \frac{\trace{~\rho ~S^x_r S^x_{r^\prime}~\rho~ S^x_r S^x_{r^\prime}}}{\trace{~\rho^2}}
\eea
Here the operator $S^x$ is odd under $\prod_i e^{i\pi S^z_i}$, so long-range order in $R^{(2)}(r,r^\prime)$ signals SWSSB of $\mathbb{Z}_2^b$.

\begin{figure}[h]
    \centering
    \includegraphics[width=0.8\linewidth]{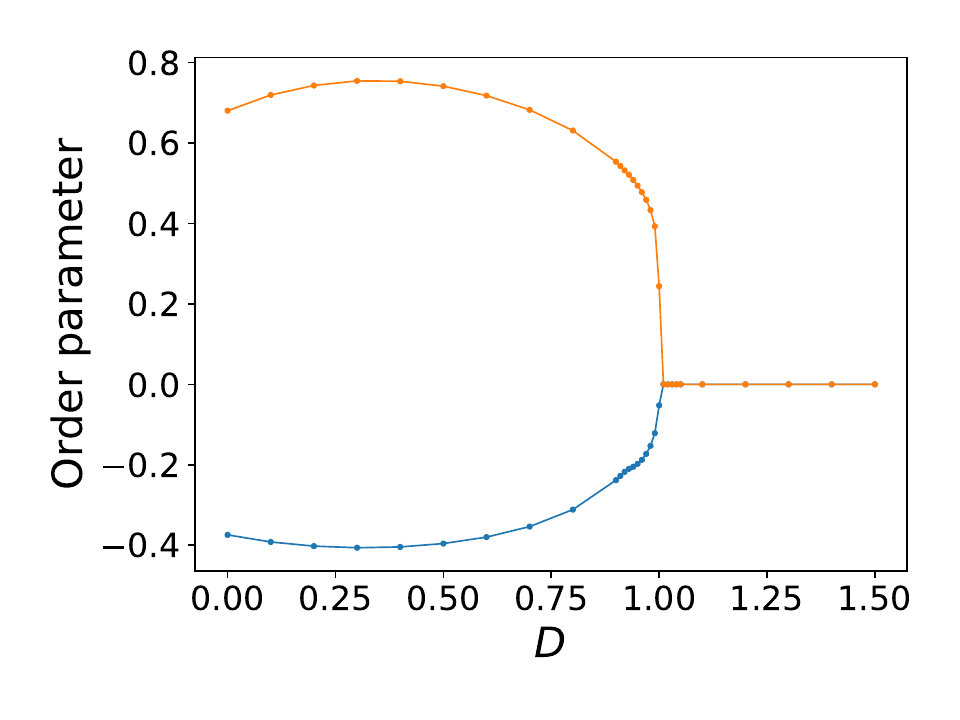}
    \caption{String order parameter (blue) for the ground state $|\psi(D)\rangle$ of the spin-1 model \eqref{eq.spin_1_hamiltonian} and R\'enyi-2 correlator (orange) for the decohered mixed state $\rho(D)$ in \eqref{eq.spin_1_mixed_state}. DMRG parameters: Open boundary condition, system size $L = 4000$, bond dimension $\chi=100$, up to 100 sweeps. $r = L / 4, r^\prime = 3L/4$.}
    \label{fig:spin_1_chain}
\end{figure}
Numerically, we use DMRG to obtain the ground state $|\psi(D)\rangle$ for varying anisotropy $D$ and then evaluate the correlator on the dephased state $\rho(D)=\mathcal{E}[|\psi\rangle\langle\psi|]$. Similar to the ZXZ model, we find that $R^{(2)}(r,r')$ develops long-range order whenever the initial pure state lies in the SPT phase. Fig.~\ref{fig:spin_1_chain} compares the pure-state string order parameter (before dephasing) with the R\'enyi-2 correlator of the mixed state (after dephasing).

\section{Charge dephasing of Quantum Spin Hall State}\label{sec:qsh}

After analyzing one-dimensional SPT states, we now turn to two dimensions and ask how local charge dephasing reshapes 2d SPT wavefunctions. The decorated-defect picture carries over directly: for SPT phases protected by a $G\times S$ symmetry, the nontrivial structure can be understood as $G$-symmetry defects decorated with $S$ charge.
 An essential qualitative difference, however, is that symmetry defects in higher dimensions come in multiple forms. In 2d, in addition to point-like defects (e.g., vortices), one can also consider extended defect objects such as line defects like domain-wall segments.
In addition, for a 2d SPT subject to partial charge dephasing, an SWSSB transition can occur already at a finite measurement rate $p$, rather than only in the fully dephasing limit. The intuition is that the Choi-double representation of a 2d wavefunction corresponds to a (2+1)d Euclidean path integral, while the dephasing acts as an interlayer coupling localized near the time boundary. Unlike the 1d case, where this boundary is effectively a (1+0)d problem, in 2d the boundary degrees of freedom form a genuine (2+0)d system, which can support symmetry-breaking transitions as the boundary coupling is tuned.

The first example we begin with is the quantum spin Hall (QSH) state, or equivalently the bosonic quantum Hall bilayer state. The QSH phase is protected by charge conservation $U_e(1)$ and spin-$S^z$ conservation $U_z(1)$\cite{sarma2025effective,wang2025fractional,bao2023mixed}. A key property of this phase is that charge and spin are intertwined: inserting a charge flux into the system induces a localized spin moment, meaning that a charge flux effectively carries spin. Ref.~\cite{klein2020nonlocal,lu2012theory} showed that the QSH state can be described as a proliferation of $U_e(1)$ vortices, with each vortex charged under the $U_z(1)$ symmetry.
Because of this decorated-defect pattern, when we project (measurement with post-selection) the QSH wavefunction’s local charge density to its eigenstate configuration $n_i$, the remaining spin degrees of freedom exhibit quasi-long-range correlations. This was first observed in Ref.~\cite{wang2025fractional, lu2022measurement} and has been used as a protocol to prepare a 2d critical state by measuring an SPT wavefunction.
In this section, we explore how the QSH wavefunction decoheres under the strong dephasing channel of the charge density.

To be specific, we start from a noninteracting spinful fermion system with decoupled spin-up and spin-down sectors. Each spin species forms a Chern insulator, but with opposite Chern numbers: the spin-$\uparrow$ sector has $C=+1$, while the spin-$\downarrow$ sector has $C=-1$. We realize these Chern-insulator bands using the Haldane model on a two-dimensional honeycomb lattice, where the electrons reside on the lattice sites. The Hamiltonian is given by:
\begin{equation}
    \begin{aligned}
        \hat{H} = &t_1\sum_{\langle i, j\rangle}  c_{i, \uparrow}^\dagger c_{j, \uparrow} + m \sum_i \xi_i c_{i, \uparrow}^\dagger c_{i, \uparrow}  +  t_2 \sum_{\langle \!\langle i, j\rangle \!\rangle} e^{i \eta_{ij}\phi}c_{i, \uparrow}^\dagger c_{j, \uparrow} \\
+&t_1\sum_{\langle i, j\rangle}  c_{i, \downarrow}^\dagger c_{j, \downarrow}  + m \sum_i \xi_i c_{i, \downarrow}^\dagger c_{i, \downarrow} + t_2 \sum_{\langle \!\langle i, j\rangle \!\rangle} e^{-i \eta_{ij}\phi}c_{i, \downarrow}^\dagger c_{j, \downarrow}
    \end{aligned}
\end{equation}
Here $\langle i,j\rangle$ and $\langle\langle i,j\rangle\rangle$ denote nearest- and next-nearest-neighbor pairs on the honeycomb lattice. The sign factors satisfy $\xi_i=+1$ ($-1$) on the $A$ ($B$) sublattice, while $\eta_{ij}=\pm1$ encodes the orientation of next-nearest-neighbor hopping following the convention of Ref.~\cite{haldane1988a} (and flips sign upon reversing direction). The parameter $m$ is a sublattice potential and $t_2$ sets the chiral next-nearest-neighbor hopping. For $|m|<3\sqrt{3}|t_2\sin\phi|$, each spin sector is in a Chern-insulator phase with $C=\pm1$. In the following, we set $m=0$ and $\phi=\pi/2$; see Refs.~\cite{haldane1988a, thonhauser2006} for further details and the phase boundary.

 The ground state of this Hamiltonian factorizes into independent contributions from the spin-up and spin-down sectors, 
\bea \label{eq:qsh}
|\Psi\rangle = |\psi_\uparrow\rangle \otimes |\psi_\downarrow\rangle
\eea
where $|\psi_\uparrow\rangle$ carries Chern number $1$ and $|\psi_\downarrow\rangle$ carries Chern number $-1$. 

We apply a local charge-dephasing quantum channel,
\bea\label{eq.qc_sqhe}
&\mathcal{E} = \prod_i \mathcal{E}_{i} \\
\label{eq.qc_sqhe_1}
&\mathcal{E}_{i}[\rho] = (1-2p)\rho + 2p(\sum_{n_i=0,1,2}  P_{n_i} \rho P_{n_i}) 
\eea
where $P_{n_i}$ projects onto the on-site charge sector with particle number $n_i$. For spinful fermions, the onsite particle density can take $n_i=0,1,2$. The parameter $p$ controls the dephasing strength: with probability $1-2p$ the state is left unchanged, while with total probability $2p$ it is projected onto a definite local charge $n_i$ and then averaged over $n_i$. In particular, at $p=\tfrac{1}{2}$ the channel reduces to full dephasing in the local charge basis, i.e., a projective measurement of the charge density without recording the outcome. In the Choi-doubled space, this enforces that the ket and bra layers have the same on-site particle number, projecting $\kket{\rho}$ onto configurations with $n_i^{(1)}=n_i^{(2)}$ on every site.

\begin{figure}[h]
    \centering
    \includegraphics[width=0.48\linewidth]{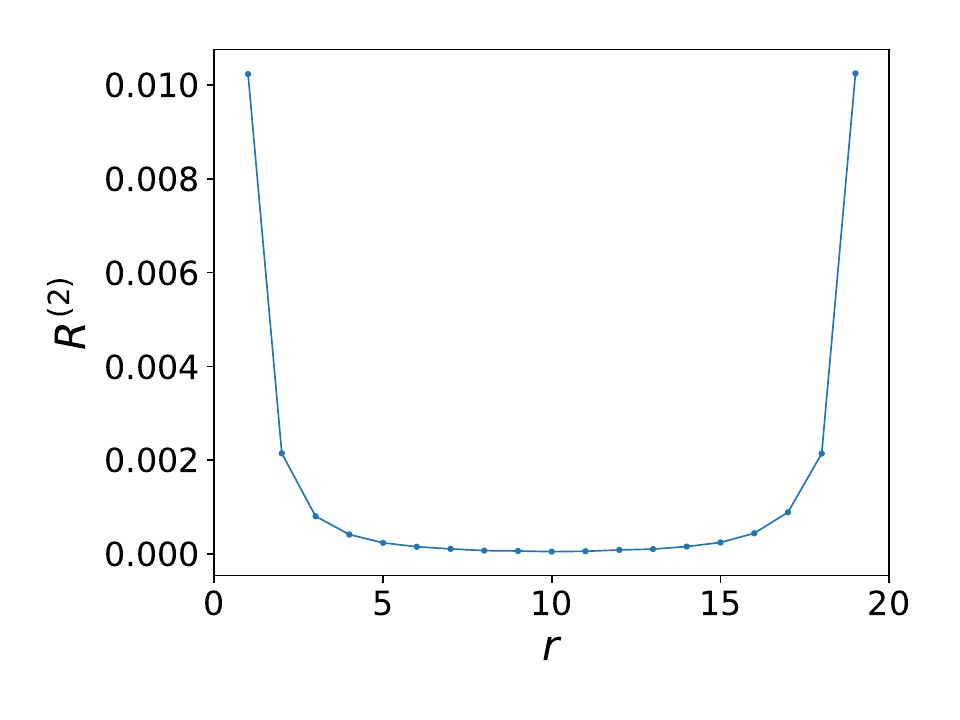}
    \includegraphics[width=0.48\linewidth]{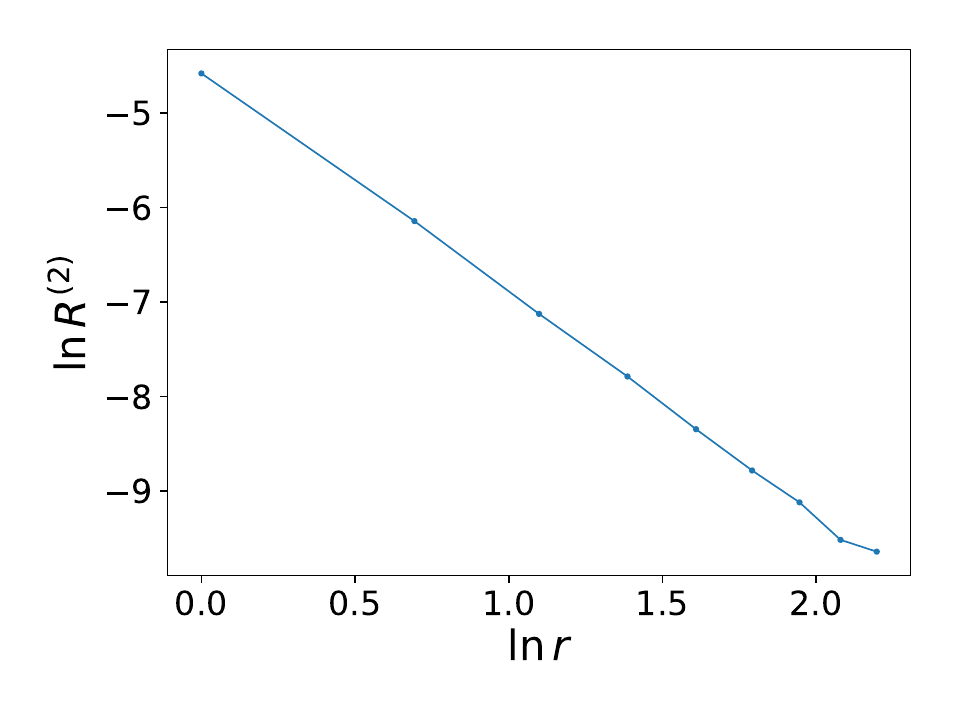}
    \caption{(a) R\'enyi-2 correlator \eqref{eq.renyi_2_qshe_S} versus distance $r$. Periodic boundary condition is used, therefore $r=20$ is equivalent to $r=0$. (b) Logarithm $\ln { R^{(2)}}$ versus logarithm of distance $\ln r$. This linearity indicates that the correlation has power-law decay. Monte Carlo simulation parameters: $L = 20$, update $200$k steps, measure for each $10$ steps. Decoherence strength $p=\frac12$, Haldane model parameters: $t_1 = 1, t_2 = 0.3, m=0,  \phi=\pi/2$. }
    \label{fig:qshe_correlation}
\end{figure}

After applying the channel \eqref{eq.qc_sqhe} with $p=1/2$ to the QSH state in Eq.~\eqref{eq:qsh}, we characterize the resulting mixed state
$\rho=\mathcal{E}[|\Psi\rangle\langle\Psi|]$
by computing its spin R\'enyi-2 correlator. Concretely, we probe spin coherence using the local spin-flip operator $c^\dagger_{r\downarrow}c_{r\uparrow}$ 
which carries one unit of $S^z$ charge:
\bea\label{eq.renyi_2_qshe_S}
R^{(2)}_S(r, r^\prime) = \frac{\trace{~\rho~ c_{r \downarrow}^\dagger c_{r \uparrow} c_{r^\prime \uparrow}^\dagger c_{r^\prime \downarrow} ~\rho~  c_{r^\prime \downarrow}^\dagger c_{r^\prime \uparrow} c_{r \uparrow}^\dagger c_{r \downarrow}}}{\trace{~\rho^2}}
\eea
Since $|\Psi\rangle$ is a free fermion ground state and the channel acts only through local particle number operators, the doubled-state weight can be efficiently sampled using fermionic Monte Carlo. We numerically evaluate the R\'enyi-2 correlator and find a clear power-law decay (Fig.~\ref{fig:qshe_correlation}), consistent with SWSSB and quasi-long-range order of the $U_z(1)$ symmetry.

Initialized as a QSH wavefunction, the SWSSB of $U_z(1)$ induced by local charge dephasing can be traced back to the string order $\hat{O}_{S}$ of the original wavefunction that exhibits quasi-long-range order \cite{kleinkvorning2020}:
\begin{align}\label{eq:string}
&\hat{O}_{S}(r,r')
= \langle c^\dagger_{r \downarrow}c_{r \uparrow}\;
c^\dagger_{r^\prime \uparrow}c_{r^\prime \downarrow}\;
\exp\!\left[i\sum_i n_i\Delta_i(r,r^\prime)\right]\rangle , \nonumber\\
&\Delta_i(r,r^\prime)
=\text{arg}(\mathbf{r}-\mathbf{r}_i)-\text{arg}(\mathbf{r}^\prime-\mathbf{r}_i),
\end{align}
where $n_i=n_{i,\uparrow}+n_{i,\downarrow}$.
This string order can be viewed as the correlator of a composite particle: a local spin-flip operator bound to a charge vortex. In the Chern-Simons description of the quantum spin Hall state, an $S_z$
 spin is attached to a charge vortex, so that the resulting spin-vortex composite field condenses and develops quasi-long-range order\cite{zhang1992chern}, which is directly manifested as quasi-long-range string order in Eq.~\eqref{eq:string}.

After applying the quantum channel in Eq.~\eqref{eq.qc_sqhe} with $p = \tfrac{1}{2}$ to the QSH state $|\Psi\rangle$, the local particle number on each site $k$ becomes identical on the bra and ket sides of the density matrix,
\bea
n_k ~\rho  = \rho ~n_k
\eea
Therefore, following the same argument as Sec.~\ref{sec:1d_zxz}, the R\'enyi-2 correlator in Eq.~\eqref{eq.renyi_2_qshe_S} can be rewritten in terms of a R\'enyi-2 string-order parameter:
\bea
R^{(2)}(r, r^\prime) = \frac{\trace {\rho ~ \hat{O}_{S} ~ \rho ~ \hat{O}_{S}^\dagger} }{\trace {\rho^2}}
\eea
This establishes a direct correspondence between the mixed-state R\'enyi-2 correlator after charge dephasing and the string order of the pre-dephasing QSH state. Notably, a closely related phenomenon was explored in Ref.~\cite{wang2025fractional}, where charge-density dephasing is introduced via a Lindblad evolution starting from a fractional quantum Hall state, and SWSSB of the $U(1)$ symmetry is observed in the steady state.

\section{charge dephasing in higher-form SPT}\label{sec:higherform}

We now consider a different class of 2d SPT states protected by higher-form symmetries. Higher-form SPT phases provide a fertile setting to explore the interplay between symmetry and entanglement, and they can host phases of matter with no counterpart in conventional (zero-form) symmetry settings. For example, a 2d $\mathbb{Z}_2$ gauge theory coupled to $\mathbb{Z}_2$ matter can be viewed as an SPT phase protected jointly by a 0-form symmetry $\mathbb{Z}_2^{(0)}$ and a 1-form symmetry $\mathbb{Z}_2^{(1)}$. Such higher-form SPTs can exhibit distinctive phenomena, including unnecessary criticality and boundary topological phase transitions \cite{verresen2022higgs}. A key difference from conventional SPT phases protected only by global (0-form) symmetries is their robustness under weak symmetry breaking: while the characteristic features of a conventional SPT typically disappear immediately upon weak symmetry breaking, a higher-form SPT can remain stable up to a finite threshold even if the one-form symmetry is explicitly broken at the UV scale \cite{su2024higher,jian2021physics,roberts2020symmetry}.

We begin with a system living on a two-dimensional square lattice with two species of spin-$\tfrac{1}{2}$ degrees of freedom: vertex spins living on lattice sites (labeled $v$) and edge-center spins living at the midpoints of links (labeled $e$). The Hamiltonian of the 2d cluster model is:
\bea  \label{eq:2dcluster}
\hat{H} = - \sum_v \left( X_v \prod_{ e \ni v}Z_e \right) -\sum_e \left(X_e \prod_{v\in e} Z_v\right)
\eea
where $v \in e$ indicates that the vertex $v$ is one of the two endpoints of the edge $e$. The first term couples each vertex spin to its four neighboring edge spins, while the second term couples each edge spin to its two adjacent vertex spins.

This model possesses two $\mathbb{Z}_2$ symmetries: a 0-form symmetry acting on all vertex spins and a 1-form symmetry acting along closed loops of edge spins,
\bea
\mathbb{Z}_2^{(0)} : X_{\text{vertex}} = \prod_v X_v \\
\mathbb{Z}_2^{(1)} : X_\gamma= \prod_{e \in \gamma} X_e
\eea
where $\gamma$ denotes an arbitrary closed loop that bounds a collection of plaquettes.
The ground state of Eq.~\eqref{eq:2dcluster} realizes a nontrivial SPT phase protected jointly by a 0-form symmetry $\mathbb{Z}_2^{(0)}$ and a 1-form symmetry $\mathbb{Z}_2^{(1)}$. This phase admits a domain-wall decoration interpretation: the ground state is a coherent superposition over domain-wall configurations of the $\mathbb{Z}_2^{(0)}$ symmetry (vertex spin), and each such domain wall is decorated by $\mathbb{Z}_2^{(1)}$ charges (edge spins) living along the wall. 
\bea
|\psi \rangle  = \sum_{\{Z_v\}} \left(\otimes_v |Z_v\rangle \otimes_e |X_e = \prod_{v \in e} Z_v  \rangle \right)
\eea

We now decohere $|\psi\rangle$ by applying a charge-dephasing channel that measures the $\mathbb{Z}_2^{(1)}$ one-form symmetry charge:
\bea
\rho = \mathcal{E}[|\psi \rangle \langle \psi|]
\eea
where
\bea
\mathcal{E} = \prod_e \mathcal{E}_e, \quad \mathcal{E}_e[\rho] = (1-p) \rho + p X_e \rho X_e
\eea
This can be viewed as a partial dephasing channel, with $p$ tuning the dephasing strength. At $p=\tfrac{1}{2}$, the channel becomes strong dephasing: it measures all edge spins in the $X$ basis and then averages over the measurement outcomes (i.e., without postselection).

Similar to the previously discussed cases, in the doubled-state representation this channel assigns a relative weight of $(1 - 2p)$ to configurations with opposite parity between the upper and lower layers, i.e. $X_{e,u} X_{e,l} = -1$.
Using the stabilizer relation $X_e = \prod_{v \in e} Z_v$, this condition can be rewritten in terms of vertex spins as
\bea
X_{e, u} X_{e, l} = \prod_{v \in e} Z_{v, u} Z_{v, l} 
\eea
where $u$ and $l$ label the upper and lower layers in the doubled space.

The mixed-state density after decoherence can therefore be written as its Choi-double form:
\bea \label{eq:tc}
&\kket{\rho} = \sum_{\{Z_{v, u}\}, \{Z_{v, l}\}} e^{\beta \sum_{\langle i,j \rangle} (Z_{i, u} Z_{i, l} Z_{j, u}Z_{j, l}-1)} \\ 
\notag
&\otimes_v |Z_{v, u}, Z_{v, l}\rangle \otimes_e |X_{e, u} = \prod_{v \in e} Z_{v, u}, X_{e, l} = \prod_{v \in e} Z_{v, l}  \rangle 
\eea
where $\langle i,j\rangle$ denotes neighboring vertex pairs on the lattice, and the parameter $\beta$ is related to the decoherence strength $p$ via
\bea
e^{-2\beta} = 1-2p
\eea
Similar density matrices have appeared previously, e.g., in studies of SWSSB from Ising paramagnet \cite{lee2022measurement,lee2023quantum,lessa2024strong} and in dephasing-induced transitions of toric-code wavefunctions \cite{bao2023mixed}. If we evaluate the fidelity correlator in Eq.~\eqref{eq:fid} using the charge operator $Z_v$ (which carries the $0$-form symmetry charge), the resulting correlator maps onto that of a random-bond Ising model on the Nishimori line and becomes finite for $p>p_c\approx0.109$ \cite{hauser2025information,lee2022,zhu2023nishimori,fan2024diagnostics}.
Therefore, dephasing the $1$-form symmetry charge in the 2d cluster state drives SWSSB of the $0$-form symmetry at a finite dephasing strength.

It is also worth noting that one can instead dephase the 0-form charge density by performing an unrecorded measurement of the vertex spin in the $X$ basis. In the strong-dephasing limit $p=1/2$, the resulting mixed state exhibits SWSSB of the 1-form symmetry. By contrast, for partial dephasing $p<1/2$, the mixed state remains strongly symmetric \cite{zhu2023nishimori,lee2022decoding,sala2024spontaneous}.

\section{Intrinsically Gapless SPT}\label{sec:gapless}

Having established how gapped SPT wavefunctions respond to decoherence, we now extend our exploration to gapless SPT states. In 1d, a particularly interesting class is the intrinsically gapless SPT (igSPT), where the infrared theory is gapless yet still carries a nontrivial 't Hooft anomaly \cite{gaplessSPT,scaffidi2017gapless,li2024,li2023intrinsically}. Much like gapped SPTs, some igSPT states admit a domain-wall decoration picture. This suggests that the same basic mechanism persists: dephasing (measuring) the charge density of one symmetry can induce SWSSB of the other symmetry.
What’s novel here is that the igSPT wavefunction is critical, with divergent correlation length and power-law correlations. Consequently, the dephasing noise cannot be treated as a purely local coupling at the time boundary. A useful perspective is to view the density matrix as a time path integral for a (1+1)d critical bulk, supplemented by an additional perturbation acting at the time boundary: because the bulk is critical, this time-boundary perturbation can mediate effectively nonlocal couplings along the boundary. As such, Refs.~\cite{garratt2023measurements,murciano2023measurement,yang2023entanglement} demonstrated that dephasing a critical wavefunction can generate new quantum critical behavior reminiscent of boundary criticality. Meanwhile, numerical results further suggest that starting from a 1d critical state, SWSSB can be induced by infinitesimally weak dephasing~\cite{lee2023quantum,guo2025}.

We begin with the igSPT model proposed in Ref.~\cite{li2025}:
\bea \label{eq.zxz_gapless_hamiltonian}
&&\hat{H} = - \sum_i( Z_{2i}X_{2i+1}Z_{2i+2}+X_{2i}(Y_{2i-1}Y_{2i+1}+ Z_{2i-1}Z_{2i+1}))\nonumber\\
&&- h \sum_{i=1}^L X_i
\eea
The first term $Z_{2i}X_{2i+1}Z_{2i+2}$ implements the decorated--domain-wall construction: domain walls of the even-sublattice $\mathbb{Z}_2$ variable are bound to symmetry charges residing on the odd sublattice.
For $0<h<1$, the model realizes an igSPT phase protected by a $\mathbb{Z}_2 \times \mathbb{Z}_4$ symmetry. Specifically, the $\mathbb{Z}_2$ and $\mathbb{Z}_4$ generators act as spin rotations about the $X$ axis on the even and odd sites, respectively:
\begin{equation}
    \begin{aligned} \label{eq.igspt_symmetry}
\mathbb{Z}^a_2 &: \prod_{i} X_{2i}\\
\mathbb{Z}^b_4 &: \prod_{i~} e^{\frac{i\pi}{4} (1-X_{2i+1})}  
    \end{aligned}
\end{equation}
Note that, upon restricting to the low-energy subspace selected by the decorated domain-wall constraint
$Z_{2i}X_{2i+1}Z_{2i+2}=1$,
the $\mathbb{Z}_4^b$ symmetry effectively reduces to a controlled-$Z$ (CZ) symmetry that acts non-onsitely:
\[
U_{\mathrm{CZ}}=\prod_i \exp\!\left[\frac{i\pi}{4}\bigl(1-Z_{2i}Z_{2i+2}\bigr)\right].
\]
Together with the onsite $\mathbb{Z}_2^a$ symmetry, this non-onsite action implies a mixed anomaly, closely analogous to the symmetry anomaly on the Levin--Gu edge \cite{li2024, levin2012, ma2025}. As such, the ground-state wavefunction is critical and exhibits an emergent anomaly \cite{thorngren2021, li2024, li2025, sun2025,ma2025, you2024,ma2023topological}.

Similar to a gapped SPT, the igSPT phase can be diagnosed by a nonlocal string order parameter:
\bea \label{eq.gapless_string}
O_{\text{str}} = \langle\psi| Z_{2r} \left(\prod_{i=r}^{r^\prime-1} X_{2i+1}\right) Z_{2r^\prime} | \psi \rangle
\eea
In the igSPT phase ($0<h<1$), $O_{\text{str}}$ approaches a finite constant at large separations, while it decays exponentially for $h>1$ when the system enters the trivial gapless phase.

We now turn to decoherence and apply a channel that partially dephases the local $\mathbb{Z}_4^b$ charge density to the ground-state wavefunction $|\psi(h)\rangle$ in Eq.~\eqref{eq.zxz_gapless_hamiltonian} :
\bea\label{eq.igspt_quantum_channel}
\mathcal{E} = \prod_{i} \mathcal{E}_{2i+1}, \quad \mathcal{E}_{2i+1}[\rho] = (1-p) \rho + p X_{2i+1} \rho X_{2i+1}
\eea
where $p$ controls the dephasing strength. At $p=\frac{1}{2}$, $\mathcal{E}$ reduces to the fully dephasing channel. 

We expect that strong dephasing of the $\mathbb{Z}_4^b$ charge can induce SWSSB of the $\mathbb{Z}_2^a$ symmetry. To test this, we evaluate the R\'enyi-2 correlator of the resulting mixed state:
\bea\label{eq.gapless_renyi_2}
R^{(2)}(r, r^\prime) = \frac{\trace{~\rho ~Z_{2r} Z_{2r^\prime}~\rho~ Z_{2r} Z_{2r^\prime} }}{\trace{~\rho^2}}
\eea

Numerically, we use DMRG to compute the R\'enyi-2 correlation function in Eq.~\eqref{eq.gapless_renyi_2} as a function of the separation $|r'-r|$. Our main goal is to resolve its long-distance behavior and thereby determine whether SWSSB occurs.

\begin{figure}[h]
    \centering
    \includegraphics[width=0.8\linewidth]{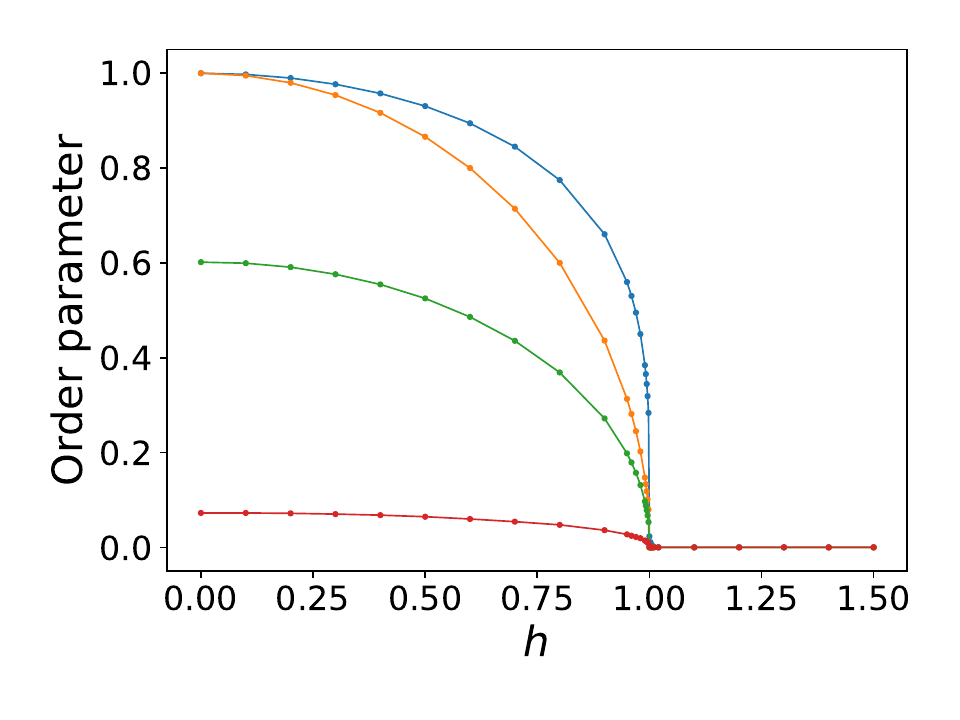}
    \caption{String order parameter (blue) for ground state of gapless SPT Hamiltonian \eqref{eq.zxz_gapless_hamiltonian} and R\'enyi-2 correlator \eqref{eq.gapless_renyi_2} for decohered state for $p=0.5$ (orange), $0.4$ (green) and $0.3$ (red) versus magnetic field strength $h$. DMRG parameters: system size $L = 1000$, bond dimension $\chi=100$, up to 100 sweeps. $2r = L / 4, 2r^\prime = 3L/4$.}
    \label{fig:zxz_gapless_correlator_h}
\end{figure}

Fig.~\ref{fig:zxz_gapless_correlator_h} shows the string order parameter \eqref{eq.gapless_string} (blue) and the R\'enyi-2 correlator \eqref{eq.gapless_renyi_2} of the decohered state for several values of $p$ and $h$. For $p=\frac12$, the R\'enyi-2 correlator remains finite throughout the igSPT regime ($h<1$), indicating SWSSB. For $p<\frac12$, the correlator also appears finite at this system size; however, we will show below that this is a \textit{finite-size effect} and that SWSSB does not survive in the thermodynamic limit.

Unlike gapped SPT states, these gapless states exhibit long-range entanglement, and DMRG is not quantitatively reliable for critical systems. Indeed, the data suggest a large correlation length comparable to the system size. We therefore rely on finite-size scaling and diagnose SWSSB using the Binder cumulant, which provides a sharper and more robust indicator.

The Binder cumulant $U$ is defined as
\begin{align}
U = 1-\frac{\langle m^4\rangle}{3\langle m^2\rangle^2} ~,\quad
\langle m^n\rangle := \frac{\mathrm{Tr}\left[\rho m^n[\rho]\right]}{\mathrm{Tr}\rho^2}
\end{align}
where the “magnetization” superoperator $m$ acts on the mixed state $\rho$ as
\bea
m[\rho]=\frac{2}{L}\sum_{i} Z_{2i}~\rho ~Z_{2i}.
\eea
Here \(m^n\) denotes \(n\)-fold composition, \(m^n(\rho)=\underbrace{m\circ\cdots\circ m}_{n~\text{times}}(\rho)\), i.e.,
\bea
m^n(\rho)=\left(\frac{2}{L}\right)^n
\sum_{i_1,\ldots,i_n=1}^{L/2}
Z_{2i_1}\cdots Z_{2i_n}\,\rho\, Z_{2i_n}\cdots Z_{2i_1}.
\eea
In particular,
\begin{align}
&m^2(\rho)=\left(\frac{2}{L}\right)^2\sum_{i,j=1}^{L/2} Z_{2i}Z_{2j}\,\rho\, Z_{2j}Z_{2i},\qquad\\
&m^4(\rho)=\left(\frac{2}{L}\right)^4\sum_{i,j,k,\ell=1}^{L/2} Z_{2i}Z_{2j}Z_{2k}Z_{2\ell}\,\rho\, Z_{2\ell}Z_{2k}Z_{2j}Z_{2i}.
\end{align}
This definition is equivalent to the conventional Binder cumulant constructed from the order parameter $Z_{2r}Z_{2r'}$ in the doubled space. Consequently, in the thermodynamic limit, $U$ distinguishes the two possibilities: $U\to 2/3$ in the SWSSB phase and $U\to 0$ when SWSSB is absent.

\begin{figure}[h]
    \centering
    \includegraphics[width=0.8\linewidth]{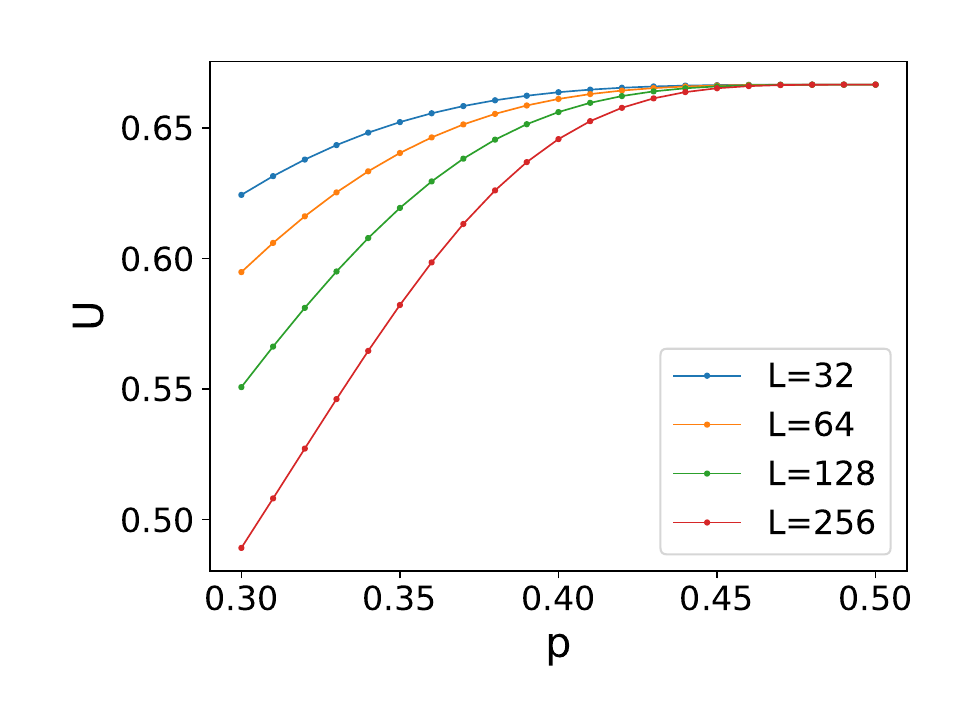}
    \caption{Binder cumulant $U$ as a function of the decoherence strength $p$ for different system sizes $L$. Data are obtained at $h=0$ in Eq.~\eqref{eq.zxz_gapless_hamiltonian}. The Hamiltonian is defined with periodic boundary conditions, while the DMRG calculation is performed using an open-boundary MPS representation. Parameters: $L=32,64,128,256$, bond dimension $\chi=100$, up to 100 sweeps.}
    \label{fig:bc_delta_0}
\end{figure}
Fig.~\ref{fig:bc_delta_0} shows $U$ versus the dephasing strength $p$ for several system sizes $L$. Although $U$ is close to $2/3$ at finite $L$ when $p$ is near $1/2$, it decreases systematically with increasing $L$. This trend indicates that SWSSB does not persist in the thermodynamic limit for $p<1/2$.

\begin{figure}[h]
    \centering
    \includegraphics[width=0.8\linewidth]{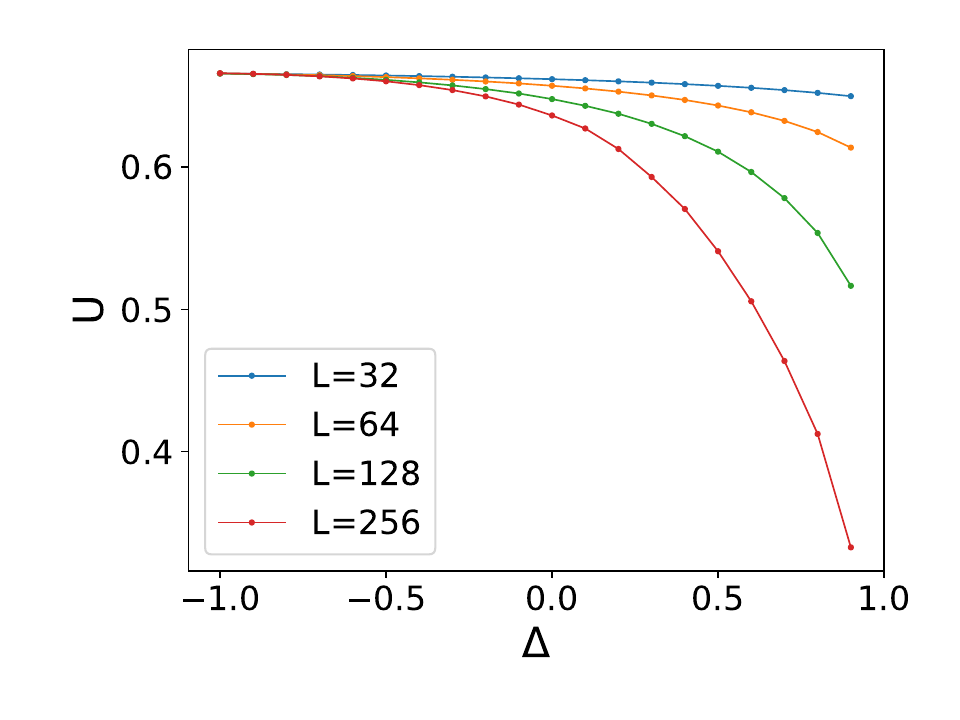}
    \caption{Binder cumulant $U$ versus $\Delta$ for different system sizes $L$ at decoherence strength $p=0.4$. Data are obtained at $h=0$ for varying $\Delta$ in Eq.~\eqref{eq.zxz_gapless_hamiltonian_delta}. The Hamiltonian is defined with periodic boundary conditions, while the DMRG calculation is performed using an open-boundary MPS representation. Parameters: $L=32,64,128,256$, bond dimension $\chi=100$, up to 100 sweeps.}
    \label{fig:bc_p_0_4}
\end{figure}

We further tune the low-energy Luttinger parameter $K$ by adding an interaction term to the Hamiltonian in Eq.~\eqref{eq.zxz_gapless_hamiltonian},
\bea\label{eq.zxz_gapless_hamiltonian_delta}
\hat{H}'=\hat{H} -\sum_{i} \Delta X_{2i-1}X_{2i+1}.
\eea
This term preserves the symmetry in Eq.~\eqref{eq.igspt_symmetry}, and for $-1<\Delta<1$ the Hamiltonian $\hat{H}'$ remains in the same igSPT phase. Fixing $h=0$ and applying the same decoherence channel \eqref{eq.igspt_quantum_channel} to the ground state of $\hat{H}'$, we compute the Binder cumulant $U$ as a function of $\Delta$ for different $L$ (Fig.~\ref{fig:bc_p_0_4}). The behavior is again consistent with the absence of SWSSB when $p<1/2$: for most $\Delta$, $U$ decreases with increasing $L$, with the only exception occurring near $\Delta=-1$. Taken together, these results support that SWSSB does not occur for $p<1/2$ and $-1<\Delta<1$.

To interpret these numerical results, we appeal to the RG analysis of boundary CFT introduced in Ref.~\cite{guo2025quantum}. The low-energy mode in the igSPT wavefunction from Eq.~\eqref{eq.zxz_gapless_hamiltonian_delta} can be characterized as a Luttinger liquid, and the quantum channel induces perturbations localized at $\tau=0$,
\bea
S=\int d\tau dx\delta(\tau)\Big[\cos \big(2\phi_1-2\phi_2\big)+\cos(4\phi_1)+\cos(4\phi_2) \Big] ~~~~~~
\eea
We analyze the weak-channel regime (small $p$), where these symmetry-allowed perturbations are initially small. If they are all irrelevant, the RG flow returns to the pure-state fixed point and the system does not flow into the SWSSB regime.

Terms of the form $\cos(n\theta)$ are forbidden by the $U(1)$ symmetry, while $\cos(2\phi)$ is forbidden by translation symmetry. The scaling dimension of $\cos(n\phi)$ is $n^2K/4$. The leading inter-replica coupling $\cos(2\phi_1-2\phi_2)$ has scaling dimension $2K$, and is therefore irrelevant when $2K>1$ (i.e., $K>\frac12$). Over the gapless parameter range $-1<\Delta<1$, we have $K>\frac12$, so the RG flow returns to $p=0$: the strong symmetry is preserved, and SWSSB occurs only at the fine-tuned point $p=1/2$. Near $\Delta\simeq -1$ (close to the antiferromagnetic phase boundary), $\cos(2\phi_1-2\phi_2)$ becomes closer to marginal, making SWSSB comparatively more stable at small $p$. This is consistent with the numerics: in Fig.~\ref{fig:bc_p_0_4}, $U$ stays closer to $2/3$ at larger $L$ near $\Delta=-1$. Overall, the Binder cumulant finite-size scaling agrees with the RG expectation.

While the mixed state exhibits SWSSB at $p=1/2$, the correlations of some charged operators (e.g., $Z_{2i+1}$) remain critical throughout the dephasing process. Consequently, the mixed state density matrix cannot be purified to a short-range entangled state in an enlarged Hilbert space. 
If we wish to drive SWSSB at finite $p$, one possibility is to introduce an additional weak dephasing of terms such as $Z_{2i-1}Z_{2i+1}$, similar to what was explored in Ref.~\cite{guo2025}. Meanwhile, at finite-$p$ dephasing there may emerge more exotic measurement-altered criticality, with critical behavior distinct from that of the corresponding pure-state transition; we defer these questions to a future work.

\section{General Discussion and Outlook} \label{sec:general}

\subsection{1d SPT with Symmetry Group $G \times S$}\label{sec:gene}

The SWSSB mechanism we describe in Sec.~\ref{sec:1d_zxz} is not specific to a particular model, but applies broadly to 1d SPT wavefunctions protected by a product symmetry $G\times S$ that admit a decorated domain-wall construction. In such states, the $G$-domain walls carry $S$ charge. Consequently, strong dephasing of the local $S$-charge density, which effectively projects onto definite $S$-charge sectors and proliferates $S$-domain walls in the Choi-doubled description, inevitably induces SWSSB of $G$ in the resulting decohered mixed state.
To illustrate this generality, we provide complementary perspectives based on an MPS formulation. A similar group-cohomology based demonstration is provided in Appendix~\ref{app:group_cohomology_swssb}.

\subsubsection{MPS Argument}

Matrix product states (MPS) \cite{schollwock2011,schollwock2005,ostlund1995}
provide a convenient framework for illustrating the symmetry fractionalization structure of 1d gapped phases. We first explain how spontaneous symmetry breaking (SSB) of symmetry $G$ can emerge after measuring the $S$ charge (with post-selection) in SPT states. An MPS is specified by a local tensor $A^{\sigma}_{\alpha\beta}$, with one physical index $\sigma$ and two virtual indices $\alpha,\beta$.
\bea
|\psi\rangle
= \sum_{\{\sigma_i\},\{\alpha_i\}}
A^{\sigma_1}_{\alpha_1\alpha_2}
A^{\sigma_2}_{\alpha_2\alpha_3}
\cdots
A^{\sigma_L}_{\alpha_L\alpha_1}
~|\sigma_1,\sigma_2,\cdots,\sigma_L\rangle .
\eea
The symmetry group $G\times S$ acts on the physical index. Because the state is invariant under the symmetry, this symmetry transformation can be pushed to the virtual bonds. Concretely, for $g\in G$ and $s\in S$,
\bea
U^g_{\sigma\sigma^\prime} A^{\sigma^\prime}_{\alpha\beta}
&= e^{i\phi_g}\,
V^g_{\alpha\alpha^\prime}\,
A^\sigma_{\alpha^\prime\beta^\prime}\,
\left(V^g_{\beta\beta^\prime}\right)^{*}, \\
U^s_{\sigma\sigma^\prime} A^{\sigma^\prime}_{\alpha\beta}
&= e^{i\phi_s}\,
V^s_{\alpha\alpha^\prime}\,
A^\sigma_{\alpha^\prime\beta^\prime}\,
\left(V^s_{\beta\beta^\prime}\right)^{*},
\eea
where $U$ acts on the physical Hilbert space and $V$ acts on the virtual space. Since we focus on an SPT state with a decorated domain-wall structure, we can choose $V^{g}$ and $V^{s}$ to furnish linear representations of $G$ and $S$, respectively, whereas their combined action realizes a projective representation of $G\times S$.
Without loss of generality, the phases $e^{i\phi_g}$ and $e^{i\phi_s}$ can be removed by a gauge choice; hence we set $\phi_g=\phi_s=0$ in what follows.

Focusing on the $S$ charge, we choose a basis in which the virtual representation $V^s$ decomposes into irreducible representations of $S$:
\bea
V^s = V^s_{\xi_1} \oplus V^s_{\xi_2} \oplus \cdots \oplus V^s_{\xi_r},
\eea
where $\xi_1,\xi_2,\ldots,\xi_r$ label the distinct irreducible representations of $S$.

Now we measure the on-site $S$ charges with post-selection, i.e., we project the state onto a sector in which the $S$ charge on each site is fixed. The symmetry constraint on the MPS tensors then simplifies to
\bea
\eta_i^s\, B^{\sigma}_{\alpha\beta,i}
= \xi^s_{\alpha,i}\, B^{\sigma}_{\alpha\beta,i}\, \bigl(\xi^s_{\beta,i+1}\bigr)^* ,
\eea
where
\bea
B^{\sigma}_{\alpha\beta,i} \equiv P_i^{\sigma\sigma^\prime} A^{\sigma^\prime}_{\alpha\beta}
\eea
is the local tensor after applying the projection operator $P_i$ on site $i$. Here $\eta_i$ and $\xi$ label irreducible representations of $S$. Throughout, $\alpha$, $\beta$, and $i$ are not summed over on the right-hand side. Since $P_i$ encodes the local measurement outcome, it may depend on $i$, and accordingly all representation labels carry an explicit site index.

This relation makes the charge constraint between physical and virtual indices manifest: whenever $B^{\sigma}_{\alpha\beta,i}\neq 0$, the corresponding representations must satisfy
\bea
\eta_i = \xi_{\alpha,i}\,\xi_{\beta,i+1}^* ,
\eea
namely, the $S$ charge carried by the physical degree of freedom on site $i$ must be consistent with the charges on the adjacent virtual links.

Using this constraint, the post-measurement MPS can be written as a superposition of MPS states built from smaller blocks,
\bea\label{eq.psi_xi}
&|\psi\rangle = \sum_{\xi} |\psi_{\xi}\rangle \\
&|\psi_{\xi}\rangle
= \sum_{\{\sigma_i\},\{\alpha_i\}}
(B_{\xi}^{1})^{\sigma_1}_{\alpha_1\alpha_2}
(B_{\xi}^{2})^{\sigma_2}_{\alpha_2\alpha_3}
\cdots
(B_{\xi}^{L})^{\sigma_L}_{\alpha_L\alpha_1}
~|\sigma_1,\sigma_2,\cdots,\sigma_L\rangle .\nonumber\\
\eea
Here $B_{\xi}^{i}$ denotes the block of $B^{\sigma}_{\alpha\beta,i}$ in which the virtual charges satisfy
\bea
\xi_{\alpha,i} = \xi \prod_{k=1}^{i-1} \eta_k,
\qquad
\xi_{\beta,i+1} = \xi \prod_{k=1}^{i} \eta_k .
\eea
Equivalently, once the $S$ charge on one virtual bond is fixed, the $S$ charges on all other virtual bonds are determined uniquely by the measured on-site charges $\{\eta_i\}$. Equation~\eqref{eq.psi_xi} thus exhibits the resulting block-diagonal (long-range entanglement) structure of the MPS.

Because the original pure state is an SPT with decorated domain-wall structure, the virtual symmetry actions $V^g$ and $V^s$ cannot be simultaneously diagonalized: in particular, $V^g$ does not preserve the $S$-charge blocks of $V^s$ and therefore maps $|\psi_{\xi}\rangle$ to a different block,
\bea
G\,|\psi_{\xi}\rangle \not\propto |\psi_{\xi}\rangle .
\eea
Consequently, within each fixed-$\{\eta_i\}$ measurement sector, the state breaks the $G$ symmetry spontaneously. We conclude that for an SPT state obtained by decorating $G$ domain walls with $S$ charges, a projective measurement of the $S$ charge induces SSB of $G$.

It is worth noting that a closely related protocol has been studied in adaptive quantum circuits \cite{bravyi2022adaptive,tantivasadakarn2024,lu2022measurement,lu2023mixed,zhang2024characterizing,sahay2025classifying}, where measurements provide a shortcut for generating long-range entangled pure states.
Building on this insight, we now turn to the mixed-state settings under a full charge-dephasing channel. This process is equivalent to performing local measurements of the S-charge and then averaging over (i.e., summing over) all measurement outcomes.
\bea
\mathcal{E}\!\left(|\psi\rangle\langle\psi|\right)
= \sum_{\{\eta\}} P^{\{\eta\}}\,|\psi\rangle\langle\psi|\,P^{\{\eta\}},
\qquad
P^{\{\eta\}} \equiv \prod_i P_i^{\eta_i} ,
\eea
where $\{\eta\}$ denotes the set of measurement outcomes on the entire chain and $P$ is the corresponding projection operator. Equivalently, the decoherence channel factorizes as
\bea\label{eq.quantum_channel_eta}
\mathcal{E} = \prod_i \mathcal{E}_i,
\qquad
\mathcal{E}_i[\rho]=\sum_{\eta} P_i^{\eta}\,\rho\,P_i^{\eta}.
\eea
This channel eliminates coherences between different on-site $S$-charge sectors.

Since each projected pure state $P^{\{\eta\}}|\psi\rangle$ exhibits SSB of $G$, there exists an order parameter $O$ carrying $G$ charge (constructed explicitly below) such that its correlator saturates at long distance within each sector:
\bea
\Bigl|\langle\psi| P^{\{\eta\}}\, O_i O_j^\dagger\, P^{\{\eta\}} |\psi\rangle\Bigr|
= C_{\{\eta\}}\, \left|\langle\psi|P^{\{\eta\}}|\psi\rangle \right|,
\eea
with $C_{\{\eta\}}$ finite. It follows that the R\'enyi-2 correlator of the mixed state remains finite and is given by a weighted average over sectors,
\bea
R^{(2)} =
\frac{\sum_{\{\eta\}} C_{\{\eta\}}^{\,2}\,\bigl|\langle\psi|P^{\{\eta\}}|\psi\rangle\bigr|^2}
{\sum_{\{\eta\}} \bigl|\langle\psi|P^{\{\eta\}}|\psi\rangle\bigr|^2}.
\eea
This leads to the conclusion that introducing 
S-charge decoherence induces SWSSB of the G-symmetry.
A complementary perspective comes from the group-cohomology construction of one-dimensional bosonic SPT states; we defer the details to the Appendix \ref{app:group_cohomology_swssb}.

\subsection{Discussion for Modulated SPT States}\label{sec:dip}

Our discussion also applies to SPT phases with generalized symmetries, e.g., 1d SPTs whose symmetry generators are spatially modulated. 
For instance, one can consider a $\mathbb{Z}^Q_N \times \mathbb{Z}^D_N$ dipolar cluster state whose charge ($Q$) and dipole ($D$) symmetries are defined as \cite{lam2024classification,lam2024topological,delfino2023anyon,delfino20232d,han2024topological}:
\begin{align}
\mathbb{Z}^Q_N: \prod_j X_j , ~~ ~\mathbb{Z}^D_N :\prod_j ( X_j )^j . \label{eq:Q-and-D}
\end{align} 
It admits an SPT state with parent Hamiltonian:
\begin{align}
H_{D} & =-\sum_{j} ( a_j + a^\dag_j ) \nonumber \\
a_j & = Z_{j-1} \big( Z^\dag_j X_j Z^\dag_j \big) Z_{j+1} ,
\label{eq:H-dipolar} 
\end{align}
The ground state of Eq.~\eqref{eq:H-dipolar} also exhibits a decorated domain-wall structure. 
Starting from this ground state, if we fully dephase the local $\mathbb{Z}_N^{Q}$ charge density,
\bea
\mathcal{E}=\prod_i\mathcal{E}_i, \qquad \mathcal{E}_{i}[\rho] = \frac{1}{2}\Big(\rho + X_{i} \rho X_{i}\Big),
\eea
the resulting state exhibits SWSSB of the $\mathbb{Z}_N^{D}$ symmetry, manifested by the R\'enyi-2 correlator:
\bea
R^{(2)}(r, r^\prime) = \frac{\trace{~\rho Z^{\dagger}_{r-1} Z_{r} Z_{r^\prime} Z^{\dagger}_{r^\prime+1}\rho Z_{r-1} Z^{\dagger}_{r} Z^{\dagger}_{r^\prime} Z_{r^\prime+1} }}{\trace{~\rho^2}}
\eea
This agrees with our general statement that dephasing the local $\mathbb{Z}_N^{Q}$ charge triggers SWSSB of $\mathbb{Z}_N^{D}$.

A related question is whether dephasing the dipole charge\footnote{Dephasing the dipole charge might require nonlocal quantum channels.} $\mathbb{Z}_N^{D}$ can trigger SWSSB of the charge symmetry $\mathbb{Z}_N^{Q}$. For a conventional $G\times S$ SPT, one would indeed expect this: the two symmetries are essentially independent and do not obstruct one another. 
In contrast, for SPT phases protected by a modulated symmetry, charge and dipole conservation are not independent. Instead, they are intertwined through lattice translations:
$$
T_x\, \mathbb{Z}_N^{D}\, T_x^{-1} \;=\; \mathbb{Z}_N^{D}\cdot \mathbb{Z}_N^{Q}\, 
$$
Notably, any correlator that develops long-range order for $\mathbb{Z}_N^{Q}$ necessarily carries charge under the strong $\mathbb{Z}_N^{D}$ symmetry, and thus explicitly breaks $\mathbb{Z}_N^{D}$. Consequently, SWSSB of $\mathbb{Z}_N^{Q}$ cannot occur unless $\mathbb{Z}_N^{D}$ is \textit{explicitly broken}.

Suppose some mechanism were to induce SWSSB of $\mathbb{Z}_N^{Q}$, it should be detectable via long-range order in the R\'enyi-2 correlator
\begin{equation}\label{eq:chare}
\frac{\trace{\rho Z_r Z_{r'}^{\dagger}\rho Z_r^{\dagger} Z_{r'}}}{\trace{\rho^2}}\sim O(1) .
\end{equation}
    However, Eq.~\eqref{eq:chare} is not invariant under the strong $\mathbb{Z}_N^{D}$ symmetry, so any long-range order in it indicates an explicit breaking of $\mathbb{Z}_N^{D}$ rather than a spontaneous one. This cannot happen if the quantum channel preserves both $\mathbb{Z}_N^{Q}$ and $\mathbb{Z}_N^{D}$. This obstruction reflects the characteristic `hierarchy structure' of modulated-symmetry SPT states, as originally discussed in Ref.~\cite{you2024intrinsic}.

\subsection{Discussion and Outlook for Generic SPT States in Higher Dimensions}

In this work, we restrict our attention to a \textit{particular subclass of SPT states} protected by $G\times S$, in which $G$ defects are decorated by $S$ charges. In higher dimensions, $G$ defects can carry $S$ charge in several ways. For concreteness, consider $2d$ (the same logic extends to general dimensions). If a $G$ defect is a point-like object, for instance a vortex of a $U(1)$ symmetry, then the defect can bind a $0$-form global $S$ charge. If instead the $G$ defect is a line-like object, such as a domain wall, then decorating it with a conserved charge naturally calls for $S$ to act as a $1$-form symmetry, so that the charge can consistently live on a one-dimensional defect. These are the cases we explored in Sec.~\ref{sec:qsh}-\ref{sec:higherform}.

The broader `zoology' of SPT phases extends well beyond this subclass.
There are also decorated-defect constructions of a different kind, in which a $G$ defect is not decorated by an $S$ charge; instead, it is decorated by a lower-dimensional SPT phase protected by $S$. A prominent example is the $\mathbb{Z}_2 \times$ SO(3) SPT in $2d$, where the domain wall of an Ising paramagnet (charged under $\mathbb{Z}_2$) is decorated by an AKLT chain (a $1d$ SPT protected by SO(3)). In this setting, fully dephasing the $\mathbb{Z}_2$ charge does not necessarily trigger SWSSB of the SO(3) symmetry. Nevertheless, dephasing might still produce interesting phenomena, which we briefly comment on below, with a more extensive exploration deferred to future work.

Consider a $2d$ honeycomb lattice with an Ising paramagnet residing at the centers of plaquettes. The domain walls of the Ising configuration then live on the edges of the honeycomb lattice and can be decorated by an AKLT chain, as studied in Ref.~\cite{yao2010symmetry,li2014topology}. After fully dephasing the Ising charge, the resulting mixed state becomes an incoherent sum over all Ising patterns $\{x_i\}$. For each fixed $\{x_i\}$, the state contains a coherent superposition of all closed domain-wall loops on the edges of the honeycomb lattice, with each loop decorated by an AKLT chain.
This superposition of decorated-loops is closely related to the resonating-valence-loop (RVL) states explored in Ref.~\cite{yao2010symmetry,savary2015quantum}, whose infrared behavior can depend sensitively on lattice geometry. For the honeycomb lattice, the RVL state exhibits quasi-long-range order in dimer-dimer correlations, whereas for the kagome lattice, PEPS simulations \cite{li2014topology} suggest a phase akin to a $\mathbb{Z}_2$ spin liquid. This opens the possibility that dephasing a general SPT wavefunction may induce a mixed-state spin liquid, a direction we leave for future investigation.

 \acknowledgments 
This research was completed in part by grant NSF PHY-2309135 to the Kavli Institute for Theoretical Physics (KITP). YY and LQS acknowledge support from NSF under Award No.~DMR-2439118 and  Cottrell Scholar Award No. CS-CSA-2026-068 from Research Corporation for Science Advancement. We thank Yuxuan Guo and Xuejia Yu for helpful discussions.

The data and code used in this work is available on Zenodo \cite{lu2025a,shuangyuanlu2025, shuangyuanlu2025a, shuangyuanlu2026}.

\appendix   

\section{Calculation of CMI and Negativity} \label{app:cmi_negativity}
In Sec.~\ref{sec:entanglement}, we present numerical results for the conditional mutual information (CMI) and entanglement negativity of the one-dimensional ZXZ chain with a transverse field in Eq.~\eqref{eq.hamiltonian_zxz}, obtained for a large system size $L=1000$. Computing the CMI and negativity for a generic one-dimensional quantum state is highly nontrivial, even when the state is represented as a matrix product state (MPS). In this appendix, we describe the procedure used in our calculations.

The computations of CMI and negativity share similar numerical techniques, which we explain together. Two key properties make the numerical evaluation of these quantities feasible. First, we employ a projective measurement quantum channel \eqref{eq.zxz_channel}. This choice renders the overall density matrix explicitly diagonal in the measurement basis of the even sites, simplifying the problem to that of pure-state calculations. Consequently, the MPS-based computation combined with Monte Carlo sampling provides a general approach for evaluating such mixed-state CMI and negativity.
Second, the ZXZ model possesses a special structural property: its wavefunction can be expressed in a `separated' form between even and odd sites. This separation allows the computation of mixed-state CMI and negativity to be reduced to that of a single pure-state CMI and negativity. In the following, we elaborate on these two points in detail, showing how they simplify the evaluation of CMI and negativity.

\subsection{General MPS States with Projective Measurement Channel}
Since the decoherence channel \eqref{eq.zxz_channel} is a projective measurement channel, we have  
\bea
\mathcal{E}_i[\rho] = P_i^+ \rho P_i^+ + P_i^- \rho P_i^-
\eea
where $P_i^+$ and $P_i^-$ are projection operators for site $i$ onto the $X$-eigenstates $|+\rangle$ and $|-\rangle$, respectively.  
The resulting mixed state can be expressed in the measurement basis as  
\bea \label{eq.rho_projection}
\rho = \sum_{\{s_\text{even}\}} P_{\{s_\text{even}\}} |\psi_s\rangle \langle \psi_s | P_{\{s_\text{even}\}}
\eea
where $P_{\{s_\text{even}\}}$ projects onto a complete even-site spin configuration $\{s_\text{even}\}$ in the $X$-basis. Each term in the summation corresponds to a distinct even-site spin configuration and is therefore orthogonal to the others. Equation~\eqref{eq.rho_projection} gives the density matrix in block-diagonal form, where each block corresponds to a pure-state sector. We rewrite it in normalized form as  
\bea
\rho = \sum_{\{s_\text{even}\}} w_{\{s_\text{even}\}}|\psi_{\{s_\text{even}\}}\rangle \langle \psi_{\{s_\text{even}\}}| = \sum_{\{s_\text{even}\}} w_{\{s_\text{even}\}}\rho_{\{s_\text{even}\}}
\eea

where $|\psi_{\{s_\text{even}\}}\rangle$ is the normalized wavefunction and $w_{\{s_\text{even}\}}$ is its corresponding weight,
\bea
&w_{\{s_\text{even}\}} = \langle \psi_s |P_{\{s_\text{even}\}} | \psi_s \rangle \\
&|\psi_{\{s_\text{even}\}}\rangle = w_{\{s_\text{even}\}} ^{-\frac12} P_{\{s_\text{even}\}} |\psi_s\rangle
\eea
With this decomposition, the CMI and negativity of the mixed state $\rho$ can be written as weighted averages of their pure-state counterparts $|\psi_{\{s_\text{even}\}}\rangle$. The CMI is given by  
\bea
&I(A: B|C) = S_{AC} (\rho) + S_{BC} (\rho) -S_{C} (\rho) - S_{ABC}(\rho)\\
& = \sum_{\{s_\text{even}\}} \left[S_{AC} (w_{\{s_\text{even}\}}\rho_{\{s_\text{even}\}}) + S_{BC} (w_{\{s_\text{even}\}} \rho_{\{s_\text{even}\}}) \right.\\
&\left. -S_{C} (w_{\{s_\text{even}\}} \rho_{\{s_\text{even}\}}) - S_{ABC}(w_{\{s_\text{even}\}}\rho_{\{s_\text{even}\}})\right]
\eea
Using the property of the von Neumann entropy (for $\mathrm{Tr}\,\rho = 1$),
\bea
S(w \rho) = - \trace{ w \rho \ln ( w \rho)} = w S(\rho ) - w \ln w
\eea
we simplify to  
\bea
&I(A:B|C) = \sum_{\{s_\text{even}\}} w_{\{s_\text{even}\}} \left[ S_{AC} (\rho_{\{s_\text{even}\}}) + S_{BC} (\rho_{\{s_\text{even}\}}) \right. \\
& \left.-S_{C} (\rho_{\{s_\text{even}\}}) - S_{ABC}(\rho_{\{s_\text{even}\}})\right]
\eea
The $w\ln w$ terms cancel. Since $\rho_{\{s_\text{even}\}}$ represents a pure state, $S_{ABC}(\rho_{\{s_\text{even}\}})=0$, and each remaining entropy $S_R$ can be replaced by its complement $\overline{R}$. Hence,  
\bea\label{eq.cmi_ave_mi}
\notag
&I_\rho(A:B|C)\\
\notag
&= \sum_{\{s_\text{even}\}} w_{\{s_\text{even}\}} \left[ S_B(\rho_{\{s_\text{even}\}}) + S_A(\rho_{\{s_\text{even}\}}) - S_{AB}(\rho_{\{s_\text{even}\}})\right]\\
& = \sum_{\{s_\text{even}\}} w_{\{s_\text{even}\}} I_{\rho_{\{s_\text{even}\}}}(A:B)
\eea
Thus, the CMI of $\rho$ can be expressed as the weighted average of the mutual information $I_{\rho_{\{s_\text{even}\}}}(A:B)$ for pure states.  
Since in our calculation both $A$ and $B$ consist of single sites, the reduced density matrices ---and therefore the mutual information ---can be directly obtained through tensor contractions once the MPS form of $|\psi_{\{s_\text{even}\}}\rangle$ is available. The CMI is then computed as a Monte Carlo average over even-site spin configurations sampled according to the weight $w_{\{s_\text{even}\}}$.

Similarly, the entanglement negativity is 
\bea
\mathcal{N}(\rho) = \sum_{\{s_\text{even}\}} w_{\{s_\text{even}\}} \mathcal{N}(\rho_{\{s_\text{even}\}})
\eea
For a pure state, the negativity can be directly obtained from its Schmidt coefficients.  
The wavefunction is  
\bea
|\psi_{\{s_\text{even}\}}\rangle = \sum_i \sqrt{\lambda_i} |i_A\rangle |i_B\rangle
\eea
where $|i_A\rangle$ and $|i_B\rangle$ are orthonormal basis states for the left ($A$) and right ($B$) halves of the chain, and $\sqrt{\lambda_i}$ are the Schmidt coefficients (which, in principle, depend on $\{s_\text{even}\}$ but are omitted for simplicity).  
The partially transposed density matrix is  
\bea
\notag
&\rho_{\{s_\text{even}\}}^{T_A} = \sum_{i, j} \sqrt{\lambda_i\lambda_j} |j_A\rangle |i_B\rangle \langle j_B| \langle i_A|\\
&= \sum_{i} \lambda_i |i_A\rangle |i_B\rangle \langle i_B| \langle i_A| \\
\notag
& + \sum_{i< j} \sqrt{\lambda_i \lambda_j} (|i_A\rangle |j_B\rangle + |j_A\rangle |i_B\rangle )(\langle i_A |\langle j_B| + \langle j_A| \langle i_B|) \\
\notag
& - \sum_{i< j} \sqrt{\lambda_i \lambda_j} (|i_A\rangle |j_B\rangle - |j_A\rangle |i_B\rangle )(\langle i_A |\langle j_B| - \langle j_A| \langle i_B|) 
\eea
The corresponding negativity is
\bea
\notag
&\mathcal{N}(\rho_{\{s_\text{even}\}}) = \sum_{i<j} \sqrt{\lambda_i \lambda_j}\\
&= \frac12 \left [ \left(\sum_i \sqrt{\lambda_i}\right)^2 - 1\right]
\eea
where we used $\sum_i \lambda_i = 1$.  
Finally, the negativity of the mixed state $\rho$ can be expressed as the weighted average of pure-state negativities and computed directly from the entanglement spectra,  
\bea\label{eq.negativity_ave_entanglement_spec}
\mathcal{N}(\rho) = \sum_{\{s_\text{even}\}} w_{\{s_\text{even}\}} \cdot \frac12 \left [ \left(\sum_i \sqrt{\lambda_{i, \{s_\text{even}\}}}\right)^2 - 1\right]
\eea
Since each projected pure state is represented in MPS form, the entanglement spectrum can be efficiently obtained by shifting the canonical center of the MPS. The averaging over different projection configurations $\{s_\text{even}\}$ is then performed using Monte Carlo sampling.

\subsection{Special Property of ZXZ Model States}

Although we have derived general expressions for the CMI and negativity and explained how they can be computed using MPS states combined with Monte Carlo sampling, the calculation for the ZXZ model with a transverse field in Eq.~\eqref{eq.hamiltonian_zxz} can be simplified even further.

We begin by considering a global transformation that separates the Hamiltonian into independent even- and odd-site parts. For each even-site spin $X$ configuration, denoted as $\{s_\text{even}\}$, we define an operator acting on the odd spins, $X_{\overline{\{s_\text{even}\}}}$. The set $\overline{\{s_\text{even}\}}$ consists of odd sites determined by the even-site spin configuration $\{s_\text{even}\}$. Specifically, we collect all even sites with $X=-1$ in the configuration $\{s_\text{even}\}$ to form an ordered set
\bea
S^- = (a_1, a_2, \cdots, a_K) ~~~~~~ a_i<a_j \iff i < j
\eea
Then, we define
\bea\notag
\overline{\{s_{\text{even}}\}} =
\{\, 2n + 1 ,\ n \in \mathbb{N} \mid \exists k \in \mathbb{N},\ 2k \le K,\ a_{2k-1} < 2n + 1 < a_{2k}\,\}.
\eea
In other words, $\overline{\{s_\text{even}\}}$ contains all odd sites that lie between the $(2k-1)$-th and $(2k)$-th occurrences of $X=-1$ in the configuration $\{s_\text{even}\}$.

Next, we define a product of $X$ operators acting on these odd sites,
\bea
X_{\overline{\{s_\text{even}\}}} = \prod_{i \in \overline{\{s_\text{even}\}}} X_i
\eea
and introduce the unitary transformation
\bea
U = \sum_{\{s_\text{even}\}} P_{\{s_\text{even}\}} X_{\overline{\{s_\text{even}\}}}
\eea
where $P_{\{s_\text{even}\}}$ projects onto the even-site configuration $\{s_\text{even}\}$. This transformation separates the Hamiltonian into independent even- and odd-site parts:
\bea
U H U^{-1} = - h\sum_i X_i - \sum_{i = 2}^{L-1} Z_{i-1} Z_{i+1}
\eea
It is evident that the transformed Hamiltonian decomposes into two decoupled transverse-field Ising models. Consequently, the ground-state wavefunction is a product of two Ising-model ground states.
The ground-state wavefunction of $H$ is: 
\bea
&\psi = U|\psi^{\text{Ising}}_{\text{even}}\rangle \otimes |\psi^{\text{Ising}}_{\text{odd}}\rangle \\
&= \sum_{\{s_{\text{even}}\}} \psi^{\text{Ising}}_{\text{even}} |\{s_{\text{even}}\}\rangle \otimes X_{\overline{\{s_\text{even}\}}} |\psi^{\text{Ising}}_{\text{odd}}\rangle
\eea
With this form of the ground-state wavefunction, we observe that after projection on the even-site spins, the resulting odd-site state is equivalent to the Ising-model ground state up to an onsite unitary transformation $X_{\overline{\{s_\text{even}\}}}$. Since onsite unitaries do not change the entanglement spectrum, quantities that depend only on the entanglement spectrum --- such as $S_R(\rho_{\{s_\text{even}\}})$ in Eq.~\eqref{eq.cmi_ave_mi} and $\lambda_{i,\{s_\text{even}\}}$ in Eq.~\eqref{eq.negativity_ave_entanglement_spec} --- are independent of $\{s_\text{even}\}$. The ensemble average therefore becomes trivial, and it suffices to consider one particular even-site configuration, for instance, the one with all even-site spins satisfying $X=1$.

Denoting the density matrix of this configuration as $\rho_0$ and its Schmidt coefficients as $\lambda_i$, Eqs.~\eqref{eq.cmi_ave_mi} and \eqref{eq.negativity_ave_entanglement_spec} reduce to
\bea
&I_\rho(A:B|C) = I_{\rho_0}(A:B)\\
& \mathcal{N}(\rho) = \frac12 \left [ \left(\sum_i \sqrt{\lambda_i}\right)^2 - 1\right]
\eea

\section{Explicit Fixed-point Construction of SPT States and SWSSB} \label{app:group_cohomology_swssb}
Following the discussion in Sec.~\ref{sec:gene}, we now give an explicit fixed-point wavefunction construction of one-dimensional bosonic SPT states. For a one-dimensional bosonic chain with an on-site unitary symmetry $G\times S$, SPT phases are classified by $H^2(G\times S,U(1))$. By the K"unneth formula, one contribution to this cohomology group takes the form:
\bea
H^1\bigl(G, H^1(S,U(1))\bigr),
\eea
which specifies a family of SPT phases admitting a domain-wall decoration description: an $S$ charge is bound to the domain wall of $G$.

To make this structure explicit, recall that for a group $G$ acting trivially on an abelian $G$-module $M$, the first cohomology group reduces to group homomorphisms,
\bea
&H^1(G,M)=\Hom(G,M) \\
&= \Bigl\{\omega:G\to M \,\Big|\, \omega(g_1)\,\omega(g_2)=\omega(g_1g_2),~\forall g_1,g_2\in G\Bigr\}.
\eea
Since $M$ is abelian, $\omega$ is insensitive to commutators: every element of the commutator subgroup
\bea
[G,G]=\{g_1g_2g_1^{-1}g_2^{-1}\mid \forall g_1,g_2\in G\}
\eea
maps to the identity in $M$. Therefore $\omega$ factors through the abelianization $G^{ab}\equiv G/[G,G]$, and equivalently
\bea
H^1(G,M)=\Hom(G^{ab},M).
\eea
Applying this to $M=H^1(S,U(1))$, we obtain the domain-wall decoration classification
\bea
H^1\bigl(G,H^1(S,U(1))\bigr)
=\Hom \bigl(G,\Hom(S^{ab},U(1))\bigr).
\eea
Here $\Hom(S^{ab},U(1))$ is the group of one-dimensional representations of $S$ (i.e., $S$ charges). When $S$ is abelian, $S^{ab}=S$. A homomorphism
\bea
\omega: G \rightarrow \Hom(S,U(1))
\eea
thus specifies the $S$ charge attached to a given $G$ domain wall, in a manner consistent with the group multiplication in $G$.

\paragraph{Lattice Hilbert space and symmetry action.}
To construct fixed-point models realizing these decorated-domain-wall SPTs, we consider a one-dimensional chain with alternating degrees of freedom: $G$ acts on even sites and $S$ acts on odd sites. The local Hilbert space on each even site is spanned by group elements $\{|g\rangle\}_{g\in G}$, while that on each odd site is spanned by $\{|s\rangle\}_{s\in S}$.

On even sites, for $g_1\in G$ we define the left-regular action
\bea
X^{g_1}\,|g\rangle = |g_1 g\rangle , \qquad \forall g,g_1\in G,
\eea
together with projectors onto the group-element basis,
\bea
P^{g_1}\,|g\rangle = \delta_{g_1,g}\,|g\rangle , \qquad \forall g,g_1\in G.
\eea
On odd sites, it is convenient to work in a charge basis labeled by one-dimensional representations $\eta\in \Hom(S,U(1))$. We define
\bea
|\eta\rangle \equiv \frac{1}{\sqrt{|S|}}\sum_{s\in S}\eta(s)^{-1}\,|s\rangle ,
\eea
with the corresponding projector
\bea
P^{\eta_1}\,|\eta\rangle = \delta_{\eta,\eta_1}\,|\eta\rangle .
\eea
We also introduce the charged operator $Z_{\eta}$ acting on the group-element basis as
\bea
Z_{\eta}\,|s\rangle = \eta(s)^{-1}\,|s\rangle , \qquad s\in S,\ \eta\in \Hom(S,U(1)).
\eea

\paragraph{Fixed-point Hamiltonian.}
The decorated-domain-wall SPTs in this subclass are in one-to-one correspondence with homomorphisms
\bea
\omega: G \rightarrow \Hom(S,U(1)),
\eea
which assign an $S$ charge to each group element in a way compatible with multiplication. A fixed-point commuting-projector Hamiltonian realizing the corresponding SPT may be written as
\begin{align}
H^\omega
=
-\sum_{i~\text{odd}}\ \sum_{g_1,g_2\in G}
P^{g_1}_{i-1}\, P^{\omega_{g_1 g_2^{-1}}}_{i}\, P^{g_2}_{i+1}
-\sum_{i~\text{even}}\ \sum_{g\in G}
Z^{\omega_{g^{-1}}}_{i-1}\, X^g_{i}\, Z^{\omega_{g}}_{i+1}.
\end{align}
The first term pins the $S$ charge on the odd site $i$ to be $\omega_{g_1 g_2^{-1}}$, which depends only on the domain wall between the neighboring $G$ variables $g_1$ and $g_2$. The second term fluctuates the $G$ configuration by applying $X^g_i$ and simultaneously updates the adjacent $S$ charges by the corresponding $Z$ operators, ensuring exact $G\times S$ symmetry. Together, these terms encode an SPT in which $S$ charges are bound to $G$ domain walls.

\paragraph{Fixed-point ground state.}
A ground state can be written explicitly as
\bea
|\psi\rangle
= C\sum_{\{g\}}
|\{g\}\rangle \otimes
\Bigl|\{\eta_i=\omega_{g_{i-1}g_{i+1}^{-1}}\}\Bigr\rangle ,
\eea
where the sum is over all configurations $\{g\}$ on even sites, $\eta_i$ labels the charge state on odd sites, and $C$ is a normalization constant.

\paragraph{$S$-charge dephasing and R\'enyi-2 correlator:}
Applying the on-site $S$-charge dephasing channel in Eq.~\eqref{eq.quantum_channel_eta} yields the mixed state $\rho=\mathcal{E}(|\psi\rangle\langle\psi|)$:
\begin{align}
\rho
=& C^2\sum_{\{g\},\{h\}}
\Biggl[\prod_{i~\text{odd}}
\delta_{\omega_{g_{i-1}g_{i+1}^{-1}},\,\omega_{h_{i-1}h_{i+1}^{-1}}}
\Biggr] \nonumber \\
&|\{g\}\rangle\Bigl|\{\eta_i=\omega_{g_{i-1}g_{i+1}^{-1}}\}\Bigr\rangle
\Bigl\langle \{\eta_i=\omega_{h_{i-1}h_{i+1}^{-1}}\}\Bigr|\langle\{h\}|.
\end{align}

We now define an order-parameter operator $O$ acting on even sites. Choose an element $s\in S$ such that the phase $\omega_g(s)$ is not identically $1$ as a function of $g$ (this is possible when $\omega$ is nontrivial). We define $O$ to be diagonal in the $|g\rangle$ basis,
\bea
O\,|g\rangle = \omega_g(s)\,|g\rangle .
\eea
Since $\omega_g(\cdot)$ is a one-dimensional representation of $S$, $\omega_g(s)\in U(1)$ is a pure phase.

The R\'enyi-2 correlator in the mixed state is
\begin{align}
&R^{(2)}(r,r^\prime)
= \frac{\trace{\rho\, O_r O_{r^\prime}^\dagger\, \rho\, O_r^\dagger O_{r^\prime}}}
{\trace{\rho^2}} \notag\\
&= \frac{
\sum_{\{g\},\{h\}}
\Bigl[\prod_{i~\text{odd}}
\delta_{\omega_{g_{i-1}g_{i+1}^{-1}},\,\omega_{h_{i-1}h_{i+1}^{-1}}}\Bigr]\,
\omega_{g_r}(s)\,\omega_{g_{r^\prime}}^*(s)\,
\omega_{h_r}^*(s)\,\omega_{h_{r^\prime}}(s)}
{
\sum_{\{g\},\{h\}}
\Bigl[\prod_{i~\text{odd}}
\delta_{\omega_{g_{i-1}g_{i+1}^{-1}},\,\omega_{h_{i-1}h_{i+1}^{-1}}}\Bigr]
}.
\end{align}
Using the identity
\bea
\frac{\omega_{g_r}(s)\,\omega_{h_r}^*(s)}
{\omega_{g_{r^\prime}}(s)\,\omega_{h_{r^\prime}}^*(s)}
=
\prod_{k=1}^{\frac{(r^\prime-r)}{2}-1}
\omega_{g_{r+2k}\, g_{r+2k+2}^{-1}}(s)\,
\omega_{h_{r+2k}\, h_{r+2k+2}^{-1}}^*(s),
\eea
we see that the product on the right-hand side equals $1$ for all configurations allowed by the $\delta$-function constraints in the sums. Therefore the numerator reduces to the denominator, and since $|\omega_{g_r}(s)\omega_{h_r}^*(s)|^2=1$, we obtain
\bea
R^{(2)}(r,r^\prime)=1.
\eea
Finally, because $\omega_g(s)$ transforms nontrivially under the $G$ action, the operator $O$ carries $G$ charge.

\section{Insight from Non-invertible Symmetries} 

In this section, we analyze how dephasing affects SPT wavefunctions from the perspective of non-invertible symmetries. As a concrete example, we revisit the 1d SPT phase discussed in Sec.~\ref{sec:1d_zxz} protected by a $\mathbb{Z}_2 \times \mathbb{Z}_2$ symmetry, and specialize to the $\mathbb{Z}_2$ cluster model:
\begin{align} H = -\sum_{j} (Z_{2j-1} X_{2j} Z_{2j+1} + Z_{2j-2}X_{2j-1} Z_{2j} )  
\label{H-cluster} 
\end{align}
 The ground state of the cluster Hamiltonian is given by
\begin{align}\label{eq:psi}
    |\psi \rangle & \propto \sum_{\bf g} \omega^{  \sum_j g_{2j} (g_{2j-1} -g_{2j+1})} |\bf g \rangle 
\end{align}

The key observation is that the cluster Hamiltonian is invariant under the transformation:
\bea
X_j \rightarrow Z_{j-1} Z_{j+1}\,, \qquad Z_{j-1}Z_{j+1} \rightarrow X_j  \,, \label{KW2}
\eea

This transformation is implemented by the following conserved operator:
\bea
D  =  T_x D^{a} D^{b} \label{D.TDD}
\eea
where
\begin{equation}
\begin{aligned}
D^{a} &= 
\frac{1+\eta^{a}}{\sqrt{2}}\,
\frac{1-i X_{2L}}{\sqrt{2}}\cdots
\frac{1-i Z_{4}Z_{2}}{\sqrt{2}}\,
\frac{1-i X_{2}}{\sqrt{2}},\\
D^{b} &= 
\frac{1+\eta^{b}}{\sqrt{2}}\,
\frac{1-i X_{2L-1}}{\sqrt{2}}\cdots
\frac{1-i Z_{3}Z_{1}}{\sqrt{2}}\,
\frac{1-i X_{1}}{\sqrt{2}}.
\end{aligned}
\end{equation}
Here $T_x$ is the lattice translation by unit cell, $\eta_a (\eta_b)$ projects the total charge parity on even (odd) sites into the net charge sector.

This is not a conventional symmetry operator: it has a nontrivial kernel and is therefore not invertible. It instead generates the following non-invertible symmetry transformation:
\begin{equation}
D X_j = Z_{j-1} Z_{j+1} D,
\qquad
D Z_{j-1} Z_{j+1} = X_j D.
\end{equation}

Since the ground-state wavefunction in Eq.~\eqref{eq:psi} is invariant under the $D$ symmetry, we can also implement the transformation to the dephasing channel:
\bea
&\mathcal{E}_{i}[\rho] = \frac{1}{2}\Big(\rho + X_{2i} \rho X_{2i}\Big) \nonumber\\
\rightarrow &\mathcal{E}_{i}[\rho] = \frac{1}{2}\Big(\rho + Z_{2i-1} Z_{2i+1}  \rho Z_{2i-1} Z_{2i+1}\Big)
\eea
This suggests that, for SPT cluster state with a non-invertible symmetry, dephasing the local charge density is effectively equivalent to introducing incoherent noise that creates charge-anticharge pairs.

Similar reasoning applies to the gapless SPT discussed in Sec.~\ref{sec:gapless}. We act with the non-invertible symmetry transformation in Eq.~\ref{D.TDD} on the Hamiltonian
\bea\label{eq:gapless}
\hat{H} = - \sum_i\Big( Z_{2i}X_{2i+1}Z_{2i+2}+X_{2i}(Y_{2i-1}Y_{2i+1}+ Z_{2i-1}Z_{2i+1})\Big)\nonumber\\
\eea
Under this transformation, the relevant operators map as
\bea
X_{2i+1} \rightarrow Z_{2i}Z_{2i+2}, \qquad Z_{2i}Z_{2i+2} \rightarrow X_{2i+1} , \nonumber\\
X_{2i} \rightarrow Z_{2i-1}Z_{2i+1}, \qquad Z_{2i-1}Z_{2i+1} \rightarrow X_{2i} ,
\nonumber\\
Y_{2i-1}Y_{2i+1} \rightarrow Z_{2i-2} X_{2i} Z_{2i+2}  .
\eea
Therefore the transformed Hamiltonian becomes
\bea\label{eq:gapless2}
\hat{H'} = - \sum_i\Big( Z_{2i}X_{2i+1}Z_{2i+2}+X_{2i}(1+Z_{2i-2}Z_{2i+2}) Z_{2i-1}Z_{2i+1}\Big)\nonumber\\
\eea
Although $\hat{H}'$ is microscopically different from $\hat{H}$, they realize the same intrinsic gapless SPT phase \cite{ma2024topological,levin05}. Consequently, applying a charge dephasing (measurement) channel on $X_{2i+1}$ to the ground state of Eq.~\eqref{eq:gapless} is equivalent to applying a channel that measures the corresponding charge-pair operator $Z_{2i}Z_{2i+2}$ on the ground state of Eq.~\eqref{eq:gapless2}.

\section{Relation to MBQC in Doubled Space} \label{sec:mbqc}

A quantum channel is a linear map acting on the density matrix $\rho$, and is therefore also linear on the doubled state $\kket{\rho}$. In the doubled Hilbert space, the quantum channel in Eq.~\eqref{eq.zxz_channel} can be represented as a projection operator onto the subspace satisfying $X_i^A X_i^B = 1$, where $A$ and $B$ label the bra and ket layers, respectively. Since measurement on a pure state is mathematically described by a projection operator, the quantum channel can thus be interpreted as a measurement process in the doubled space.
It is well known that measuring a pure SPT state can generate long-range entanglement together with a spontaneously symmetry-breaking pattern.
Viewing decoherence as an effective measurement in the Choi-double picture, one can relate decoherence-driven SWSSB in the resulting mixed state to measurement-induced long-range order in the corresponding pure-state setting.
Moreover, this correspondence between quantum channels and pure-state measurements extends beyond the fully decohered limit and can be formulated more generally through purification. The quantum channel we consider is
\bea\label{eq.zxz_channel_p}
\mathcal{E} = \prod_{i} \mathcal{E}_{2i} \quad \mathcal{E}_{2i}[\rho] = (1-p) \rho +p X_{2i} \rho X_{2i}
\eea
which reduces to Eq.~\eqref{eq.zxz_channel} when $p=\tfrac{1}{2}$, corresponding to complete decoherence.

To construct an explicit purification, we realize the channel unitarily by introducing ancilla qubits. Such a local unitary transformation exists for any local quantum channel by construction. For the 1d ZXZ model, we attach a spin-$\frac12$ ancilla to each even site, initialized in the product state
\begin{equation}
|\widetilde{\psi}_a\rangle = \otimes_{2i} |\widetilde{ +}\rangle_{2i}
\end{equation}
We use subscripts $s$ and $a$ to distinguish system and ancilla degrees of freedom, and we denote ancilla operators and states with a tilde. Let $|\psi_s\rangle$ be the ground state of the Hamiltonian in Eq.~\eqref{eq.hamiltonian_zxz}. We couple system and ancilla via a local unitary
\bea
U = \prod_{i} U_{2i}
\eea
where each gate acts on an even site and its attached ancilla as
\bea
\notag
&U_i = \frac12 ( \mathbb{I}_i + X_i) \otimes (\cos \theta ~\widetilde{\mathbb{I}}_i + i\sin \theta~ \widetilde{Z}_i) \\
&+ \frac12 (\mathbb{I}_i - X_i) \otimes (\sin \theta~ \widetilde{\mathbb{I}}_i + i\cos \theta ~ \widetilde{Z}_i)
\eea
Tracing out the ancillas yields a mixed state on the system, and this construction realizes the channel
\bea \label{eq.Psi_rho_relation}
\text{Tr}_a \left[ U ~\left(|\psi_s\rangle \otimes |\widetilde{\psi}_a \rangle \langle \widetilde{\psi}_a | \otimes \langle \psi_s| \right)~ U^\dagger \right] = \mathcal{E} [|\psi_s\rangle \langle \psi_s|] 
\eea
The resulting map is of the form \eqref{eq.zxz_channel_p}, with the channel parameter related to the rotation angle by
\bea
1-2p = \sin 2\theta
\eea
In particular, $p=\tfrac12$ corresponds to $\theta=0$ and reproduces the complete dephasing channel in Eq.~\eqref{eq.zxz_channel}. Varying $\theta$ therefore interpolates continuously between different decoherence strengths.

The pure state
\bea
|\Psi\rangle=U\big(|\psi_s\rangle\otimes|\widetilde{\psi}_a\rangle\big)
\eea
is a purification of $\rho=\mathcal{E}[|\psi_s\rangle\langle\psi_s|]$, so SWSSB can be equivalently discussed in terms of $|\Psi\rangle$. To connect the R\'enyi-2 correlator to an ordinary correlator in a pure-state setting, we introduce two copies labeled by $A$ and $B$. The relation between the doubled mixed state $\kket{\rho}$ and the doubled purified state $|\Psi\rangle^A\otimes|\Psi^*\rangle^B$ follows directly from Eq.~\eqref{eq.Psi_rho_relation}: $\rho$ is obtained by tracing out the ancillas. In the doubled space, tracing out corresponds to projecting the ancilla degrees of freedom onto a product of EPR pairs,
\bea
|\widetilde{\text{EPR}}_a\rangle = \otimes_{2i} ~ \frac{1}{\sqrt{2}} \left(| \widetilde{+} \rangle_{2i}^A  \otimes |\widetilde{+}\rangle_{2i}^B + | \widetilde{-}\rangle^A_{2i} \otimes |\widetilde{-}\rangle^B_{2i} \right)
\eea
which entangles the two replicas site by site. Equivalently, the Choi-double $\kket{\rho}$ is obtained by projecting the ancilla sector of $|\Psi\rangle^A\otimes|\Psi^*\rangle^B$ onto this state:
\bea \label{eq.rho_epr_Psi}
\kket{\rho} = \langle \widetilde{\text{EPR}}_a| ~ \left(|\Psi\rangle^A \otimes |\Psi^*\rangle^B \right)
\eea
Physically, this corresponds to measuring each ancilla pair in the EPR basis and post-selecting the outcome in which all pairs collapse onto the symmetric EPR state above; the post-measurement state on the system layers is precisely $\kket{\rho}$.

Consequently, for an operator of the form $O_r O_{r^\prime}^\dagger$ acting on the system, the R\'enyi-2 correlator can be written as an ordinary expectation value in the doubled purified state with an ancilla projection inserted:
\begin{align}\label{eq.epr_correlator}
&C(r, r^\prime) = R^{(2)}(r, r^\prime) = \\
&\frac{{}^B\langle \Psi^* |\otimes{}^A\langle \Psi |~ \left( \widetilde{P}_{\text{EPR}, a} \otimes \left( O_r^A {O_{r^\prime}^\dagger}^A\otimes O_r^B {O_{r^\prime}^\dagger}^B \right)\right)~| \Psi \rangle^A\otimes| \Psi^* \rangle^B}{{}^B\langle \Psi^* |\otimes{}^A\langle \Psi |~ \left ( \widetilde{P}_{\text{EPR}, a} \otimes \mathbb{I}_s^{AB} \right)~| \Psi \rangle^A\otimes| \Psi^* \rangle^B}\nonumber
\end{align} 
where $\widetilde{P}_{\mathrm{EPR},a}=|\widetilde{\mathrm{EPR}}_a\rangle\langle\widetilde{\mathrm{EPR}}_a|$ projects only the ancilla spins and $\mathbb{I}_s^{AB}$ is the identity on the system spins in both layers. For the 1d ZXZ model, taking $O_r = Z_{2r-1}, O_{r^\prime} = Z_{2r^\prime+1}$ reproduces $R^{(2)}$ in Eq.~\eqref{eq.zxz_renyi_2}.

This establishes a direct relation between the R\'enyi-2 correlator $R^{(2)}(r, r^\prime)$ \eqref{eq.zxz_renyi_2} of the mixed state $\rho$ and the ordinary correlator $C(r, r^\prime)$ in the corresponding pure-state measurement problem defined by Eq.\eqref{eq.rho_epr_Psi}.

\section{R\'enyi-2 Correlator and Strange Correlator}\label{sec:sc}

In this section, we relate the R\'enyi-2 correlator of the mixed-state density matrix to a strange correlator of the purified state.

We build this relation starting from Sec.~\ref{sec:mbqc}, where we realized the quantum channel via a local unitary $U$ acting on the system and ancillas, and thus obtained an explicit purification of the mixed state $\rho$. This construction led to an equivalent expression for the R\'enyi-2 correlator, Eq.~\eqref{eq.epr_correlator}. We now insert a resolution of the identity on the doubled system Hilbert space,
\bea
\mathbb{I}_s^{AB} = \sum_{s^A, s^B} | s^A, s^B\rangle \langle  s^A, s^B|
\eea
where $|s^A,s^B\rangle$ are product basis states on the $A$ and $B$ system layers, chosen to respect the symmetries that protect the SPT phase. The correlator can then be rewritten as
\begin{widetext}

\begin{equation}
\begin{aligned}\label{eq.renyi_2_strange}
C({x, y})& = \frac{ \sum_{s^A, s^B} {}^A \langle \Psi | \otimes {}^B\langle \Psi^*|\left(| \widetilde{\text{EPR}}_a; s^A, s^B \rangle ~\langle \widetilde{\text{EPR}}_a; s^A, s^B|O_r^A {O_{r^\prime}^\dagger}^A\otimes O_r^B {O_{r^\prime}^\dagger}^B \right)| \Psi\rangle ^A \otimes|\Psi^* \rangle ^B}
{\sum_{s^A, s^B}\left|\langle \widetilde{\text{EPR}}_a; s^A, s^B|| \Psi\rangle ^A \otimes|\Psi^* \rangle ^B\right|^2} \\
& =\frac{ \sum_{s^A, s^B} \left|\langle \widetilde{\text{EPR}}_a; s^A, s^B|| \Psi\rangle ^A \otimes|\Psi^* \rangle ^B\right|^2 C_{\text{strange}, s^A, s^B} (r, r^\prime)  }
{\sum_{s^A, s^B}\left|\langle \widetilde{\text{EPR}}_a; s^A, s^B|| \Psi\rangle ^A \otimes|\Psi^* \rangle ^B\right|^2}
\end{aligned}    
\end{equation}
where $| \widetilde{\text{EPR}}_a; s^A, s^B \rangle = | \widetilde{\text{EPR}}_a\rangle \otimes | s^A, s^B \rangle$, and $C_{\text{strange}, s^A, s^B} (r, r^\prime)$ denotes the strange correlator:
\bea
C_{\text{strange}, s^A, s^B} (r, r^\prime) = \frac{\langle \widetilde{\text{EPR}}_a; s^A, s^B|O_r^A {O_{r^\prime}^\dagger}^A\otimes O_r^B {O_{r^\prime}^\dagger}^B| \Psi\rangle ^A \otimes|\Psi^* \rangle ^B}{\langle \widetilde{\text{EPR}}_a; s^A, s^B|| \Psi\rangle ^A \otimes|\Psi^* \rangle ^B}
\eea
\end{widetext}
This quantity represents the strange correlator of the pure state $| \Psi\rangle ^A \otimes|\Psi^* \rangle ^B$ with respect to the symmetric, trivial reference state $| \widetilde{\text{EPR}}_a; s^A, s^B \rangle$. We thus find that the correlator $C(r, r^\prime)$ (equivalent to the R\'enyi-2 correlator) is a weighted average of strange correlators $C_{\text{strange}, s^A, s^B} (r, r^\prime)$ over all product-state configurations $|s^A, s^B\rangle$, with weights $\left|\langle \widetilde{\text{EPR}}_a; s^A, s^B|| \Psi\rangle ^A \otimes|\Psi^* \rangle ^B\right|^2$.

\section{Monte Carlo Simulation of the 2d QSH Model}\label{sec:MCappen}

In this section we describe how to compute the R\'enyi-2 correlator of a two-dimensional QSH state under decoherence using a Monte Carlo (MC) algorithm. The approach is a standard ``wavefunction Monte Carlo'' method for evaluating observables by sampling the configuration weight of a many-body wavefunction. Closely related techniques appear in variational Monte Carlo and in determinant-based Monte Carlo methods for free-fermion and auxiliary-field formulations. We first outline the method in a model-independent way and then specialize to our QSH wavefunction.

\paragraph*{MC evaluation of observables from a wavefunction.}
Suppose we are given a normalized (or unnormalized) wavefunction $|\psi\rangle$ and a basis $\{|i\rangle\}$ such that the amplitudes
\bea
\psi_i=\langle i|\psi\rangle
\eea
can be computed efficiently for any configuration $i$. The expectation value of an operator $O$ is
\begin{align}
\langle O\rangle
&=\frac{\langle\psi|O|\psi\rangle}{\langle\psi|\psi\rangle}
=\frac{\sum_{i,j}\psi_j^*\,O_{ji}\,\psi_i}{\sum_i|\psi_i|^2}\nonumber\\
&=\frac{\sum_i |\psi_i|^2\,O_{\mathrm{loc}}(i)}{\sum_i|\psi_i|^2},
\end{align}
where we defined the local estimator
\bea
O_{\mathrm{loc}}(i):=\sum_j \frac{\psi_j^*}{\psi_i^*}\,O_{ji}.
\eea
If $O$ is local (or sparse) in the chosen basis, then for a given $i$ only a finite number of $j$ contribute and $O_{\mathrm{loc}}(i)$ can be evaluated efficiently. One may then sample configurations $i$ from the probability distribution
\bea
\pi(i)=\frac{|\psi_i|^2}{\sum_{i'}|\psi_{i'}|^2}
\eea
using a Markov chain, and estimate the expectation value as the MC average
\bea
\langle O\rangle = \overline{O_{\mathrm{loc}}(i)}\qquad (i\sim \pi).
\eea

\paragraph*{Wavefunction amplitudes for the doubled-space state.}
In our problem, the R\'enyi-2 correlator of the QSH state under decoherence can be written as an expectation value in the doubled (Choi) space. Accordingly, we need to evaluate amplitudes of the doubled-space wavefunction $\kket\rho$ in the product basis $|i\rangle\otimes|i'\rangle$. We use the convention
\bea
 \brakket{i,i'}{\rho} \equiv \rho_{i,i'} = p_{i,i'}\,\psi_i^*\,\psi_{i'},
\eea
where $\psi$ is the many-body wavefunction of the (pure) QSH free-fermion state (including both spin-up and spin-down layers), and $i,i'$ label occupation-number (Fock) configurations. The factor $p_{i,i'}$ is contributed by the decoherence channel in Eq.~\eqref{eq.qc_sqhe_1}. Since the channel is diagonal in the occupation basis, it only contributes a configuration-dependent scalar factor and can be computed directly once $i,i'$ are specified.

The dominant numerical cost is therefore the evaluation (and update) of the free-fermion amplitudes $\psi_i$.

\paragraph*{Determinant form and fast ratio updates.}
For a Slater-determinant (free-fermion) state, the amplitude in an occupation configuration $i$ is given by a determinant. Let $N$ be the number of occupied single-particle orbitals and let $\Phi$ be the $M\times N$ matrix whose columns are the occupied single-particle wavefunctions in the real-space basis. For a configuration $i$ specifying a set of occupied sites $\{r_1,\dots,r_N\}$, we form the $N\times N$ submatrix $\Phi_i$ by selecting the corresponding rows, and
\bea
\psi_i = \det(\Phi_i).
\eea
Computing $\det(\Phi_i)$ from scratch costs $O(N^3)$, which would be too expensive if done at every MC proposal. However, in a local-update Markov chain, successive configurations typically differ by moving a single particle (or exchanging an occupied and an unoccupied site). In that case, $\Phi_i$ changes by a rank-one modification, and the determinant ratio
\bea
\phi=\frac{\psi_{i^{(2)}}}{\psi_{i^{(1)}}}
\eea
can be computed in $O(N^2)$ time using standard determinant-update identities (equivalently, Sherman--Morrison/Woodbury updates for the inverse matrix). Concretely, if $A$ denotes the current $N\times N$ matrix and $A\to A+u v^T$ under a single local move, then
\bea
\frac{\det(A+u v^T)}{\det(A)} = 1+ v^T A^{-1}u,
\eea
and the inverse can be updated as
\bea
(A+u v^T)^{-1}=A^{-1}-\frac{A^{-1}u v^T A^{-1}}{1+v^T A^{-1}u}.
\eea
Thus each accepted local update costs $O(N^2)$ once $A^{-1}$ is maintained, while an occasional recomputation at $O(N^3)$ can be used for numerical stabilization.

\paragraph*{Overall scaling.}
With $N_{\mathrm{MC}}$ MC sweeps and $O(N)$ local proposals per sweep, the total runtime scales as
\bea
T = O\!\left(N_{\mathrm{MC}}\,N^3\right),
\eea
assuming determinant ratios are evaluated via $O(N^2)$ updates. This determinant-based MC framework allows us to estimate general expectation values of the doubled-space wavefunction $|\rho\rangle\rangle$, including the R\'enyi-2 correlator studied in this work.

\bibliography{ref}

@article{wen03,
  author = {Wen, Xiao-Gang},
  title = {Quantum Orders in an Exact Soluble Model},
  journal = {Physical Review Letters},
  year = {2003},
  volume = {90},
  number = {1},
  pages = {016803},
  doi = {10.1103/PhysRevLett.90.016803},
  url = {https://link.aps.org/doi/10.1103/PhysRevLett.90.016803},
  publisher = {American Physical Society},
  month = jan
}

@article{chen11a,
  author = {Chen, X. and Gu, Z.-C. and Wen, X.-G.},
  title = {Classification of Gapped Symmetric Phases in One-Dimensional Spin Systems},
  journal = {Physical Review B},
  year = {2011},
  volume = {83},
  number = {3},
  pages = {035107},
  doi = {10.1103/PhysRevB.83.035107},
  url = {http://link.aps.org/doi/10.1103/PhysRevB.83.035107},
  publisher = {American Physical Society},
  month = jan
}

@article{chen11b,
  author = {Chen, X. and Gu, Z.-C. and Wen, X.-G.},
  title = {Complete Classification of One-Dimensional Gapped Quantum Phases in Interacting Spin Systems},
  journal = {Physical Review B},
  year = {2011},
  volume = {84},
  number = {23},
  pages = {235128},
  doi = {10.1103/PhysRevB.84.235128},
  url = {http://link.aps.org/doi/10.1103/PhysRevB.84.235128},
  publisher = {American Physical Society},
  month = dec
}

@article{chen2011two,
  author = {Chen, Xie and Liu, Zheng-Xin and Wen, Xiao-Gang},
  title = {Two-Dimensional Symmetry-Protected Topological Orders and Their Protected Gapless Edge Excitations},
  journal = {Physical Review B},
  year = {2011},
  volume = {84},
  number = {23},
  pages = {235141},
  publisher = {APS}
}

@article{Chen:2011pg,
  author = {Chen, Xie and Gu, Zheng-Cheng and Liu, Zheng-Xin and Wen, Xiao-Gang},
  title = {Symmetry Protected Topological Orders and the Group Cohomology of Their Symmetry Group},
  journal = {Physical Review B},
  year = {2013},
  volume = {87},
  number = {15},
  pages = {155114},
  doi = {10.1103/PhysRevB.87.155114},
  eprint = {1106.4772},
  archiveprefix = {arXiv},
  primaryclass = {cond-mat.str-el}
}

@article{pollmann2012a,
  author = {Pollmann, Frank and Turner, Ari M.},
  title = {Detection of Symmetry-Protected Topological Phases in One Dimension},
  journal = {Physical Review B},
  year = {2012},
  volume = {86},
  number = {12},
  pages = {125441},
  doi = {10.1103/PhysRevB.86.125441},
  url = {https://link.aps.org/doi/10.1103/PhysRevB.86.125441},
  month = sep
}

@article{pollmann2012symmetry,
  author = {Pollmann, Frank and Berg, Erez and Turner, Ari M and Oshikawa, Masaki},
  title = {Symmetry Protection of Topological Phases in One-Dimensional Quantum Spin Systems},
  journal = {Physical review b},
  year = {2012},
  volume = {85},
  number = {7},
  pages = {075125},
  publisher = {APS}
}

@article{else2014classifying,
  author = {Else, Dominic V and Nayak, Chetan},
  title = {Classifying Symmetry-Protected Topological Phases through the Anomalous Action of the Symmetry on the Edge},
  journal = {Physical Review B},
  year = {2014},
  volume = {90},
  number = {23},
  pages = {235137},
  publisher = {APS}
}

@unpublished{lee2023quantum,
  author = {Lee, Jong Yeon and Jian, Chao-Ming and Xu, Cenke},
  title = {Quantum Criticality under Decoherence or Weak Measurement},
  year = {2023},
  eprint = {2301.05238},
  archiveprefix = {arXiv}
}

@unpublished{ma2023topological,
  author = {Ma, Ruochen and Zhang, Jian-Hao and Bi, Zhen and Cheng, Meng and Wang, Chong},
  title = {Topological Phases with Average Symmetries: The Decohered, the Disordered, and the Intrinsic},
  year = {2023},
  eprint = {2305.16399},
  archiveprefix = {arXiv}
}

@article{sarma2025effective,
  author = {Sarma, Abhijat and Bao, Yimu and Myerson-Jain, Nayan and Kiely, Thomas and Xu, Cenke},
  title = {Effective Conformal Field Theory Generated from Pure and Dephased {{Chern}} Insulator},
  journal = {Physical Review B},
  year = {2025},
  volume = {112},
  number = {8},
  pages = {085130},
  publisher = {APS}
}

@unpublished{sala2025entanglement,
  author = {Sala, Pablo and Pollmann, Frank and Oshikawa, Masaki and You, Yizhi},
  title = {Entanglement Holography in Quantum Phases via Twisted {{R}}\'enyi-{{N}} Correlators},
  year = {2025},
  eprint = {2506.10076},
  archiveprefix = {arXiv}
}

@article{wang2024anomaly,
  author = {Wang, Zijian and Li, Linhao},
  title = {Anomaly in Open Quantum Systems and Its Implications on Mixed-State Quantum Phases},
  journal = {PRX Quantum},
  year = {2025},
  volume = {6},
  number = {1},
  pages = {010347},
  doi = {10.1103/PRXQuantum.6.010347},
  url = {https://link.aps.org/doi/10.1103/PRXQuantum.6.010347},
  publisher = {American Physical Society},
  month = mar
}

@misc{wang2023intrinsic,
  author = {Wang, Zijian and Wu, Zhengzhi and Wang, Zhong},
  title = {Intrinsic Mixed-State Topological Order without Quantum Memory},
  year = {2023},
  eprint = {2307.13758},
  archiveprefix = {arXiv},
  primaryclass = {quant-ph}
}

@misc{xu2024average,
  author = {Xu, Yichen and Jian, Chao-Ming},
  title = {Average-Exact Mixed Anomalies and Compatible Phases},
  year = {2024},
  eprint = {2406.07417},
  archiveprefix = {arXiv},
  primaryclass = {cond-mat.str-el},
  url = {https://arxiv.org/abs/2406.07417}
}

@article{you2024intrinsic,
  author = {You, Yizhi and Oshikawa, Masaki},
  title = {Intrinsic Symmetry-Protected Topological Mixed State from Modulated Symmetries and Hierarchical Structure of Boundary Anomaly},
  journal = {Physical Review B},
  year = {2024},
  volume = {110},
  number = {16},
  pages = {165160},
  publisher = {APS}
}

@article{zang2024detecting,
  author = {Zang, Yunlong and Gu, Yingfei and Jiang, Shenghan},
  title = {Detecting Quantum Anomalies in Open Systems},
  journal = {Physical Review Letters},
  year = {2024},
  volume = {133},
  number = {10},
  pages = {106503},
  publisher = {APS}
}

@misc{gu2024spontaneous,
  author = {Gu, Ding and Wang, Zijian and Wang, Zhong},
  title = {Spontaneous Symmetry Breaking in Open Quantum Systems: Strong, Weak, and Strong-to-Weak},
  year = {2024},
  eprint = {2406.19381},
  archiveprefix = {arXiv},
  primaryclass = {quant-ph},
  url = {https://arxiv.org/abs/2406.19381}
}

@misc{guo2024new,
  author = {Guo, Yuchen and Ding, Ke and Yang, Shuo},
  title = {A New Framework for Quantum Phases in Open Systems: {{Steady}} State of Imaginary-Time Lindbladian Evolution},
  year = {2024},
  eprint = {2408.03239},
  archiveprefix = {arXiv},
  primaryclass = {quant-ph},
  url = {https://arxiv.org/abs/2408.03239}
}

@article{guo2025strong,
  author = {Guo, Yuchen and Yang, Shuo},
  title = {Strong-to-Weak Spontaneous Symmetry Breaking Meets Average Symmetry-Protected Topological Order},
  journal = {Physical Review B},
  year = {2025},
  volume = {111},
  number = {20},
  pages = {L201108},
  publisher = {APS}
}

@article{moharramipour2024symmetry,
  author = {Moharramipour, Amin and Lessa, Leonardo A and Wang, Chong and Hsieh, Timothy H and Sahu, Subhayan},
  title = {Symmetry-Enforced Entanglement in Maximally Mixed States},
  journal = {PRX Quantum},
  year = {2024},
  volume = {5},
  number = {4},
  pages = {040336},
  publisher = {APS}
}

@article{salo_steady_state_strong_sym_2025,
  author = {Li, Yahui and Pollmann, Frank and Read, Nicholas and Sala, Pablo},
  title = {Highly Entangled Stationary States from Strong Symmetries},
  journal = {Physical Review X},
  year = {2025},
  volume = {15},
  number = {1},
  pages = {011068},
  doi = {10.1103/PhysRevX.15.011068},
  url = {https://link.aps.org/doi/10.1103/PhysRevX.15.011068},
  publisher = {American Physical Society},
  month = mar
}

@unpublished{liu2025parent,
  author = {Liu, Yuhan and Ruiz-de-Alarcón, Alberto and Styliaris, Georgios and Sun, Xiao-Qi and P\'{e}rez-Garc\'{\i}a, David and Cirac, J Ignacio},
  title = {Parent {{Lindbladians}} for Matrix Product Density Operators},
  year = {2025},
  eprint = {2501.10552},
  archiveprefix = {arXiv}
}

@unpublished{lu2025holographic,
  author = {Lu, Tsung-Cheng and Liu, Yu-Jie and Gopalakrishnan, Sarang and You, Yizhi},
  title = {Holographic Duality between Bulk Topological Order and Boundary Mixed-State Order},
  year = {2025},
  eprint = {2511.19597},
  archiveprefix = {arXiv}
}

@unpublished{bao2023mixed,
  author = {Bao, Yimu and Fan, Ruihua and Vishwanath, Ashvin and Altman, Ehud},
  title = {Mixed-State Topological Order and the Errorfield Double Formulation of Decoherence-Induced Transitions},
  year = {2023},
  eprint = {2301.05687},
  archiveprefix = {arXiv}
}

@unpublished{fan2023diagnostics,
  author = {Fan, Ruihua and Bao, Yimu and Altman, Ehud and Vishwanath, Ashvin},
  title = {Diagnostics of Mixed-State Topological Order and Breakdown of Quantum Memory},
  year = {2023},
  eprint = {2301.05689},
  archiveprefix = {arXiv}
}

@article{LeeYouXu2022,
  author = {Lee, Jong Yeon and Jian, Chao-Ming and Xu, Cenke},
  title = {Quantum Criticality under Decoherence or Weak Measurement},
  journal = {PRX Quantum},
  year = {2023},
  volume = {4},
  number = {3},
  pages = {030317},
  doi = {10.1103/PRXQuantum.4.030317},
  url = {https://link.aps.org/doi/10.1103/PRXQuantum.4.030317},
  publisher = {American Physical Society},
  month = aug
}

@unpublished{zhang2022strange,
  author = {Zhang, Jian-Hao and Qi, Yang and Bi, Zhen},
  title = {Strange Correlation Function for Average Symmetry-Protected Topological Phases},
  year = {2022},
  eprint = {2210.17485},
  archiveprefix = {arXiv}
}

@article{lessa2024mixedstate,
  author = {Lessa, Leonardo A. and Cheng, Meng and Wang, Chong},
  title = {Mixed-State Quantum Anomaly and Multipartite Entanglement},
  journal = {Physical Review X},
  year = {2025},
  volume = {15},
  number = {1},
  pages = {011069},
  doi = {10.1103/PhysRevX.15.011069},
  url = {https://link.aps.org/doi/10.1103/PhysRevX.15.011069},
  publisher = {American Physical Society},
  month = mar
}

@article{huang2025interaction,
  author = {Huang, Ze-Min and Diehl, Sebastian},
  title = {Interaction-Induced Topological Phase Transition at Finite Temperature},
  journal = {Physical Review Letters},
  year = {2025},
  volume = {134},
  number = {5},
  pages = {053002},
  publisher = {APS}
}

@unpublished{ding2024boundary,
  author = {Ding, Ke and Zhang, Hao-Ran and Liu, Bai-Ting and Yang, Shuo},
  title = {Boundary Anomaly Detection in Two-Dimensional Subsystem Symmetry-Protected Topological Phases},
  year = {2024},
  eprint = {2412.07563},
  archiveprefix = {arXiv}
}

@article{hsin2024anomalies,
  author = {Hsin, Po-Shen and Luo, Zhu-Xi and Sun, Hao-Yu},
  title = {Anomalies of Average Symmetries: {{Entanglement}} and Open Quantum Systems},
  journal = {Journal of High Energy Physics},
  year = {2024},
  volume = {2024},
  number = {10},
  pages = {1--42},
  publisher = {Springer}
}

@article{sala2024spontaneous,
  author = {Sala, Pablo and Gopalakrishnan, Sarang and Oshikawa, Masaki and You, Yizhi},
  title = {Spontaneous Strong Symmetry Breaking in Open Systems: {{Purification}} Perspective},
  journal = {Physical Review B},
  year = {2024},
  volume = {110},
  number = {15},
  pages = {155150},
  publisher = {APS}
}

@article{kawabata2024lieb,
  author = {Kawabata, Kohei and Sohal, Ramanjit and Ryu, Shinsei},
  title = {Lieb-{{Schultz-Mattis}} Theorem in Open Quantum Systems},
  journal = {Physical review letters},
  year = {2024},
  volume = {132},
  number = {7},
  pages = {070402},
  publisher = {APS}
}

@article{zhou2025reviving,
  author = {Zhou, Yi-Neng and Li, Xingyu and Zhai, Hui and Li, Chengshu and Gu, Yingfei},
  title = {Reviving the {{Lieb}}–{{Schultz}}–{{Mattis}} Theorem in Open Quantum Systems},
  journal = {National Science Review},
  year = {2025},
  volume = {12},
  number = {1},
  pages = {nwae287},
  publisher = {Oxford University Press}
}

@unpublished{sun2024holographic,
  author = {Sun, Shijun and Zhang, Jian-Hao and Bi, Zhen and You, Yizhi},
  title = {Holographic View of Mixed-State Symmetry-Protected Topological Phases in Open Quantum Systems},
  year = {2024},
  eprint = {2410.08205},
  archiveprefix = {arXiv}
}

@unpublished{ma2022average,
  author = {Ma, Ruochen and Wang, Chong},
  title = {Average Symmetry-Protected Topological Phases},
  year = {2022},
  eprint = {2209.02723},
  archiveprefix = {arXiv}
}

@article{lee2025symmetry,
  author = {Lee, Jong Yeon and You, Yi-Zhuang and Xu, Cenke},
  title = {Symmetry Protected Topological Phases under Decoherence},
  journal = {Quantum},
  year = {2025},
  volume = {9},
  pages = {1607},
  publisher = {Verein zur Förderung des Open Access Publizierens in den Quantenwissenschaften}
}

@article{ma2024topological,
  author = {Ma, Ruochen and Zhang, Jian-Hao and Bi, Zhen and Cheng, Meng and Wang, Chong},
  title = {Topological Phases with Average Symmetries: {{The}} Decohered, the Disordered, and the Intrinsic},
  journal = {Physical Review X},
  year = {2025},
  volume = {15},
  number = {2},
  pages = {021062},
  doi = {10.1103/PhysRevX.15.021062},
  url = {https://link.aps.org/doi/10.1103/PhysRevX.15.021062},
  publisher = {American Physical Society},
  month = may
}

@unpublished{ando2024gauge,
  author = {Ando, Takamasa and Ryu, Shinsei and Watanabe, Masataka},
  title = {Gauge Theory and Mixed State Criticality},
  year = {2024},
  eprint = {2411.04360},
  archiveprefix = {arXiv}
}

@unpublished{zhang2025probing,
  author = {Zhang, Yuxuan and Hsieh, Timothy H and Kim, Yong Baek and Zou, Yijian},
  title = {Probing Mixed-State Phases on a Quantum Computer via {{Renyi}} Correlators and Variational Decoding},
  year = {2025},
  eprint = {2505.02900},
  archiveprefix = {arXiv}
}

@article{murciano2023measurement,
  author = {Murciano, Sara and Sala, Pablo and Liu, Yue and Mong, Roger SK and Alicea, Jason},
  title = {Measurement-Altered {{Ising}} Quantum Criticality},
  journal = {Physical Review X},
  year = {2023},
  volume = {13},
  number = {4},
  pages = {041042},
  publisher = {APS}
}

@article{garratt2023measurements,
  author = {Garratt, Samuel J and Weinstein, Zack and Altman, Ehud},
  title = {Measurements Conspire Nonlocally to Restructure Critical Quantum States},
  journal = {Physical Review X},
  year = {2023},
  volume = {13},
  number = {2},
  pages = {021026},
  publisher = {APS}
}

@article{briegel2009measurement,
  author = {Briegel, Hans J and Browne, David E and Dür, Wolfgang and Raussendorf, Robert and Van den Nest, Maarten},
  title = {Measurement-Based Quantum Computation},
  journal = {Nature Physics},
  year = {2009},
  volume = {5},
  number = {1},
  pages = {19--26},
  publisher = {Nature Publishing Group UK London}
}

@unpublished{lee2022measurement,
  author = {Lee, Jong Yeon and Ji, Wenjie and Bi, Zhen and Fisher, Matthew},
  title = {Measurement-{{Prepared Quantum Criticality}}: From {{Ising}} Model to Gauge Theory, and Beyond},
  year = {2022},
  eprint = {2208.11699},
  archiveprefix = {arXiv}
}

@article{raussendorf2003measurement,
  author = {Raussendorf, Robert and Browne, Daniel E and Briegel, Hans J},
  title = {Measurement-Based Quantum Computation on Cluster States},
  journal = {Physical review A},
  year = {2003},
  volume = {68},
  number = {2},
  pages = {022312},
  publisher = {APS}
}

@article{Stephen_2017,
  author = {Stephen, David T. and Wang, Dong-Sheng and Prakash, Abhishodh and Wei, Tzu-Chieh and Raussendorf, Robert},
  title = {Computational Power of Symmetry-Protected Topological Phases},
  journal = {Physical Review Letters},
  year = {2017},
  volume = {119},
  number = {1},
  pages = {010504},
  doi = {10.1103/PhysRevLett.119.010504},
  publisher = {American Physical Society (APS)},
  month = jul
}

@article{Lee_2023,
  author = {Lee, Jong Yeon and Jian, Chao-Ming and Xu, Cenke},
  title = {Quantum Criticality under Decoherence or Weak Measurement},
  journal = {PRX Quantum},
  year = {2023},
  volume = {4},
  number = {3},
  doi = {10.1103/prxquantum.4.030317},
  url = {http://dx.doi.org/10.1103/PRXQuantum.4.030317},
  publisher = {American Physical Society (APS)},
  month = aug
}

@article{lessa2025,
  author = {Lessa, Leonardo A. and Ma, Ruochen and Zhang, Jian-Hao and Bi, Zhen and Cheng, Meng and Wang, Chong},
  title = {Strong-to-{{Weak Spontaneous Symmetry Breaking}} in {{Mixed Quantum States}}},
  journal = {PRX Quantum},
  year = {2025},
  volume = {6},
  number = {1},
  pages = {010344},
  doi = {10.1103/PRXQuantum.6.010344},
  url = {https://link.aps.org/doi/10.1103/PRXQuantum.6.010344},
  month = mar
}

@unpublished{song2025strong,
  author = {Song, Zijian and Zhang, Jian-Hao},
  title = {Strong-to-Weak Spontaneous Symmetry Breaking of Higher-Form Non-Invertible Symmetries in {{Kitaev}}'s Quantum Double Model},
  year = {2025},
  eprint = {2509.24179},
  archiveprefix = {arXiv}
}

@article{zhang2024strong,
  author = {Zhang, Carolyn and Xu, Yichen and Zhang, Jian-Hao and Xu, Cenke and Bi, Zhen and Luo, Zhu-Xi},
  title = {Strong-to-Weak Spontaneous Breaking of 1-Form Symmetry and Intrinsically Mixed Topological Order},
  journal = {Physical Review B},
  year = {2025},
  volume = {111},
  number = {11},
  pages = {115137},
  doi = {10.1103/PhysRevB.111.115137},
  url = {https://link.aps.org/doi/10.1103/PhysRevB.111.115137},
  publisher = {American Physical Society},
  month = mar
}

@article{lee2025,
  author = {Lee, Jong Yeon and You, Yi-Zhuang and Xu, Cenke},
  title = {Symmetry Protected Topological Phases under Decoherence},
  journal = {Quantum},
  year = {2025},
  volume = {9},
  pages = {1607},
  doi = {10.22331/q-2025-01-23-1607},
  url = {https://quantum-journal.org/papers/q-2025-01-23-1607/},
  month = jan
}

@unpublished{sang2023mixed,
  author = {Sang, Shengqi and Zou, Yijian and Hsieh, Timothy H},
  title = {Mixed-State Quantum Phases: {{Renormalization}} and Quantum Error Correction},
  year = {2023},
  eprint = {2310.08639},
  archiveprefix = {arXiv}
}

@unpublished{sang2025mixed,
  author = {Sang, Shengqi and Lessa, Leonardo A and Mong, Roger SK and Grover, Tarun and Wang, Chong and Hsieh, Timothy H},
  title = {Mixed-State Phases from Local Reversibility},
  year = {2025},
  eprint = {2507.02292},
  archiveprefix = {arXiv}
}

@article{su2025spin,
  author = {Su, Kaixiang and Sarma, Abhijat and Bintz, Marcus and Kiely, Thomas and Bao, Yimu and Fisher, Matthew PA and Xu, Cenke},
  title = {Spin Liquid and Superconductivity Emerging from Steady States and Measurements},
  journal = {Physical Review Letters},
  year = {2025},
  volume = {135},
  number = {5},
  pages = {050403},
  publisher = {APS}
}

@article{chen2014,
  author = {Chen, Xie and Lu, Yuan-Ming and Vishwanath, Ashvin},
  title = {Symmetry-Protected Topological Phases from Decorated Domain Walls},
  journal = {Nature Communications},
  year = {2014},
  volume = {5},
  number = {1},
  pages = {3507},
  doi = {10.1038/ncomms4507},
  url = {https://www.nature.com/articles/ncomms4507},
  month = mar
}

@article{you2016decorated,
  author = {You, Yizhi},
  title = {Decorated Defect Condensate: {{A}} Window to Unconventional Quantum Phases in {{Weyl}} Semimetals},
  journal = {Physical Review B},
  year = {2016},
  volume = {94},
  number = {19},
  pages = {195112},
  publisher = {APS}
}

@article{li2024,
  author = {Li, Linhao and Oshikawa, Masaki and Zheng, Yunqin},
  title = {Decorated Defect Construction of Gapless-{{SPT}} States},
  journal = {SciPost Physics},
  year = {2024},
  volume = {17},
  number = {1},
  pages = {013},
  doi = {10.21468/SciPostPhys.17.1.013},
  url = {https://scipost.org/10.21468/SciPostPhys.17.1.013},
  month = jul
}

@article{ma2023average,
  author = {Ma, Ruochen and Wang, Chong},
  title = {Average Symmetry-Protected Topological Phases},
  journal = {Physical Review X},
  year = {2023},
  volume = {13},
  number = {3},
  pages = {031016},
  publisher = {APS}
}

@unpublished{SpectralSequence,
  author = {Wang, Qing-Rui and Ning, Shang-Qiang and Cheng, Meng},
  title = {Domain Wall Decorations, Anomalies and Spectral Sequences in Bosonic Topological Phases},
  year = {2021},
  eprint = {2104.13233},
  archiveprefix = {arXiv},
  primaryclass = {cond-mat.str-el},
  month = apr
}

@article{lu2022measurement,
  author = {Lu, Tsung-Cheng and Lessa, Leonardo A and Kim, Isaac H and Hsieh, Timothy H},
  title = {Measurement as a Shortcut to Long-Range Entangled Quantum Matter},
  journal = {PRX Quantum},
  year = {2022},
  volume = {3},
  number = {4},
  pages = {040337},
  publisher = {APS}
}

@article{verresen2021prediction,
  author = {Verresen, Ruben and Lukin, Mikhail D and Vishwanath, Ashvin},
  title = {Prediction of Toric Code Topological Order from {{Rydberg}} Blockade},
  journal = {Physical Review X},
  year = {2021},
  volume = {11},
  number = {3},
  pages = {031005},
  publisher = {APS}
}

@unpublished{verresen2021efficiently,
  author = {Verresen, Ruben and Tantivasadakarn, Nathanan and Vishwanath, Ashvin},
  title = {Efficiently Preparing Schr{\"o}dinger's Cat, Fractons and Non-{{Abelian}} Topological Order in Quantum Devices},
  year = {2021},
  eprint = {2112.03061},
  archiveprefix = {arXiv}
}

@article{tantivasadakarn2024,
  author = {Tantivasadakarn, Nathanan and Thorngren, Ryan and Vishwanath, Ashvin and Verresen, Ruben},
  title = {Long-{{Range Entanglement}} from {{Measuring Symmetry-Protected Topological Phases}}},
  journal = {Physical Review X},
  year = {2024},
  volume = {14},
  number = {2},
  pages = {021040},
  doi = {10.1103/PhysRevX.14.021040},
  url = {https://link.aps.org/doi/10.1103/PhysRevX.14.021040},
  month = jun
}

@misc{lee2022,
  author = {Lee, Jong Yeon and Ji, Wenjie and Bi, Zhen and Fisher, Matthew P. A.},
  title = {Decoding {{Measurement-Prepared Quantum Phases}} and {{Transitions}}: {{From Ising}} Model to Gauge Theory, and Beyond},
  year = {2022},
  doi = {10.48550/ARXIV.2208.11699},
  url = {https://arxiv.org/abs/2208.11699}
}

@misc{zonzo2018,
  author = {Zonzo, G and Giampaolo, S M},
  title = {$N$ -Cluster Models in a Transverse Magnetic Field},
  year = {2018},
  volume = {2018},
  number = {6},
  pages = {063103},
  doi = {10.1088/1742-5468/aac443},
  url = {https://iopscience.iop.org/article/10.1088/1742-5468/aac443}
}

@article{you2014,
  author = {You, Yi-Zhuang and Bi, Zhen and Rasmussen, Alex and Slagle, Kevin and Xu, Cenke},
  title = {Wave {{Function}} and {{Strange Correlator}} of {{Short-Range Entangled States}}},
  journal = {Physical Review Letters},
  year = {2014},
  volume = {112},
  number = {24},
  pages = {247202},
  doi = {10.1103/PhysRevLett.112.247202},
  url = {https://link.aps.org/doi/10.1103/PhysRevLett.112.247202},
  month = jun
}

@article{lepori2023strange,
  author = {Lepori, Luca and Burrello, Michele and Trombettoni, Andrea and Paganelli, Simone},
  title = {Strange Correlators for Topological Quantum Systems from Bulk-Boundary Correspondence},
  journal = {Physical Review B},
  year = {2023},
  volume = {108},
  number = {3},
  pages = {035110},
  publisher = {APS}
}

@misc{sang2023mixedstate,
  author = {Sang, Shengqi and Zou, Yijian and Hsieh, Timothy H.},
  title = {Mixed-State Quantum Phases: {{Renormalization}} and Quantum Error Correction},
  year = {2023},
  eprint = {2310.08639},
  archiveprefix = {arXiv},
  primaryclass = {quant-ph}
}

@unpublished{chen2023separability,
  author = {Chen, Yu-Hsueh and Grover, Tarun},
  title = {Separability Transitions in Topological States Induced by Local Decoherence},
  year = {2023},
  eprint = {2309.11879},
  archiveprefix = {arXiv}
}

@article{chen2023symmetryenforced,
  author = {Chen, Yu-Hsueh and Grover, Tarun},
  title = {Symmetry-Enforced Many-Body Separability Transitions},
  journal = {PRX Quantum},
  year = {2024},
  volume = {5},
  number = {3},
  pages = {030310},
  doi = {10.1103/PRXQuantum.5.030310},
  url = {https://link.aps.org/doi/10.1103/PRXQuantum.5.030310},
  publisher = {American Physical Society},
  month = jul
}

@misc{lessa2024strong,
  author = {Lessa, Leonardo A. and Ma, Ruochen and Zhang, Jian-Hao and Bi, Zhen and Cheng, Meng and Wang, Chong},
  title = {Strong-to-Weak Spontaneous Symmetry Breaking in Mixed Quantum States},
  year = {2024},
  eprint = {2405.03639},
  archiveprefix = {arXiv},
  primaryclass = {quant-ph},
  url = {https://arxiv.org/abs/2405.03639}
}

@article{piroli2021quantum,
  author = {Piroli, Lorenzo and Styliaris, Georgios and Cirac, J Ignacio},
  title = {Quantum Circuits Assisted by Local Operations and Classical Communication: {{Transformations}} and Phases of Matter},
  journal = {Physical Review Letters},
  year = {2021},
  volume = {127},
  number = {22},
  pages = {220503},
  publisher = {APS}
}

@unpublished{tantivasadakarn2021long,
  author = {Tantivasadakarn, Nathanan and Thorngren, Ryan and Vishwanath, Ashvin and Verresen, Ruben},
  title = {Long-Range Entanglement from Measuring Symmetry-Protected Topological Phases},
  year = {2021},
  eprint = {2112.01519},
  archiveprefix = {arXiv}
}

@unpublished{bravyi2022adaptive,
  author = {Bravyi, Sergey and Kim, Isaac and Kliesch, Alexander and Koenig, Robert},
  title = {Adaptive Constant-Depth Circuits for Manipulating Non-Abelian Anyons},
  year = {2022},
  eprint = {2205.01933},
  archiveprefix = {arXiv}
}

@article{tantivasadakarn2023hierarchy,
  author = {Tantivasadakarn, Nathanan and Vishwanath, Ashvin and Verresen, Ruben},
  title = {Hierarchy of Topological Order from Finite-Depth Unitaries, Measurement, and Feedforward},
  journal = {PRX Quantum},
  year = {2023},
  volume = {4},
  number = {2},
  pages = {020339},
  publisher = {APS}
}

@unpublished{iqbal2023creation,
  author = {Iqbal, Mohsin and Tantivasadakarn, Nathanan and Verresen, Ruben and Campbell, Sara L and Dreiling, Joan M and Figgatt, Caroline and Gaebler, John P and Johansen, Jacob and Mills, Michael and Moses, Steven A and others},
  title = {Creation of Non-Abelian Topological Order and Anyons on a Trapped-Ion Processor},
  year = {2023},
  eprint = {2305.03766},
  archiveprefix = {arXiv}
}

@unpublished{foss2023experimental,
  author = {Foss-Feig, Michael and Tikku, Arkin and Lu, Tsung-Cheng and Mayer, Karl and Iqbal, Mohsin and Gatterman, Thomas M and Gerber, Justin A and Gilmore, Kevin and Gresh, Dan and Hankin, Aaron and others},
  title = {Experimental Demonstration of the Advantage of Adaptive Quantum Circuits},
  year = {2023},
  eprint = {2302.03029},
  archiveprefix = {arXiv}
}

@unpublished{iqbal2023topological,
  author = {Iqbal, Mohsin and Tantivasadakarn, Nathanan and Gatterman, Thomas M and Gerber, Justin A and Gilmore, Kevin and Gresh, Dan and Hankin, Aaron and Hewitt, Nathan and Horst, Chandler V and Matheny, Mitchell and others},
  title = {Topological Order from Measurements and Feed-Forward on a Trapped Ion Quantum Computer},
  year = {2023},
  eprint = {2302.01917},
  archiveprefix = {arXiv}
}

@article{satzinger2021realizing,
  author = {family=Satzinger, given=KJ, given-i=KJ and Liu, Y-J and Smith, A and Knapp, C and Newman, M and Jones, C and Chen, Z and Quintana, C and Mi, X and Dunsworth, A and others},
  title = {Realizing Topologically Ordered States on a Quantum Processor},
  journal = {Science},
  year = {2021},
  volume = {374},
  number = {6572},
  pages = {1237--1241},
  publisher = {American Association for the Advancement of Science}
}

@article{seiberg2024majorana,
  author = {Seiberg, Nathan and Shao, Shu-Heng},
  title = {Majorana Chain and {{Ising}} Model-(Non-Invertible) Translations, Anomalies, and Emanant Symmetries},
  journal = {SciPost Physics},
  year = {2024},
  volume = {16},
  number = {3},
  pages = {064}
}

@article{seifnashri2024cluster,
  author = {Seifnashri, Sahand and Shao, Shu-Heng},
  title = {Cluster State as a Noninvertible Symmetry-Protected Topological Phase},
  journal = {Physical Review Letters},
  year = {2024},
  volume = {133},
  number = {11},
  pages = {116601},
  publisher = {APS}
}

@unpublished{kim2025noninvertible,
  author = {Kim, Jintae and You, Yizhi and Han, Jung Hoon},
  title = {Noninvertible Symmetry and Topological Holography for Modulated {{SPT}} in One Dimension},
  year = {2025},
  eprint = {2507.02324},
  archiveprefix = {arXiv}
}

@article{guo2025quantum,
  author = {Guo, Yuxuan and Yang, Sheng and Yu, Xue-Jia},
  title = {Quantum Strong-to-Weak Spontaneous Symmetry Breaking in Decohered One-Dimensional Critical States},
  journal = {PRX Quantum},
  year = {2025},
  volume = {6},
  number = {4},
  pages = {040311},
  publisher = {APS}
}

@article{haldane1983,
  author = {Haldane, F. D. M.},
  title = {Nonlinear {{Field Theory}} of {{Large-Spin Heisenberg Antiferromagnets}}: {{Semiclassically Quantized Solitons}} of the {{One-Dimensional Easy-Axis N\'{e}el State}}},
  journal = {Physical Review Letters},
  year = {1983},
  volume = {50},
  number = {15},
  pages = {1153--1156},
  doi = {10.1103/PhysRevLett.50.1153},
  url = {https://link.aps.org/doi/10.1103/PhysRevLett.50.1153},
  month = apr
}

@article{haldane1983a,
  author = {Haldane, F.D.M.},
  title = {Continuum Dynamics of the 1-{{D Heisenberg}} Antiferromagnet: {{Identification}} with the {{O}}(3) Nonlinear Sigma Model},
  journal = {Physics Letters A},
  year = {1983},
  volume = {93},
  number = {9},
  pages = {464--468},
  doi = {10.1016/0375-9601(83)90631-X},
  url = {https://linkinghub.elsevier.com/retrieve/pii/037596018390631X},
  month = feb
}

@article{hu2011,
  author = {Hu, Shijie and Normand, B. and Wang, Xiaoqun and Yu, Lu},
  title = {Accurate Determination of the {{Gaussian}} Transition in Spin-1 Chains with Single-Ion Anisotropy},
  journal = {Physical Review B},
  year = {2011},
  volume = {84},
  number = {22},
  pages = {220402},
  doi = {10.1103/PhysRevB.84.220402},
  url = {https://link.aps.org/doi/10.1103/PhysRevB.84.220402},
  month = dec
}

@unpublished{wang2025fractional,
  author = {Wang, Zijian and Fan, Ruihua and Wang, Tianle and Garratt, Samuel J and Altman, Ehud},
  title = {Fractional Quantum {{Hall}} States under Density Decoherence},
  year = {2025},
  eprint = {2510.08490},
  archiveprefix = {arXiv}
}

@article{klein2020nonlocal,
  author = {Klein Kvorning, Thomas and Spånslätt, Christian and Chan, AtMa PO and Ryu, Shinsei},
  title = {Nonlocal Order Parameters for States with Topological Electromagnetic Response},
  journal = {Physical Review B},
  year = {2020},
  volume = {101},
  number = {20},
  pages = {205101},
  publisher = {APS}
}

@article{lu2012theory,
  author = {Lu, Yuan-Ming and Vishwanath, Ashvin},
  title = {Theory and Classification of Interacting Integer Topological Phases in Two Dimensions: {{A Chern-Simons}} Approach},
  journal = {Physical Review B},
  year = {2012},
  volume = {86},
  number = {12},
  pages = {125119},
  publisher = {APS}
}

@article{haldane1988a,
  author = {Haldane, F. D. M.},
  title = {Model for a {{Quantum Hall Effect}} without {{Landau Levels}}: {{Condensed-Matter Realization}} of the "{{Parity Anomaly}}"},
  journal = {Physical Review Letters},
  year = {1988},
  volume = {61},
  number = {18},
  pages = {2015--2018},
  doi = {10.1103/PhysRevLett.61.2015},
  url = {https://link.aps.org/doi/10.1103/PhysRevLett.61.2015},
  month = oct
}

@article{thonhauser2006,
  author = {Thonhauser, T. and Vanderbilt, David},
  title = {Insulator/{{Chern-insulator}} Transition in the {{Haldane}} Model},
  journal = {Physical Review B},
  year = {2006},
  volume = {74},
  number = {23},
  pages = {235111},
  doi = {10.1103/PhysRevB.74.235111},
  url = {https://link.aps.org/doi/10.1103/PhysRevB.74.235111},
  month = dec
}

@article{kleinkvorning2020,
  author = {Klein Kvorning, Thomas and Spånslätt, Christian and Chan, AtMa P. O. and Ryu, Shinsei},
  title = {Nonlocal Order Parameters for States with Topological Electromagnetic Response},
  journal = {Physical Review B},
  year = {2020},
  volume = {101},
  number = {20},
  pages = {205101},
  doi = {10.1103/PhysRevB.101.205101},
  url = {https://link.aps.org/doi/10.1103/PhysRevB.101.205101},
  month = may
}

@article{zhang1992chern,
  author = {Zhang, Shou Cheng},
  title = {The {{Chern}}–{{Simons}}–{{Landau}}–{{Ginzburg}} Theory of the Fractional Quantum {{Hall}} Effect},
  journal = {International Journal of Modern Physics B},
  year = {1992},
  volume = {6},
  number = {01},
  pages = {25--58},
  publisher = {World Scientific}
}

@unpublished{verresen2022higgs,
  author = {Verresen, Ruben and Borla, Umberto and Vishwanath, Ashvin and Moroz, Sergej and Thorngren, Ryan},
  title = {Higgs Condensates Are Symmetry-Protected Topological Phases: {{I}}. Discrete Symmetries},
  year = {2022},
  eprint = {2211.01376},
  archiveprefix = {arXiv}
}

@article{su2024higher,
  author = {Su, Kaixiang and Myerson-Jain, Nayan and Wang, Chong and Jian, Chao-Ming and Xu, Cenke},
  title = {Higher-Form Symmetries under Weak Measurement},
  journal = {Physical Review Letters},
  year = {2024},
  volume = {132},
  number = {20},
  pages = {200402},
  publisher = {APS}
}

@article{jian2021physics,
  author = {Jian, Chao-Ming and Wu, Xiao-Chuan and Xu, Yichen and Xu, Cenke},
  title = {Physics of Symmetry Protected Topological Phases Involving Higher Symmetries and Its Applications},
  journal = {Physical Review B},
  year = {2021},
  volume = {103},
  number = {6},
  pages = {064426},
  publisher = {APS}
}

@article{roberts2020symmetry,
  author = {Roberts, Sam and Bartlett, Stephen D},
  title = {Symmetry-Protected Self-Correcting Quantum Memories},
  journal = {Physical Review X},
  year = {2020},
  volume = {10},
  number = {3},
  pages = {031041},
  publisher = {APS}
}

@unpublished{hauser2025information,
  author = {Hauser, Jacob and Lavasani, Ali and Vijay, Sagar and Fisher, Matthew},
  title = {Information Dynamics and Symmetry Breaking in Generic Monitored {{Z}}$_2$-Symmetric Open Quantum Systems},
  year = {2025},
  eprint = {2512.03031},
  archiveprefix = {arXiv}
}

@article{zhu2023nishimori,
  author = {Zhu, Guo-Yi and Tantivasadakarn, Nathanan and Vishwanath, Ashvin and Trebst, Simon and Verresen, Ruben},
  title = {Nishimori's Cat: {{Stable}} Long-Range Entanglement from Finite-Depth Unitaries and Weak Measurements},
  journal = {Physical Review Letters},
  year = {2023},
  volume = {131},
  number = {20},
  pages = {200201},
  doi = {10.1103/PhysRevLett.131.200201},
  url = {https://link.aps.org/doi/10.1103/PhysRevLett.131.200201},
  publisher = {American Physical Society},
  month = nov
}

@misc{fan2024diagnostics,
  author = {Fan, Ruihua and Bao, Yimu and Altman, Ehud and Vishwanath, Ashvin},
  title = {Diagnostics of Mixed-State Topological Order and Breakdown of Quantum Memory},
  year = {2023},
  eprint = {2301.05689},
  archiveprefix = {arXiv},
  primaryclass = {quant-ph}
}

@misc{lee2022decoding,
  author = {Lee, Jong Yeon and Ji, Wenjie and Bi, Zhen and Fisher, Matthew P. A.},
  title = {Decoding Measurement-Prepared Quantum Phases and Transitions: From Ising Model to Gauge Theory, and Beyond},
  year = {2022},
  eprint = {2208.11699},
  archiveprefix = {arXiv},
  primaryclass = {cond-mat.str-el}
}

@article{gaplessSPT,
  author = {Thorngren, Ryan and Vishwanath, Ashvin and Verresen, Ruben},
  title = {Intrinsically Gapless Topological Phases},
  journal = {Physical Review B},
  year = {2021},
  volume = {104},
  number = {7},
  pages = {075132},
  doi = {10.1103/PhysRevB.104.075132},
  publisher = {American Physical Society (APS)},
  month = aug
}

@article{scaffidi2017gapless,
  author = {Scaffidi, Thomas and Parker, Daniel E and Vasseur, Romain},
  title = {Gapless Symmetry-Protected Topological Order},
  journal = {Physical Review X},
  year = {2017},
  volume = {7},
  number = {4},
  pages = {041048},
  publisher = {APS}
}

@unpublished{li2023intrinsically,
  author = {Li, Linhao and Oshikawa, Masaki and Zheng, Yunqin},
  title = {Intrinsically/Purely Gapless-Spt from Non-Invertible Duality Transformations},
  year = {2023},
  eprint = {2307.04788},
  archiveprefix = {arXiv}
}

@article{yang2023entanglement,
  author = {Yang, Zhou and Mao, Dan and Jian, Chao-Ming},
  title = {Entanglement in a One-Dimensional Critical State after Measurements},
  journal = {Physical Review B},
  year = {2023},
  volume = {108},
  number = {16},
  pages = {165120},
  publisher = {APS}
}

@article{guo2025,
  author = {Guo, Yuxuan and Yang, Sheng and Yu, Xue-Jia},
  title = {Quantum {{Strong-To-Weak Spontaneous Symmetry Breaking}} in {{Decohered One-Dimensional Critical States}}},
  journal = {PRX Quantum},
  year = {2025},
  volume = {6},
  number = {4},
  pages = {040311},
  doi = {10.1103/4vs5-l54f},
  url = {https://link.aps.org/doi/10.1103/4vs5-l54f},
  month = oct
}

@article{li2025,
  author = {Li, Linhao and Oshikawa, Masaki and Zheng, Yunqin},
  title = {Intrinsically/Purely Gapless-{{SPT}} from Non-Invertible Duality Transformations},
  journal = {SciPost Physics},
  year = {2025},
  volume = {18},
  number = {5},
  pages = {153},
  doi = {10.21468/SciPostPhys.18.5.153},
  url = {https://scipost.org/10.21468/SciPostPhys.18.5.153},
  month = may
}

@article{levin2012,
  author = {Levin, Michael and Gu, Zheng-Cheng},
  title = {Braiding Statistics Approach to Symmetry-Protected Topological Phases},
  journal = {Physical Review B},
  year = {2012},
  volume = {86},
  number = {11},
  pages = {115109},
  doi = {10.1103/PhysRevB.86.115109},
  url = {https://link.aps.org/doi/10.1103/PhysRevB.86.115109},
  month = sep
}

@article{ma2025,
  author = {Ma, Ruochen and Zhang, Jian-Hao and Bi, Zhen and Cheng, Meng and Wang, Chong},
  title = {Topological {{Phases}} with {{Average Symmetries}}: {{The Decohered}}, the {{Disordered}}, and the {{Intrinsic}}},
  journal = {Physical Review X},
  year = {2025},
  volume = {15},
  number = {2},
  pages = {021062},
  doi = {10.1103/PhysRevX.15.021062},
  url = {https://link.aps.org/doi/10.1103/PhysRevX.15.021062},
  month = may
}

@article{thorngren2021,
  author = {Thorngren, Ryan and Vishwanath, Ashvin and Verresen, Ruben},
  title = {Intrinsically Gapless Topological Phases},
  journal = {Physical Review B},
  year = {2021},
  volume = {104},
  number = {7},
  pages = {075132},
  doi = {10.1103/PhysRevB.104.075132},
  url = {https://link.aps.org/doi/10.1103/PhysRevB.104.075132},
  month = aug
}

@article{sun2025,
  author = {Sun, Shijun and Zhang, Jian-Hao and Bi, Zhen and You, Yizhi},
  title = {Holographic {{View}} of {{Mixed-State Symmetry-Protected Topological Phases}} in {{Open Quantum Systems}}},
  journal = {PRX Quantum},
  year = {2025},
  volume = {6},
  number = {2},
  pages = {020333},
  doi = {10.1103/PRXQuantum.6.020333},
  url = {https://link.aps.org/doi/10.1103/PRXQuantum.6.020333},
  month = may
}

@article{you2024,
  author = {You, Yizhi and Oshikawa, Masaki},
  title = {Intrinsic Symmetry-Protected Topological Mixed State from Modulated Symmetries and Hierarchical Structure of Boundary Anomaly},
  journal = {Physical Review B},
  year = {2024},
  volume = {110},
  number = {16},
  pages = {165160},
  doi = {10.1103/PhysRevB.110.165160},
  url = {https://link.aps.org/doi/10.1103/PhysRevB.110.165160},
  month = oct
}

@misc{schollwock2011,
  author = {Schollwöck, Ulrich},
  title = {The Density-Matrix Renormalization Group in the Age of Matrix Product States},
  year = {2011},
  volume = {326},
  number = {1},
  pages = {96--192},
  doi = {10.1016/j.aop.2010.09.012},
  url = {https://linkinghub.elsevier.com/retrieve/pii/S0003491610001752}
}

@misc{schollwock2005,
  author = {Schollwöck, U.},
  title = {The Density-Matrix Renormalization Group},
  year = {2005},
  volume = {77},
  number = {1},
  pages = {259--315},
  doi = {10.1103/RevModPhys.77.259},
  url = {https://link.aps.org/doi/10.1103/RevModPhys.77.259}
}

@misc{ostlund1995,
  author = {Östlund, Stellan and Rommer, Stefan},
  title = {Thermodynamic {{Limit}} of {{Density Matrix Renormalization}}},
  year = {1995},
  volume = {75},
  number = {19},
  pages = {3537--3540},
  doi = {10.1103/PhysRevLett.75.3537},
  url = {https://link.aps.org/doi/10.1103/PhysRevLett.75.3537}
}

@article{lu2023mixed,
  author = {Lu, Tsung-Cheng and Zhang, Zhehao and Vijay, Sagar and Hsieh, Timothy H.},
  title = {Mixed-State Long-Range Order and Criticality from Measurement and Feedback},
  journal = {PRX Quantum},
  year = {2023},
  volume = {4},
  number = {3},
  pages = {030318},
  doi = {10.1103/PRXQuantum.4.030318},
  url = {https://link.aps.org/doi/10.1103/PRXQuantum.4.030318},
  publisher = {American Physical Society},
  month = aug
}

@unpublished{zhang2024characterizing,
  author = {Zhang, Yifan and Gopalakrishnan, Sarang and Styliaris, Georgios},
  title = {Characterizing {{MPS}} and {{PEPS}} Preparable via Measurement and Feedback},
  year = {2024},
  eprint = {2405.09615},
  archiveprefix = {arXiv}
}

@article{sahay2025classifying,
  author = {Sahay, Rahul and Verresen, Ruben},
  title = {Classifying one-dimensional quantum states prepared by a single round of measurements},
  journal = {PRX Quantum},
  year = {2025},
  volume = {6},
  number = {1},
  pages = {010329},
  publisher = {APS}
}

@article{lam2024classification,
  author = {Lam, Ho Tat},
  title = {Classification of Dipolar Symmetry-Protected Topological Phases: {{Matrix}} Product States, Stabilizer {{Hamiltonians}}, and Finite Tensor Gauge Theories},
  journal = {Physical Review B},
  year = {2024},
  volume = {109},
  number = {11},
  pages = {115142},
  publisher = {APS}
}

@unpublished{lam2024topological,
  author = {Lam, Ho Tat and Han, Jung Hoon and You, Yizhi},
  title = {Topological Dipole Insulator},
  year = {2024},
  eprint = {2403.13880},
  archiveprefix = {arXiv}
}

@unpublished{delfino2023anyon,
  author = {Delfino, Guilherme and You, Yizhi},
  title = {Anyon Condensation Web and Multipartite Entanglement in {{2D}} Fracton-like Theories},
  year = {2023},
  eprint = {2310.09490},
  archiveprefix = {arXiv}
}

@unpublished{delfino20232d,
  author = {Delfino, Guilherme and Chamon, Claudio and You, Yizhi},
  title = {{{2D}} Fractons from Gauging Exponential Symmetries},
  year = {2023},
  eprint = {2306.17121},
  archiveprefix = {arXiv}
}

@article{han2024topological,
  author = {Han, Jung Hoon and Lake, Ethan and Lam, Ho Tat and Verresen, Ruben and You, Yizhi},
  title = {Topological Quantum Chains Protected by Dipolar and Other Modulated Symmetries},
  journal = {Physical Review B},
  year = {2024},
  volume = {109},
  number = {12},
  pages = {125121},
  publisher = {APS}
}

@unpublished{yao2010symmetry,
  author = {Yao, Hong and Fu, Liang and Qi, Xiao-Liang},
  title = {Symmetry Fractional Quantization in Two Dimensions},
  year = {2010},
  eprint = {1012.4470},
  archiveprefix = {arXiv}
}

@article{li2014topology,
  author = {Li, Wei and Yang, Shuo and Cheng, Meng and Liu, Zheng-Xin and Tu, Hong-Hao},
  title = {Topology and Criticality in the Resonating {{Affleck-Kennedy-Lieb-Tasaki}} Loop Spin Liquid States},
  journal = {Physical Review B},
  year = {2014},
  volume = {89},
  number = {17},
  pages = {174411},
  publisher = {APS}
}

@unpublished{savary2015quantum,
  author = {Savary, Lucile},
  title = {Quantum Loop States in Spin-Orbital Models on the Honeycomb Lattice},
  year = {2015},
  eprint = {1511.01505},
  archiveprefix = {arXiv}
}

@misc{lu2025a,
  author = {Lu, Shuangyuan},
  title = {{{SPT}}\_{{SWSSB}}\_data},
  year = {2025},
  doi = {10.5281/ZENODO.17252759},
  url = {https://zenodo.org/doi/10.5281/zenodo.17252759},
  publisher = {Zenodo},
  month = dec
}

@misc{shuangyuanlu2025,
  author = {Lu, Shuangyuan},
  title = {{{ShuangyuanLu}}/{{DMRG}}\_mixed\_phase\_transition},
  year = {2025},
  doi = {10.5281/ZENODO.18014706},
  url = {https://zenodo.org/doi/10.5281/zenodo.18014706},
  month = dec,
  organization = {Zenodo}
}

@misc{shuangyuanlu2025a,
  author = {Lu, Shuangyuan},
  title = {{{ShuangyuanLu}}/{{QSHE}}\_{{MC}}},
  year = {2025},
  doi = {10.5281/ZENODO.18014502},
  url = {https://zenodo.org/doi/10.5281/zenodo.18014502},
  month = dec,
  organization = {Zenodo}
}

@misc{shuangyuanlu2026,
  author = {Lu, Shuangyuan},
  title = {{{ShuangyuanLu}}/{{MC}}\_{{Qauntum}}\_{{Sampling}}: {{V0}}.1.0},
  year = {2026},
  doi = {10.5281/ZENODO.18112471},
  url = {https://zenodo.org/doi/10.5281/zenodo.18112471},
  month = jan,
  organization = {Zenodo}
}

@article{levin05,
  author = {Levin, Michael and Gu, Zheng-Cheng},
  title = {Braiding Statistics Approach to Symmetry-Protected Topological Phases},
  journal = {Physical Review B},
  year = {2012},
  volume = {86},
  number = {11},
  pages = {115109},
  doi = {10.1103/PhysRevB.86.115109},
  url = {https://link.aps.org/doi/10.1103/PhysRevB.86.115109},
  publisher = {American Physical Society},
  month = sep
}

\end{document}